\documentclass[trackchanges]{aastex701}
\def\apjl{ApJ}
\def\apjs{ApJS}
\def\ao{Appl.~Opt.}
\def\aap{A\&A}

\newcommand{\be}{\begin{equation}}
\newcommand{\ee}{\end{equation}}
\newcommand{\bary}{\begin{eqnarray}}
\newcommand{\eary}{\end{eqnarray}}

\usepackage{graphicx}	
\usepackage{amsmath}	
\usepackage{multirow}

\begin{document}

\title{Synchrotron self-Compton Reverse Shock in Energy-Injection and radiative Scenarios}

\author[orcid=0000-0002-0173-6453]{Nissim Fraija}
\affiliation{Instituto de Astronom\' ia, Universidad Nacional Aut\'onoma de M\'exico, Circuito Exterior, C.U., A. Postal 70-264, 04510 M\'exico City, M\'exico}
\email[show]{nifraija@astro.unam.mx}  

\author[orcid=0000-0002-2516-5739]{B. Betancourt Kamenetskaia} 
\affiliation{Cosmology, Gravity, and Astroparticle Physics Group, Center for Theoretical Physics of the Universe, Institute for Basic Science (IBS), Daejeon, 34126, Korea}
\email{boris_betancourt@ciencias.unam.mx}

\author[orcid=0000-0001-5193-3693]{Antonio Galv\'an}
\affiliation{Instituto de F\'isica, Universidad Nacional Aut\'onoma de M\'exico, Circuito Exterior, C.U., A. Postal 70-264, 04510 M\'exico City, M\'exico}
\email{edwin@fisica.unam.mx}

\author[orcid=0000-0003-4442-8546]{Maria G. Dainotti}
\affiliation{Division of Science, National Astronomical Observatory of Japan, 2-21-1 Osawa, Mitaka, Tokyo 181-8588, Japan}
\affiliation{The Graduate University for Advanced Studies (SOKENDAI), Shonankokusaimura, Hayama, Miura District, Kanagawa 240-0115, Japan}
\email{mariagiovannadainotti@yahoo.it}

\begin{abstract}

Gamma-ray bursts (GRBs) are among the most energetic extragalactic transients, providing a unique laboratory to investigate the physical origin of the diverse morphologies observed in high-energy light curves and spectra. These morphologies, interpreted through synchrotron and synchrotron self-Compton (SSC) closure relations (CRs), may depend on the circumburst density profile, energy injection, and the dynamical evolution of the reverse shock in the adiabatic or radiative regime. We derive the SSC closure relations produced in the reverse-shock region for both thick- and thin-shell cases, considering constant-density and stratified circumburst media. The analysis assumes two power-law (PL) distributions for the electron spectral index and includes scenarios with and without energy injection from the progenitor. We show that, depending on the physical parameters, SSC emission from the reverse shock can reproduce the initial steep-decay and plateau phases commonly observed in GRB afterglows. We then compare the predicted spectral and temporal indices with those derived from PL and broken power-law (BPL) fits reported in the Second Fermi-LAT Gamma-ray Burst Catalog (2FLGC). Our results indicate that PL (BPL) light-curve fits preferentially favor scenarios with (without) energy injection. In scenarios with energy injection, the data preferentially support a constant-density environment over a stellar-wind profile.

\end{abstract}


\keywords{Gamma-ray bursts: individual (GRB )  --- Physical data and processes: acceleration of particles  --- Physical data and processes: radiation mechanism: nonthermal --- ISM: general - magnetic fields}

\section{Introduction}

The most energetic transient events in the Universe are gamma-ray bursts (GRBs), with energy outputs ranging from \(10^{49}\) to \(10^{55}\) erg. They produce temporally prolonged emission across different wavelengths as well as non-repeating gamma-ray pulses.   Extragalactic sources typically exhibit an initial phase known as the prompt phase, followed by a subsequent phase called afterglow. The prompt event is often detected across a wide range of the energy spectrum, from a few keV to several MeV, and can show diverse light-curve shapes and variability. The afterglow, which occurs after the prompt emission, is observed at various wavelengths, from radio waves to gamma rays, for extended durations, ranging from seconds to days. The fireball model is the most effective theoretical framework for explaining the multi-wavelength data associated with GRBs \citep{1986ApJ...308L..43P,1986ApJ...308L..47G,1978MNRAS.183..359C}.  In the current scenario, the prompt episode refers to the loss of kinetic energy that occurs during internal shocks. In contrast, the afterglow is produced when the expanding blast wave interacts with the surrounding medium, causing a significant portion of its energy to be transferred to that medium \citep{2004RvMP...76.1143P}.  The fireball produces a forward shock that propagates into the circumburst medium~\citep{1995ApJ...455L.143S, 1998ApJ...497L..17S, 1999ApJ...513..669K} and a reverse shock that travels back into the ejecta \citep{2000ApJ...545..807K, 2003ApJ...597..455K, 2016ApJ...818..190F}.   Electrons are accelerated to relativistic energies during the external forward shock. They then cool through synchrotron radiation and the synchrotron self-Compton (SSC) mechanism, which can last from seconds to years. This results in a gradual softening of the light curves during the afterglow episode \citep{1998ApJ...503..314P, 1998ApJ...497L..17S, 2007MNRAS.379..331P}.  As the reverse shock traverses the shell, shock-accelerated electrons emit photons through synchrotron and SSC processes, resulting in transient flares detectable across various bands.

In the conventional GRB afterglow scenario, the reverse shock is usually assumed to be completely adiabatic; however, it may display partial or complete radiative characteristics \citep{1998MNRAS.298...87D, 1998ApJ...497L..17S, 2000ApJ...532..281B, 2010MNRAS.403..926G}.  When the dynamical timescale exceeds the cooling timescale, the electrons accelerated by the shock enter the fast-cooling regime. In this case, the afterglow phase should occur in the radiative regime rather than the adiabatic regime, particularly when the microphysical parameter \(\varepsilon_e\) is greater than 0.1 and the accelerated electrons are in the fast-cooling scenario \citep{2000ApJ...532..281B,2002MNRAS.330..955L, 2019ApJ...886..106P}. When the GRB fireball is in a radiative phase, it decelerates more rapidly in the circumburst environment compared to when it is in an adiabatic phase. Consequently, the shock energetics and the temporal light curves are altered \citep{2000ApJ...532..281B, 2005ApJ...619..968W}. \cite{2010MNRAS.403..926G} analyzed the eleven GRBs observed by the Fermi Large Area Telescope (LAT) up to October 2009, specifically those with energies greater than 100 MeV. The study reported that the temporal decay indices were consistent with synchrotron afterglow emission in the fully radiative regime. This suggests that a radiatively efficient fireball could address the efficiency issues observed in the early afterglow phase \citep{2007ApJ...655..989Z}.\\
 
 A typical X-ray light curve exhibits a distinct structure comprising four separate power-law (PL) segments $\propto t^{-\alpha}$, along with a significant flare \citep[e.g., see][]{2006ApJ...642..354Z, 2006ApJ...642..389N}. One of them corresponds to a shallower decay than the usual phase described with a PL index  $-0.1\lesssim\alpha\lesssim$0.5,   the so-called ``plateau" phase.  The plateau phase and the giant flare have been interpreted in the scenario of an energy injection from the central engine into the blastwave. For instance, the continuous injection is associated with the energy of a spinning magnetized neutron star (NS) or a central black hole (BH)  \citep{2005ApJ...635L.133B,  2005ApJ...630L.113K, 2006Sci...311.1127D, 2006ApJ...636L..29P, 2006MNRAS.370L..61P, 2005Sci...309.1833B, 2007ApJ...671.1903C, 2017MNRAS.464.4399D, 2019ApJ...872..118B, 2019ApJ...887..254B}, and the instantaneous injection is identified with stratified ejecta \citep{2006ApJ...640L.139T, 2007ApJ...656L..57J, 2017MNRAS.472L..94H} and ejecta with a wide range of Lorentz factors \citep{1998ApJ...496L...1R,2000ApJ...532..286K, 2000ApJ...535L..33S, 2002ApJ...566..712Z, 2019ApJ...871..200F}.  Extensive investigations have been conducted on multi-wavelength afterglows, including giant flares, that involve energy injection models \citep[e.g., see][]{2015ApJ...814....1L, 2021ApJ...906...60Z, 2019ApJ...872..118B, 2022MNRAS.511.6205P, 2021ApJ...918...12F}. 

This manuscript investigates the CRs generated by the SSC mechanism in the reverse shock region during the radiative regime, particularly when the progenitor injects energy into the circumburst medium over a specific time interval. We consider an electron population characterized by spectral indices in the ranges of $1 < p < 2$ and $p > 2$. The study examines both thick- and thin-shell reverse shocks as well as jets decelerated in a stellar wind and in a constant-density medium.  We compare these CRs with the observed spectral and temporal features of the bursts recorded in the second catalog of LAT-detected GRBs (2FLGC).  The paper is organized as follows. In Section \ref{sec2}, we introduce the SSC emission from reverse shocks in both the radiative regime and during energy injection. Section \ref{sec3} presents the data analysis, evaluates the closure relations, and discusses the results. Section \ref{sec4} provides a summary of our findings.

\section{SSC Light curves and Closure relations from reverse shock}\label{sec2}

When a relativistic jet with a bulk Lorentz factor ($\Gamma_0$) interacts with the circumburst medium, whether it is constant or a stellar wind, it produces a reverse shock that propagates back into the outflow \citep{1997ApJ...476..232M, 1999ApJ...517L.109S, 2000ApJ...542..819K,2019ApJ...881...12B,2019ApJ...887..254B}.   The density profile of the circumburst region is characterised by $n(r) = A_{\rm k} r^{-\rm k}$, where ${\rm k=2}$ corresponds to the stellar wind driven stratified medium and ${\rm k=0}$ corresponds to the homogeneous density interstellar medium (ISM). For the wind case, ${\rm A_2}= 3.0\times 10^{35}\,{\rm cm^{-1}}\,A_{\rm W}$, where $A_{\rm W}$ is the density parameter. For the ISM, ${\rm A_0 = n}$, the number density of the ISM in units of ${\rm cm^{-3}}$.  

During the evolution of the afterglow, a forward shock propagating into the circumburst medium and a reverse shock propagating back into the ejecta are formed. These two shocks, together with the contact discontinuity, divide the system into four regions: (i) the unshocked circumburst medium (region 1), (ii) the shocked circumburst medium (region 2), (iii) the shocked ejecta (region 3), and (iv) the unshocked ejecta (region 4). In this work, we focus on region 3, which corresponds to the downstream of the reverse shock, where the ejecta has already been shocked by the reverse shock. The bulk Lorentz factor of this region, $\gamma_3$, characterizes the dynamical state of the shocked ejecta and is typically comparable to that of the forward-shocked region (region 2) after shock crossing. Before shock crossing, $\gamma_3$ evolves as the reverse shock propagates into the ejecta, while after shock crossing it approaches the bulk Lorentz factor of the forward shock.

In addition to the characteristics of the surrounding medium, and whether it is a thick- or thin-shell scenario, the evolution of the reverse shock depends on several factors, including the adiabatic or radiative regime and energy injection.  The dynamics of the reverse shock is determined by  the Sedov length \citep{1995ApJ...455L.143S, 2015AdAst2015E..13G}
\be\label{Sedov}
\ell=\left[\frac{(3-k)E}{4 \pi m_{p} c^2\,A_{\rm k}\,}\right]^{\frac{1}{3-k}}\,,
\ee
and the width of the shell $\Delta_{\rm x}\approx c (1+z)^{-1}\,t_{\rm x}$ via the bulk Lorentz factor $\Gamma_0=\left[\frac{\ell}{\Delta_{\rm x}}\right]^{\frac{3-k}{2(4-k)}}$, where $z$ is the redshift and $E$ the isotropic-equivalent kinetic energy.\footnote{The term $m_p$ corresponds to the proton mass and $c$ the speed of light.} The timescale $t_{\rm x}$ corresponds to the shock-crossing time, defined for the thin-shell regime as 

{\small
\begin{eqnarray}
\label{tx_eps}
t_{\rm x}\equiv \begin{cases} 
1.7\times 10^2\,{\rm s}\left(\frac{1+z}{1.2}\right)\, n^{-\frac13}_{}\, E^{\frac13}_{0,52.3}\, \Gamma_{0,2}^{-\frac{8}{3}}\hspace{0.76cm}{\rm for}\hspace{0.2cm}{\rm ISM}\,\,(\rm k=0) \cr
1.4\times 10^2\,{\rm s}\left(\frac{1+z}{1.2}\right)\, A_{\rm W,-2}^{-1}\, E_{0,52.3}\,\Gamma_{0,2}^{-4} \hspace{0.55cm}{\rm for}\hspace{0.16cm}{\rm wind}\,\,(\rm k=2)\,,
\end{cases}
\end{eqnarray}
}
in the radiative regime and

{\small
\begin{eqnarray}
\label{tx_q}
t_{\rm x}\equiv \begin{cases} 
2.9\times 10^2\,{\rm s}\left(\frac{1+z}{1.2}\right)^{\frac{3}{2+q}}\, n^{-\frac{1}{2+q}}_{-1}\, E^{\frac{1}{2+q}}_{\rm inj,52.3}\, \Gamma^{-\frac{8}{2+q}}_{0,2}\hspace{0.7cm}{\rm for}\hspace{0.2cm}{\rm ISM}\,\,(\rm k=0) \cr
7.1\times 10\,{\rm s}\left(\frac{1+z}{1.2}\right)^{\frac{1}{q}}\, A_{\rm W,-2}^{-\frac{1}{q}}\, E_{\rm inj,52.3}^{\frac{1}{q}}\, \Gamma^{-\frac{4}{q}}_{0,2}\hspace{1.3cm}{\rm for}\hspace{0.16cm}{\rm wind}\,\,(\rm k=2)\,,
\end{cases}
\end{eqnarray}
}
during energy injection.  The terms $\Gamma_0$ and $E_0$ correspond to the initial bulk Lorentz factor and isotropic-equivalent kinetic energy, and $E_{\rm inj}$ is the  energy injection. The energy injection index $q$ quantifies the rate at which additional energy is supplied to the blast wave over time, modifying its dynamical evolution.

On the other hand,  the critical Lorentz factor represents the dynamical boundary between thin- and thick-shell regimes, encoding the balance between ejecta inertia and the external medium, and thereby setting the relativistic nature of the reverse shock. The critical Lorentz factor is not a fixed quantity, but rather a dynamical threshold that depends on the ejecta kinetic energy and the density of the external medium. It is defined as the Lorentz factor at which the reverse shock transitions from Newtonian to relativistic, corresponding to the condition where the shock crossing time becomes comparable to the deceleration timescale. The critical Lorentz factor is defined as 
{\small
\begin{eqnarray}
\label{Gc_eps}
\Gamma_c\equiv \begin{cases} 
2.3\times 10^2\left(\frac{1+z}{1.2} \right)^{\frac{3}{8-\epsilon}} E^{\frac{1}{8-\epsilon}}_{0,53.3} n_{}^{-\frac{1}{8-\epsilon}}\Gamma_{0,2.5}^{-\frac{\epsilon}{8-\epsilon}} \left(\frac{T_{90}}{30\,{\rm s}}\right)^{-\frac{3}{8-\epsilon}}\hspace{0.5cm}{\rm for}\hspace{0.2cm}{\rm ISM}\,\,(\rm k=0) \cr
2.2\times 10^2\left(\frac{1+z}{1.2} \right)^{\frac{1}{4-\epsilon}} E^{\frac{1}{4-\epsilon}}_{0,53.3} A_{\rm W,-2}^{-\frac{1}{4-\epsilon}}\Gamma_{0,2.5}^{-\frac{\epsilon}{4-\epsilon}} \left(\frac{T_{90}}{30\,{\rm s}}\right)^{-\frac{1}{4-\epsilon}}\hspace{0.5cm}{\rm for}\hspace{0.16cm}{\rm wind}\,\,(\rm k=2)\,,
\end{cases}
\end{eqnarray}
}

in the radiative regime and 

{\small
\begin{eqnarray}
\label{Gc_q}
\Gamma_c\equiv \begin{cases} 
1.9\times 10^2\left(\frac{1+z}{1.2} \right)^{\frac38} E^{\frac18}_{\rm inj,53.3} n_{-1}^{-\frac18} \left(\frac{T_{90}}{30\,s}\right)^{-\frac{2+q}{8}}\hspace{0.5cm}{\rm for}\hspace{0.2cm}{\rm ISM}\,\,(\rm k=0) \cr
1.1\times 10^2\left(\frac{1+z}{1.2} \right)^{\frac14} E^{\frac14}_{\rm inj,53.3} A_{\rm W,-2}^{-\frac14} \left(\frac{T_{90}}{30\,s}\right)^{-\frac{q}{4}}\hspace{0.5cm}{\rm for}\hspace{0.16cm}{\rm wind}\,\,(\rm k=2)\,,
\end{cases}
\end{eqnarray}
}
during energy injection, with $\epsilon$ the radiative parameter.  The energy fraction that is transferred to the reverse shock is distributed to the electrons and the magnetic field through the microphysical parameters $\varepsilon_{e, r}$ and $\varepsilon_{B, r}$, respectively.\footnote{The subscript ``r" corresponds to quantities derived in the reverse-shock region.}  Therefore, relativistic electrons immersed in a region with a magnetic field are shock-accelerated and cooled primarily by synchrotron and SSC processes. Depending on the range of the spectral index of the electron distributions ($1<p<2$ or $p>2$), the Lorentz factor of electrons with the lowest energy becomes $\gamma_{\rm m, r}:\simeq\left[ 1.8\times 10^3\tilde{g}(p)\varepsilon_{e, r} \chi_{\rm e}^{-p+1} \gamma_{\rm max}\Gamma\right]^{\frac{1}{p-1}}$ with $\tilde{g}(p)=\frac{2-p}{p-1}$ for $1<p<2$ \citep{2001ApJ...558L.109D} and $\simeq 1.8\times 10^3 g(p)\varepsilon_{e, r}\chi_{\rm e}^{-1}\Gamma$ with $g(p)=\frac{p-2}{p-1}$ for $p>2$, where $\chi_{\rm e}$ corresponds to the fraction of shock-accelerated electrons in the reverse-shock region \citep{2006MNRAS.369..197F}, and $\gamma_{\rm max}$ is the electron Lorentz factor of maximum energy.\\

We estimate the CRs produced by the SSC reverse shock process after the shock-crossing time for two cases: (i) adiabatic and radiative regimes, and (ii) when the central engine continuously injects energy in both the thick and thin-shell domains.  Note that the evolution of the shell (thick or thin) will be affected by the critical Lorentz factor ($\Gamma_c$),  the shock crossing time ($t_{\rm x}$) and the burst duration ($T_{90}$). For $\Gamma_c \lesssim \Gamma$ ($t_{\rm x} \lesssim T_{90}$), the reverse shock evolves in the thick-shell domain and for $\Gamma < \Gamma_c$ ($T_{90} < t_{\rm x}$) in the thin-shell domain.

\subsection{SSC Light curves in the radiative regime}

The dynamics of the reverse shock in the radiative scenario depends on the radiative parameter ($\epsilon$), and the evolution of the initial quantities such as the bulk Lorentz factor ($\Gamma_0$) and the isotropic-equivalent kinetic energy ($E_0$).  The isotropic-equivalent kinetic energy  which is associated to the isotropic gamma-ray energy, $E_{\rm \gamma, iso}$, via the kinetic efficiency $\eta=E_{\rm \gamma, iso}/(E+E_{\rm \gamma, iso})$ becomes $E= E_0 \left( \frac{\Gamma}{\Gamma_0}\right)^\epsilon$ \citep{2000ApJ...532..281B, 2005ApJ...619..968W}. The parameter $\epsilon$ provides the details of the evolution; $\epsilon=1$ corresponds to fully radiative, $\epsilon=0$ fully adiabatic, and $0 <\epsilon < 1$ partially radiative/adiabatic. The SSC scenario represents the inherent high-energy extension of the synchrotron radiation model. Photons generated by synchrotron radiation in the reverse-shock region can be upscattered by the same electron population via inverse Compton scattering. 

\subsubsection{Thick-shell scenario}

We consider the thick-shell case in the post shock-crossing regime. This choice is motivated by the fact that, after the reverse shock has crossed the ejecta, the system enters a more stable and self-similar evolution, allowing for the derivation of robust CRs that can be directly compared with high-energy observations, such as those reported in the \textit{Fermi}-LAT GRB catalog. In contrast, the pre shock-crossing phase is characterized by a rapidly evolving and non-self-similar dynamics, where the emission depends sensitively on the initial conditions and the detailed structure of the ejecta. Since the primary goal of this work is to establish SSC closure relations and compare them with high-energy observations, we restrict our analysis to the post shock-crossing phase.

In this scenario, the reverse shock is extremely relativistic and can substantially decelerate the shell.  The critical Lorentz factor is given by Eq. \ref{Gc_eps}. Hereafter, we consider the value of  radiative parameter $\epsilon=0.5$, unless otherwise specified.  We adopt the convention $Q_{\rm x}=Q/10^{\rm x}$ in c.g.s. units and also the unprimed and primed variables for the observer and comoving frames, respectively.

\paragraph{Constant-density medium.}
Following the reverse shock that traverses the shell in region 3, the scaling of the hydrodynamic variables progresses as $\gamma_3\propto t^{-\frac{56-\epsilon}{16(8-\epsilon)}}$, $n_{3}\propto t^{-\frac{104+5\epsilon}{16(8-\epsilon)}}$, $p_{3}\propto t^{-\frac{26-\epsilon}{3(8-\epsilon)}}$ and $N_{\rm e}\propto t^{-\frac{3\epsilon}{8-\epsilon}}$. For a homogeneous medium, the comoving magnetic field evolves as $B'\propto t^{-\frac{26-\epsilon}{6(8-\epsilon)}}$. 

\subparagraph{\bf For $1<p<2$:}  The Lorentz factors of low-energy electrons and the characteristic spectral break scale as $\gamma_{\rm m,r}\propto t^{-\frac{68(8-\epsilon)-3p(56-\epsilon)}{96(p-1)(8-\epsilon)}}$ and $\nu^{\rm syn}_{\rm m, r}\propto t^{-\frac{168-57\epsilon+8p(26-\epsilon)}{48(p-1)(8-\epsilon)}}$.   The characteristic spectral break ($h\nu^{\rm ssc}_{\rm m, r}\simeq\gamma^2_{\rm m, r} h\nu^{\rm syn}_{\rm m, r}$) is

{\small
\begin{eqnarray}\label{break_thick_aft_k0_e}
h \nu^{\rm ssc}_{\rm m, r} &=& 
6.8\times 10^{-8}\,{\rm GeV}  \, \left(\frac{1+z}{1.2} \right)^{\frac{1096-173\epsilon-43p(8-\epsilon)}{48(p-1)(8-\epsilon)}}\,\tilde{g}^{\frac{4}{p-1}}(1.9)\,\chi_{\rm e,-0.3}^{-4} \varepsilon^{\frac{4}{p-1}}_{\rm e_r,-1} \varepsilon^{\frac{3-p}{2(p-1)}}_{\rm B_r,-2}n^{\frac{28-3\epsilon-p(8-\epsilon)}{2(p-1)(8-\epsilon)}}_{}\,\Delta_{11.8}^{\frac{5(25+p)}{48(p-1)}}\,\Gamma_{0,2.5}^{\frac{2(16-\epsilon)}{(p-1)(8-\epsilon)}}\,E^{-\frac{2}{(p-1)(8-\epsilon)}}_{0,53.3}\cr
&&\hspace{13cm} t^{-\frac{712-125\epsilon+5p(8-\epsilon)}{48(p-1)(8-\epsilon)}}_{1.5}\,.
\end{eqnarray}

The Lorentz factors of the limit over which electrons dissipate energy effectively and the cooling spectral break scale as $\gamma_{\rm c,r}\propto t^{\frac{200+29\epsilon}{48(8-\epsilon)}}$ and
$\nu^{\rm syn}_{\rm c, r}\propto t^{\frac{8+23\epsilon}{16(8-\epsilon)}}$, respectively.  The cutoff break after the shock-crossing time is determined by analyzing the adiabatic expansion of the fluid (\citet{2000ApJ...545..807K}. Therefore, the cutoff spectral break  can be written as

\begin{eqnarray}
h \nu^{\rm ssc}_{\rm cut, r} &\simeq&  \gamma^2_{\rm i, r} h\nu^{\rm syn}_{\rm c, r} (t_{\rm x})\,\left(\frac{t}{t_{\rm x}} \right)^{-\frac{33}{16}}\cr
&\simeq& 2.5\times10^{-2}\,{\rm GeV}\, \left(\frac{1+z}{1.2} \right)^{\frac{17}{16}}\,\varepsilon^{-\frac72}_{\rm B_r,-2} n^{-\frac{36-7\epsilon}{2(8-\epsilon)}}_{}\,   \left(\frac{1+\tilde{Y}_{\rm r}}{2}\right)^{-4}\,  \Delta^{\frac{232+31\epsilon}{16(8-\epsilon)}}_{11.8}\,\Gamma_{0,2.5}^{\frac{10\epsilon}{8-\epsilon}}\,E^{-\frac{10}{8-\epsilon}}_{0,53.3}\,t^{-\frac{33}{16}}_{1.5},
\end{eqnarray}

with $\tilde{Y}_{\rm r}$ the Compton parameter, which for limiting cases $\tilde{Y}_{\rm r}:=\eta \frac{\varepsilon_{\rm e_r}}{\varepsilon_{\rm B_r}}$ if $\frac{\eta\varepsilon_{\rm e_r}}{\varepsilon_{\rm B_r}}\ll 1$ and  $\left( \eta \frac{\varepsilon_{\rm e_r}}{\varepsilon_{\rm B_r}} \right)^{\frac12}$ if $\frac{\eta\varepsilon_{\rm e_r}}{\varepsilon_{\rm B_r}}\gg 1$ with $\eta=1$ for fast-cooling regime and $\left( \frac{\gamma_{\rm c,r}}{\gamma_{\rm m,r}} \right)^{p-2}$ for slow-cooling regime. It should be noted that the Compton parameter is modified due to the Klein-Nishina (KN) effects \citep[e.g., see][]{2009ApJ...703..675N, 2010ApJ...712.1232W}. The observed attenuation of up-scattered synchrotron photons in the SSC spectrum is a direct result of the influence of the KN effect. Nonetheless, the primary characteristics of SSC photons and the cooling processes of certain injected electrons with different Lorentz factors are dependent phenomena. The two phenomena are interconnected. 
Considering the critical photon energy above which inverse Compton scatterings with electrons of a given energy begin to enter the KN regime ($h\nu^{\rm syn}_{\rm KN, e, r}\simeq \frac{\gamma_3}{(1+z)} \frac{m_e c^2}{\gamma_{\rm e,r}}$) \citep{2010ApJ...712.1232W}, the spectral break in the KN regime for $\nu^{\rm ssc}_{\rm c, r}<\nu^{\rm ssc}_{\rm m, r}$ is

{\small
\bary
h \nu^{\rm ssc}_{\rm KN, m, r} =
3.6\,{\rm GeV}\,\left(\frac{1+z}{1.2}\right)^{\frac{976-158\epsilon-p(600-93\epsilon)}{96(p-1)(8-\epsilon)}} \tilde{g}^{\frac{1}{p-1}}(1.9)\,\chi_{e,-0.3}^{-1}\,\varepsilon^{\frac{1}{p-1}}_{\rm e_r,-1}\,\,\varepsilon^{\frac{2-p}{4(p-1)}}_{\rm B_r,-2}\,n^{\frac{(2-p)(10-\epsilon)}{4(p-1)(8-\epsilon)}}_{}\,\Gamma_{0,2.5}^{\frac{16-\epsilon p}{2(p-1)(8-\epsilon)}}\,E^{\frac{p-2}{2(p-1)(8-\epsilon)}}_{0,53.3}\Delta^{\frac{62+3p}{96(p-1)}}_{11.8}\cr
\hspace{9.9cm}\,t^{-\frac{208-62\epsilon+3p(56-\epsilon)}{96(p-1)(8-\epsilon)}}_{1.5}\,, \cr
\eary

and for $\nu^{\rm ssc}_{\rm m, r}<\nu^{\rm ssc}_{\rm c, r}$ is

\bary
h \nu^{\rm ssc}_{\rm KN, c, r}\simeq 8.9\times 10\,{\rm GeV}\,\left(\frac{1+z}{1.2}\right)^{-\frac{26-\epsilon}{3(8-\epsilon)}} \left(\frac{1+\tilde{Y}_{\rm r}}{2} \right)^{-1}\,\varepsilon^{-1}_{\rm B_r,-2}\,n^{-\frac{6-\epsilon}{8-\epsilon}}_{}\,\Gamma_{0,2.5}^{\frac{2\epsilon}{8-\epsilon}}\,E^{-\frac{2}{8-\epsilon}}_{0, 53.3}\,\Delta^{-\frac{1}{3}}_{11.8}\,t^{\frac{2(1+\epsilon)}{3(8-\epsilon)}}_{1.5}\,.
\eary
}

The maximum Lorentz factor ($\gamma_{\rm max, r}$) and the respective spectral break ($h\nu^{\rm ssc}_{\rm max, r}\simeq\gamma^2_{\rm max, r} h\nu^{\rm syn}_{\rm max, r}$) become

\begin{equation}
    \gamma_{\rm max, r}\simeq 1.8\times 10^7\left(\frac{1+z}{1.2} \right)^{-\frac{26-\epsilon}{12(8-\epsilon)}}\,\varepsilon^{-\frac14}_{\rm B_r,-2} n^{-\frac{6-\epsilon}{4(8-\epsilon)}}_{}\,\Delta^{-\frac{1}{12}}_{11.8}\,\Gamma_{0,2.5}^{\frac{\epsilon}{2(8-\epsilon)}}\,E^{-\frac{1}{2(8-\epsilon)}}_{0,53.3}\,t^{\frac{26-\epsilon}{12(8-\epsilon)}}_{1.5}\, ,
\end{equation}

and 

\begin{equation}
h \nu^{\rm ssc}_{\rm max, r}\simeq\,1.2\times 10^{15}\,{\rm GeV}\,\left(\frac{1+z}{1.2} \right)^{-\frac{53}{48}}\,\varepsilon^{-\frac12}_{\rm B_r,-2}\,  n_{}^{-\frac{1}{2}}\,  \Delta_{11.8}^{-\frac{5}{48}}\,t^{\frac{5}{48}}_{1.5}\,, 
\end{equation}

respectively.  The maximum SSC flux for a constant-density medium ($F^{\rm ssc}_{\rm max,r}\sim\, \frac{4\sigma_Tn r}{3\tilde{g}(p)}\,F^{\rm syn}_{\rm max,r}$) is given by

\begin{equation}
F^{\rm ssc}_{\rm max, r}\simeq 3.3\times 10^{-1}\,{\rm mJy}\,\left(\frac{1+z}{1.2} \right)^{\frac{712+127\epsilon}{48(8-\epsilon)}}\,\tilde{g}^{-1}(2.2)\,\chi_{e,-0.3}\,   \varepsilon^{\frac12}_{\rm B_r,-2}\,  n_{}^{\frac{16-5\epsilon}{2(8-\epsilon)}}\,  \Delta_{11.8}^{\frac{17}{48}} \,d^{-2}_{\rm z,27.9}\,\Gamma_{0,2.5}^{-\frac{8+11\epsilon}{8-\epsilon}} \,E^{\frac{12}{8-\epsilon}}_{0,53.3}\,t^{-\frac{328+175\epsilon}{48(8-\epsilon)}}_{1.5}\,, 
\end{equation}

where $d_{\rm z}$ corresponds to the luminosity distance \citep{1972gcpa.book.....W}.\\

\subparagraph{\bf For $p>2$:}  The Lorentz factors of low-energy electrons and the characteristic spectral break scale as $\gamma_{\rm m,r}\propto t^{-\frac{104-31\epsilon}{48(8-\epsilon)}}$ and $\nu^{\rm syn}_{\rm m, r}\propto t^{-\frac{73}{48}}$.   The characteristic spectral break is

\begin{eqnarray}\label{nu_m_eps_2}
h \nu^{\rm ssc}_{\rm m, r} =5.8\times 10^{-5}\,{\rm GeV}  \, \left(\frac{1+z}{1.2} \right)^{\frac{136-29\epsilon}{16(8-\epsilon)}}\,g^4(2.2)\,\chi_{\rm e,-0.3}^{-4}\, \,\varepsilon^4_{\rm e_r,-1} \varepsilon^{\frac12}_{\rm B_r,-2}  n^{\frac{12-\epsilon}{2(8-\epsilon)}}_{}\,\Delta_{11.8}^{\frac{45}{16}}\,\Gamma_{0,2.5}^{\frac{2(16-\epsilon)}{8-\epsilon}}\,E^{-\frac{2}{8-\epsilon}}_{0,53.3}t^{-\frac{3(88-15\epsilon)}{16(8-\epsilon)}}_{1.4}\,.
\end{eqnarray}

The cooling Lorentz factor and the respective spectral break evolve similar to hard-spectral index ($1<p<2$).  Therefore, the cutoff spectral break  can be written as

\begin{eqnarray}
h \nu^{\rm ssc}_{\rm cut, r} \simeq 2.5\times10^{-2}\,{\rm GeV}\, \left(\frac{1+z}{1.2} \right)^{\frac{17}{16}}\,\varepsilon^{-\frac72}_{\rm B_r,-2} n^{-\frac{36-7\epsilon}{2(8-\epsilon)}}_{}\,   \left(\frac{1+ Y_{\rm r}}{2}\right)^{-4}\,  \Delta^{\frac{232+31\epsilon}{16(8-\epsilon)}}_{11.8}\,\Gamma_{0,2.5}^{\frac{10\epsilon}{8-\epsilon}}\,E^{-\frac{10}{8-\epsilon}}_{0,53.3}\,t^{-\frac{33}{16}}_{1.5},
\end{eqnarray}

with $Y_{\rm r}$ the Compton parameter. In this case, the parameter $\eta= \left( \frac{\gamma_{\rm c,r}}{\gamma_{\rm m,r}} \right)^{p-2}$ for slow-cooling regime, where $\gamma_{\rm m,r}$ is given by Eq. (\ref{nu_m_eps_2}). The spectral breaks in the KN regime for $\nu^{\rm ssc}_{\rm c, r}<\nu^{\rm ssc}_{\rm m, r}$ is
\bary
h \nu^{\rm ssc}_{\rm KN, m, r}=2.0\times 10\,{\rm GeV}\,\left(\frac{1+z}{1.2}\right)^{-\frac{7}{24}} g(2.2)\,\chi_{e,-0.3}^{-1}\,\varepsilon_{\rm e_r,-1}\,\Gamma_{0,2.5}\Delta^{\frac{17}{24}}_{11.8}\,t^{-\frac{17}{24}}_{1.5},\hspace{5.9cm}{\rm for} 
\eary

and for $\nu^{\rm ssc}_{\rm m, r}<\nu^{\rm ssc}_{\rm c, r}$ is

\bary
h \nu^{\rm ssc}_{\rm KN, c, r}\simeq 8.9\times 10\,{\rm GeV}\,\left(\frac{1+z}{1.2}\right)^{-\frac{26-\epsilon}{3(8-\epsilon)}} \left(\frac{1+Y_{\rm r}}{2} \right)^{-1}\,\varepsilon^{-1}_{\rm B_r,-2}\,n^{-\frac{6-\epsilon}{8-\epsilon}}_{}\,\Gamma_{0,2.5}^{\frac{2\epsilon}{8-\epsilon}}\,E^{-\frac{2}{8-\epsilon}}_{0, 53.3}\,\Delta^{-\frac{1}{3}}_{11.8}\,t^{\frac{2(1+\epsilon)}{3(8-\epsilon)}}_{1.5}.\,\,
\eary

The maximum SSC flux for a constant-density medium is given by

\bary
F^{\rm ssc}_{\rm max, r}\simeq 3.3\times 10^{-1}\,{\rm mJy}\,\left(\frac{1+z}{1.2} \right)^{\frac{712+127\epsilon}{48(8-\epsilon)}}\,g^{-1}(2.2)\,\chi_{e,-0.3}\,   \varepsilon^{\frac12}_{\rm B_r,-2}\,  n_{}^{\frac{16-5\epsilon}{2(8-\epsilon)}}\,  \Delta_{11.8}^{\frac{17}{48}} \,d^{-2}_{\rm z,27.9}\,\Gamma_{0,2.5}^{-\frac{8+11\epsilon}{8-\epsilon}} \,E^{\frac{12}{8-\epsilon}}_{0,53.3}\,t^{-\frac{328+175\epsilon}{48(8-\epsilon)}}_{1.5}\,. 
\eary

\paragraph{Stellar-wind environment.} The evolution of the hydrodynamic quantities in region 3 becomes $\gamma_3\propto t^{-\frac{12-\epsilon}{8(4-\epsilon)}}$, $n_{3}\propto t^{-\frac{36-11\epsilon}{8(4-\epsilon)}}$, $p_{3}\propto t^{-\frac{2(3-\epsilon)}{4-\epsilon}}$ and $N_{\rm e}\propto t^{-\frac{\epsilon}{4-\epsilon}}$. The evolution of the comoving magnetic field, the lowest and the cooling Lorentz factors of electrons, for $1<p<2$ ($p > 2$) are ${B'}_{\rm r}\propto t^{-\frac{3-\epsilon}{4-\epsilon}}$, $\gamma_{\rm m,r}\propto t^{-\frac{64-24\epsilon-p(20-7\epsilon)}{16(p-1)(4-\epsilon)}}\,(t^{-\frac{12-5\epsilon}{8(4-\epsilon)}})$ and $\gamma_{\rm c,r}\propto t^{\frac{28-9\epsilon}{8(4-\epsilon)}}$, respectively. The evolution of the synchrotron spectral breaks and the respective maximum flux for $1<p<2$ ($p > 2$) are $\nu^{\rm syn}_{\rm m, r}\propto t^{-\frac{28-15\epsilon-2p(8-\epsilon)}{8(p-1)(4-\epsilon)}}\,(t^{-\frac{60-19\epsilon}{8(4-\epsilon)}})$, $\nu^{\rm syn}_{\rm c, r}\propto t^{\frac{20-9\epsilon}{8(4-\epsilon)}}$ and $F^{\rm syn}_{\rm max,r}\propto t^{-\frac{36-\epsilon}{8(4-\epsilon)}}$, respectively.  The characteristic spectral break for $1<p<2$ is

\begin{eqnarray}\label{break_thick_aft_k2_e}
h \nu^{\rm ssc}_{\rm m, r} &=&
1.8\times 10^{-7}\,{\rm GeV}\,\left(\frac{1+z}{1.2}\right)^{\frac{124-47\epsilon-p(36-13\epsilon)}{8(p-1)(4-\epsilon)}} \tilde{g}^{\frac{4}{p-1}}(1.9) \chi_{\rm e,-1}^{-4} \varepsilon^{\frac{4}{p-1}}_{\rm e_r,-1} \varepsilon^{\frac{3-p}{2(p-1)}}_{\rm B_r,-2} A^{\frac{28-3\epsilon-p(8-\epsilon)}{2(p-1)(4-\epsilon)}}_{\rm W,-2} \Delta^{\frac{3(5+p)}{8(p-1)}}_{11.8}\,\Gamma_{0,2.5}^\frac{2[8+\epsilon(2-p)]}{(p-1)(4-\epsilon)}\,E^{-\frac{2(4-p)}{(p-1)(4-\epsilon)}}_{0,53.3}\cr
&&\hspace{12.4cm} t^{-\frac{92-39\epsilon-p(4-5\epsilon)}{8(p-1)(4-\epsilon)}}_{1.5}\,,
\eary

and $ p > 2$ is

\begin{eqnarray}\nonumber
h \nu^{\rm ssc}_{\rm m, r} = 1.3\times 10^{-4}\,{\rm GeV} \left(\frac{1+z}{1.2}\right)^{\frac{52-21\epsilon}{8(4-\epsilon)}}\,g^{4}(2.2)\,\chi_{\rm e,-1}^{-4}\,\epsilon^4_{\rm e_r,-1} \epsilon^{\frac12}_{\rm B_r,-2} \,A^{\frac{12-\epsilon}{2(4-\epsilon)}}_{\rm W,-2}\,\Delta_{11.8}^{\frac{21}{8}}\, \Gamma_{0,2.5}^{\frac{16}{4-\epsilon}}\,  E^{-\frac{4}{4-\epsilon}}_{\rm 0,53.3}\,t^{-\frac{84-29\epsilon}{8(4-\epsilon)}}_{1.5}\,.
\eary

The cutoff spectral breaks with $\nu^{\rm ssc}_{\rm cut, r}=\nu^{\rm ssc}_{\rm c, r}(t_{\rm x})\,\left(\frac{t}{t_{\rm x}} \right)^{-\frac{21}{8}}$ can be written as

\begin{eqnarray} \nonumber
h \nu^{\rm ssc}_{\rm cut, r}&\simeq& 1.0\times 10^{-4}\,{\rm GeV} \left(\frac{1+z}{1.2}\right)^{\frac{13}{8}}\,\left(\frac{1+\mathcal{Y_{\rm r}}}{2} \right)^{-4}  \epsilon^{-\frac72}_{\rm B_r,-2}\,A^{-\frac{36-7\epsilon}{2(4-\epsilon)}}_{\rm W,-2}\,\Delta_{11.8}^{\frac{148-45\epsilon}{8(4-\epsilon)}}\,\Gamma_0^{-\frac{4\epsilon}{4-\epsilon}}\, E_{\rm 0,53.3}^{\frac{4}{4-\epsilon}}\, t^{-\frac{21}{8}}_{1.5}\,,
\eary

where hereafter $\mathcal{Y_{\rm r}}:=\tilde{Y}_{\rm r}$ for $1<p<2$ and $Y_{\rm r}$ for $p>2$. The maximum flux in the stellar-wind environment

\begin{eqnarray} \nonumber
F^{\rm ssc}_{\rm max,r} &\simeq&  9.7\times 10^{-1}\,{\rm mJy}\,\left(\frac{1+z}{1.2}\right)^{\frac{76-15\epsilon}{8(4-\epsilon)}}\,G^{-1}(2.2)\,\chi_{\rm e,-1}\,  \epsilon^{\frac12}_{\rm B_r,-2} \, A^{\frac{16-5\epsilon}{2(4-\epsilon)}}_{W,-2}\, \Delta_{11.8}^{-\frac18}\, d^{-2}_{\rm z,27.9}\,\Gamma_{0,2.5}^{-\frac{4+\epsilon}{4-\epsilon}}\,E^{\frac{2}{4-\epsilon}}_{\rm 0,53.3}\, t^{-\frac{44-7\epsilon}{8(4-\epsilon)}}_{1.5}\,,
\eary

where hereafter $G(2.2):=\tilde{g}(2.2)$ for $1<p<2$ and $g(2.2)$ for $p>2$. The maximum Lorentz factor and the respective spectral break for $1<p<2$ are

\begin{equation}
    \gamma_{\rm max, r}\simeq 1.3\times 10^7\left(\frac{1+z}{1.2} \right)^{-\frac{3-\epsilon}{2(4-\epsilon)}}\,\varepsilon^{-\frac14}_{\rm B_r,-2} A_{W,-2}^{-\frac{6-\epsilon}{4(4-\epsilon)}}\,\Gamma_{0,2.5}^{-\frac{\epsilon}{2(4-\epsilon)}}\,E^{\frac{1}{2(4-\epsilon)}}_{0,53.3}\,t^{\frac{3-\epsilon}{2(4-\epsilon)}}_{1.5}\, ,
\end{equation}

and

\begin{equation}
h \nu^{\rm ssc}_{\rm max, r}\simeq\,5.6\times 10^{14}\,{\rm GeV}\,\left(\frac{1+z}{1.2} \right)^{-\frac{44-15\epsilon}{8(4-\epsilon)}}\,\epsilon^{-\frac12}_{\rm B_r,-2}\,  A_{W,-2}^{-\frac{8-\epsilon}{2(4-\epsilon)}}\,  \Delta_{11.8}^{\frac{1}{8}}\,\Gamma_{0,2.5}^{-\frac{2\epsilon}{4-\epsilon}}\,E^{\frac{2}{4-\epsilon}}_{\rm 0,53.3}\,t^{\frac{12-7\epsilon}{8(4-\epsilon)}}_{1.5}\,,
\end{equation}

respectively.  In the case $\nu^{\rm ssc}_{\rm c,r} < \nu^{\rm ssc}_{\rm m,r}$, the spectral breaks in the KN regime for $1<p<2$ can be expressed as

\bary
h \nu^{\rm ssc}_{\rm KN, m, r} &\simeq&
3.8\,{\rm GeV}\,\left(\frac{1+z}{1.2}\right)^{\frac{104-38\epsilon-p(60-21\epsilon)}{16(p-1)(4-\epsilon)}} \tilde{g}^{\frac{1}{p-1}}(1.9) \chi_{\rm e,-1}^{-1} \varepsilon^{\frac{1}{p-1}}_{\rm e_r,-1}\,\varepsilon^{\frac{2-p}{4(p-1)}}_{\rm B_r,-2} A^{\frac{(2-p)(10-\epsilon)}{4(p-1)(4-\epsilon)}}_{\rm W,-2} \Delta^{\frac{3(p+2)}{16(p-1)}}_{11.8}\,\Gamma_{0,2.5}^{\frac{8+\epsilon(4-3p)}{2(p-1)(4-\epsilon)}}\,E^{-\frac{3(2-p)}{2(p-1)(4-\epsilon)}}_{0,53.3}\cr
&&\hspace{11cm} t^{-\frac{40-22\epsilon+p(4+5\epsilon)}{16(p-1)(4-\epsilon)}}_{1.5}\,.
\eary

and $ p > 2$ is

\bary
h \nu^{\rm ssc}_{\rm KN, m, r}=1.9\times 10\,{\rm GeV}\,\left(\frac{1+z}{1.2}\right)^{-\frac14} g(2.2)\,\chi_{e,-0.3}^{-1}\,\varepsilon_{\rm e_r,-1}\,\Gamma_{0,2.5}\,\Delta^{\frac34}_{11.8}\,t^{-\frac34}_{1.5}\,.
\eary

In the case $\nu^{\rm ssc}_{\rm m, r}<\nu^{\rm ssc}_{\rm c, r}$, the spectral break in the KN regime yields

\bary
h \nu^{\rm ssc}_{\rm KN, c, r}\simeq 1.8\times 10\,{\rm GeV}\,\left(\frac{1+z}{1.2}\right)^{-\frac{2(3-\epsilon)}{4-\epsilon}} \left(\frac{1+\mathcal{Y_{\rm r}}}{2} \right)^{-1}\,\varepsilon^{-1}_{\rm B_r,-2}\,A^{-\frac{6-\epsilon}{4-\epsilon}}_{\rm W,-2}\,\Gamma_{0,2.5}^{-\frac{2\epsilon}{4-\epsilon}}\,E^{\frac{2}{4-\epsilon}}_{0,53.3}\,t^{\frac{2-\epsilon}{4-\epsilon}}_{1.5}\,.
\eary

\subsubsection{Thin-shell scenario after the shock-crossing time}

This scenario highlights the dynamics of a mildly relativistic reverse shock that cannot decelerate the shell effectively. In this context, the time it takes for the shock to traverse a constant-density medium and stellar wind is critically important and is given by Eq. \ref{tx_eps}.

\paragraph{Constant-density medium.} In this scenario, the timescales of the post-shock magnetic field, and the breaks of the electron Lorentz factors are ${B'}_{\rm r} \propto t^{-\frac47}$, $\gamma_{\rm m, r}\propto t^{-\frac{24-7p}{35(p-1)}}\,(t^{-\frac27})$, $\gamma_{\rm c,r}\propto  t^{\frac{19}{35}}$ and $\nu^{\rm syn}_{\rm m, r}\propto t^{-\frac{2(10p+7)}{35(p-1)}}\,(t^{-\frac{54}{35}})$, respectively.  The corresponding timescales of the synchrotron breaks and the maximum synchrotron flux are  $\nu^{\rm syn}_{\rm c, r}\propto t^{\frac{4}{35}}$ and $F^{\rm syn}_{\rm max,r }\propto t^{-\frac{34}{35}}$ for $1<p<2$ ($p > 2$), respectively.   The characteristic spectral break for $1<p<2$ is

\bary\label{break_thin_aft_k0_e}
h \nu^{\rm ssc}_{\rm m, r} =
1.2\,\times 10^{-8}\,{\rm GeV} \left(\frac{1+z}{1.2}\right)^{\frac{97-29p}{35(p-1)}}\,\tilde{g}^{\frac{4}{p-1}}(1.9)\,\chi_{\rm e,-1}^{-4}\,\varepsilon^{\frac{4}{p-1}}_{\rm e_r,-1} \varepsilon^{\frac{3-p}{2(p-1)}}_{\rm B_r,-2} \, n^{\frac{191-117p}{210(p-1)}}_{}\,\Gamma^{-\frac{2(143+24p)}{105(p-1)}}_{0,2}\,  E^{\frac{2(3p+31)}{105(p-1)}}_{0,52.3}t^{-\frac{2(31+3p)}{35(p-1)}}_{2.5}\,,
\eary

and for $p>2$ is 

\bary
h \nu^{\rm ssc}_{\rm m, r}=1.4\,\times 10^{-7}\,{\rm GeV} \left(\frac{1+z}{1.2}\right)^{\frac{39}{35}}\,g^4(2.2)\,\chi_{\rm e,-1}^{-4}\,\varepsilon^4_{\rm e_r,-1} \varepsilon^{\frac12}_{\rm B_r,-2}\,n^{-\frac{43}{210}}_{}\, \Gamma^{-\frac{382}{105}}_{0,2}\,  E^{\frac{74}{105}}_{0,52.3}\,t^{-\frac{74}{35}}_{2.5}\,.
\eary

Considering the adiabatic expansion of the fluid,  the cutoff spectral breaks with $\nu^{\rm ssc}_{\rm cut, r}=\nu^{\rm ssc}_{\rm c, r}(t_{\rm x})\,\left(\frac{t}{t_{\rm x}} \right)^{-\frac{74}{35}}$ can be written as

\bary
h \nu^{\rm ssc}_{\rm cut, r}\simeq 7.6\times 10^{-3}\,{\rm GeV} \left(\frac{1+z}{1.2}\right)^{\frac{39}{35}}\,\left(\frac{1+\mathcal{Y_{\rm r}}}{2}\right)^{-4}  \varepsilon^{-\frac72}_{\rm B_r,-2}\,n^{-\frac{201}{70}}_{-1} \, \Gamma^{-\frac{174}{35}}_{0,2}\, E^{-\frac{22}{35}}_{0,52.3}\, t^{-\frac{74}{35}}_{2.5}\,.
\eary

The peak flux associated with the SSC emission is given by

\bary
F^{\rm ssc}_{\rm max,r} \simeq  8.5\times 10^{-8}\,{\rm mJy}\, \left(\frac{1+z}{1.2}\right)^{\frac{62}{35}} \,G^{-1}(2.2)\,\chi_{\rm e,-1}\,\varepsilon^{\frac12}_{\rm B_r,-2} \, n^{\frac{191}{210}}_{} \,\Gamma^{-\frac{181}{105}}_{0,2}\,d^{-2}_{\rm z,27.5}E^{\frac{167}{105}}_{0,52.3}\, t^{-\frac{27}{35}}_{2.5}\,.
\eary

The maximum Lorentz factor, along with the corresponding spectral break for $1<p<2$, can be expressed as

\begin{equation}\nonumber
    \gamma_{\rm max, r}\simeq 3.1\times 10^7 \left(\frac{1+z}{1.2} \right)^{-\frac{2}{7}}\,\varepsilon^{-\frac14}_{\rm B_r,-2} n^{-\frac{13}{84}}_{-1}\,\Delta^{-\frac{1}{12}}_{11.8}\,\Gamma_{0,2.5}^{\frac{11}{42}}\,E^{-\frac{2}{21}}_{0,53.3}\,t^{\frac{2}{7}}_{1.5}\, ,
\end{equation}

and

\begin{equation}\nonumber
h \nu^{\rm ssc}_{\rm max, r}\simeq\,1.5\times 10^{15}\,{\rm GeV}\,\left(\frac{1+z}{1.2} \right)^{-\frac{41}{35}}\,\varepsilon^{-\frac12}_{\rm B_r,-2}\,  n_{-1}^{-\frac{31}{70}}\,\Gamma^{\frac{16}{35}}_{0,2.5}\,E^{-\frac{2}{35}}_{0,53.3}\,t^{\frac{6}{35}}_{1.5}\,,
\end{equation}

respectively. For $\nu^{\rm ssc}_{\rm c, r}<\nu^{\rm ssc}_{\rm m, r}$,  the KN spectral breaks for $1<p<2$ are given by

\bary
h\nu^{\rm ssc}_{\rm KN, m, r}\simeq
1.5\,{\rm GeV}  \,\left(\frac{1+z}{1.2}\right)^{\frac{45-28p}{35(p-1)}}\tilde{g}^{\frac{1}{p-1}}(1.9)\chi_{e,-0.3}^{-1}\,\varepsilon^{\frac{1}{p-1}}_{\rm e_r,-1}\,\epsilon^{\frac{2-p}{4(p-1)}}_{\rm B_r,-2}\,n^{\frac{170-133p}{420(p-1)}}_{\rm -1} \,\Gamma^{-\frac{160+7p}{210(p-1)}}_{0,2}\,E^{\frac{10+7p}{105(p-1)}}_{0,52.3}t^{-\frac{10+7p}{35(p-1)}}_{2.5}
\eary

 and $p>2$ yield 

\bary
h\nu^{\rm ssc}_{\rm KN, m, r}\simeq 2.9\,{\rm GeV}  \,\left(\frac{1+z}{1.2}\right)^{-\frac{11}{35}}g(2.2)\chi_{e,-0.3}^{-1}\,\varepsilon_{\rm e_r,-1}\,n^{-\frac{8}{35}}_{\rm -1} \,\Gamma^{-\frac{29}{35}}_{0,2}\,E^{\frac{8}{35}}_{0,52.3}\,t^{-\frac{24}{35}}_{2.5}\,.
\eary

For $\nu^{\rm ssc}_{\rm m, r}<\nu^{\rm ssc}_{\rm c, r}$, the spectral break becomes

\bary
h\nu^{\rm ssc}_{\rm KN, c, r}\simeq 3.7\times10\,{\rm GeV}  \,\left(\frac{1+z}{1.2}\right)^{-\frac{8}{7}}\, \left(\frac{1+\mathcal{Y_{\rm r}}}{2} \right)^{-1}\,\epsilon^{-1}_{\rm B_r,-2}\,n^{-\frac{13}{21}}_{} \,\Gamma^{\frac{22}{21}}_{0,2}\,E^{-\frac{8}{21}}_{\rm  0,53}\,t^{\frac{1}{7}}_{2.5}.
\eary

%
%
\paragraph{Stellar-wind medium.}
The scaling of the hydrodynamic variables in the region 3 evolve as $\gamma_3\propto t^{-\frac13}$, $n_{3}\propto t^{-\frac87}$, $p_{3}\propto t^{-\frac{32}{21}}$ and $N_{\rm e}\propto t^0$.  The timescale of the comoving magnetic field, the Lorentz factor for the minimum and cooling electrons for $1<p<2$ ($p > 2$) are ${B'}_{\rm r}\propto t^{-\frac{16}{21}}$, $\gamma_{\rm m,r}\propto t^{-\frac{22-7p}{21(p-1)}}\,(t^{-\frac{8}{21}})$ and $\gamma_{\rm c,r}\propto t^\frac67$. The timescale of the synchrotron spectral breaks, and the maximum synchrotron flux for $1<p<2$ ($p > 2$) are  $\nu^{\rm syn}_{\rm m, r}\propto t^{-\frac{7+3p}{7(p-1)}}\,(t^{-\frac{13}{7}})$ and $\nu^{\rm syn}_{\rm c, r}\propto t^{\frac{13}{21}}$ and $F^{\rm syn}_{\rm max,r}\propto t^{-\frac{23}{21}}$, respectively.   The characteristic spectral break for $1<p<2$ is

\bary\label{break_thin_aft_k2_e}
h \nu^{\rm ssc}_{\rm m, r} =
1.6\times 10^{-11}\,{\rm GeV}\, \left(\frac{1+z}{1.2}\right)^{\frac{86-26p}{21(p-1)}}\,\tilde{g}^{\frac{4}{p-1}}(1.9)\,\chi_{\rm e,-0.5}^{-4}\,\varepsilon^{\frac{4}{p-1}}_{\rm e_r,-1} \varepsilon^{\frac{3-p}{2(p-1)}}_{\rm B_r,-2} \,A^{\frac{59-53p}{42(p-1)}}_{\rm W,-2}\, \Gamma^{-\frac{2(46+11p)}{21(p-1)}}_{0,2}\,  E^{\frac{2(8p+1)}{21(p-1)}}_{0,52.3}t^{-\frac{5(13-p)}{21(p-1)}}_{2.5}
\eary

\bary
h \nu^{\rm ssc}_{\rm m, r} =3.1\times 10^{-10}\,{\rm GeV}\, \left(\frac{1+z}{1.2}\right)^{\frac{34}{21}}\,g^4(2.2)\,\chi_{\rm e,-1}^{-4}\,\varepsilon^4_{\rm e_r,-1} \varepsilon^{\frac12}_{\rm B_r,-2}\,A^{-\frac{47}{42}}_{\rm W,-2}  \,\Gamma_{0,2}^{-\frac{136}{21}}\,  E^{\frac{34}{21}}_{0,52.3}\,t^{-\frac{55}{21}}_{2.5}
\eary

The SSC spectral breaks with $\nu^{\rm ssc}_{\rm cut, r}=\nu^{\rm ssc}_{\rm c, r}(t_{\rm x})\,\left(\frac{t}{t_{\rm x}} \right)^{-\frac{55}{21}}$, yield

\bary
h \nu^{\rm ssc}_{\rm cut, r}&\simeq& 2.2\times 10^{-8}\,{\rm GeV} \left(\frac{1+z}{1.2}\right)^{\frac{34}{21}}\,\left(\frac{1+\mathcal{Y_{\rm r}}}{2}\right)^{-4}  \varepsilon^{-\frac72}_{\rm B_r,-2}\,A^{-\frac{383}{42}}_{\rm W,-2}  \,\Gamma^{-\frac{388}{21}}_{0,2}\, E^{\frac{118}{21}}_{0,52.3}\, t^{-\frac{55}{21}}_{2.5}\,.
\eary

The maximum flux produced by the SSC mechanism yields

\bary
F^{\rm ssc}_{\rm max,r} &\simeq&  2.5\times 10^{-8}\,{\rm mJy}\, \left(\frac{1+z}{1.2}\right)^{\frac{17}{7}} \,G^{-1}(2.2)\,\chi_{\rm e,-1}\,\varepsilon^{\frac12}_{\rm B_r,-2} \, A^{\frac{29}{14}}_{\rm W,-2}\, \Gamma_{0,2}^{-\frac{5}{7}}\,d^{-2}_{\rm z,27.5}E^{\frac{3}{7}}_{0,52.3}\, t^{-\frac{10}{7}}_{2.5}\,.
\eary

For $1<p<2$, the maximum Lorentz factor and the corresponding spectral break can be expressed as

\begin{equation}\nonumber
    \gamma_{\rm max, r}\simeq 0.9\times 10^7 \left(\frac{1+z}{1.2} \right)^{-\frac{8}{21}}\,\varepsilon^{-\frac14}_{\rm B_r,-2} A_{W,-2}^{-\frac{31}{84}}\,\Gamma_{0,2.5}^{\frac{1}{42}}\,E^{\frac{5}{42}}_{0,53.3}\,t^{\frac{8}{21}}_{2.5}\, ,
\end{equation}

and

\begin{equation}\nonumber
h \nu^{\rm ssc}_{\rm max, r}\simeq\,1.5\times 10^{14}\,{\rm GeV}\,\left(\frac{1+z}{1.2} \right)^{-\frac{10}{7}}\,\varepsilon^{-\frac12}_{\rm B_r,-2}\,  A_{\rm W,-2}^{-\frac{15}{14}}\,\Gamma^{-\frac{2}{7}}_{0,2.5}\,E^{\frac{4}{7}}_{0,53.3}\,t^{\frac{3}{7}}_{2.5}\,, 
\end{equation}

respectively.   For $\nu^{\rm ssc}{\rm c,r} < \nu^{\rm ssc}{\rm m,r}$, the spectral breaks in the KN regime for $1<p<2$ are given by

\bary
h\nu^{\rm ssc}_{\rm KN, m, r}\simeq 
1.9\times 10^{-1}\,{\rm GeV}  \,\left(\frac{1+z}{1.2}\right)^{\frac{12-7p}{7(p-1)}}\tilde{g}^{\frac{1}{p-1}}(1.9)\chi_{e,-0.3}^{-1}\,\varepsilon^{\frac{1}{p-1}}_{\rm e_r,-1}\,\varepsilon^{\frac{2-p}{4(p-1)}}_{\rm B_r,-2}\,A^{\frac{22-21p}{28(p-1)}}_{\rm W,-2}E^{-\frac{4-7p}{14(p-1)}}_{0,52.3}\,\Gamma_{0,2}^{-\frac{12+7p}{14(p-1)}}\,t^{-\frac{5}{7(p-1)}}_{2.5}
\eary

and for $p>2$ yields

\bary
h\nu^{\rm ssc}_{\rm KN, m, r}\simeq3.8\times10^{-1}\,{\rm GeV}  \,\left(\frac{1+z}{1.2}\right)^{-\frac{2}{7}}g(2.2)\chi_{e,-0.3}^{-1}\,\varepsilon_{\rm e_r,-1}\,A^{-\frac{5}{7}}_{\rm W,-2} \,\Gamma_{0,2}^{-\frac{13}{7}}\,E^{\frac{5}{7}}_{0,52.3}\,t^{-\frac{5}{7}}_{2.5}\,.
\eary

For $\nu^{\rm ssc}_{\rm m, r}<\nu^{\rm ssc}_{\rm c, r}$, the spectral break  becomes

\bary
h\nu^{\rm ssc}_{\rm KN, c, r}\simeq 7.8\times 10^{-1}\,{\rm GeV}  \,\left(\frac{1+z}{1.2}\right)^{-\frac{32}{21}} \left(\frac{1+\mathcal{Y_{\rm r}}}{2} \right)^{-1}\,\varepsilon^{-1}_{\rm B_r,-2}\,A^{-\frac{31}{21}}_{\rm W,-2}\,\Gamma_{0,2}^{\frac{2}{21}}\,E^{\frac{10}{21}}_{0,52.3}\,t^{\frac{11}{21}}_{2.5},\,\,\,
\eary


\subsection{SSC Light curves with energy injection}

In this section, the energy-injection scenario is treated under the assumption of adiabatic evolution ($\epsilon \approx 0$), in order to isolate the dynamical impact of energy injection on the reverse-shock emission. We note that energy injection does not necessarily imply a fully radiative evolution, as the effective radiative efficiency depends on the balance between energy losses and continuous energy replenishment.

The energy injected by the progenitor into the surrounding medium has the potential to generate revitalized shock waves, significantly impacting the dynamics of the environment. The continuous luminosity released into the blastwave can be described by $L_{\rm inj}= L_0\,\left(\frac{t}{t_0}\right)^{-q}$ where $q$ is the energy injection index, $t_0$ is the injection timescale and $L_0$ the initial luminosity \citep[e.g.,][]{2006ApJ...642..354Z}. The energy injected into the circumburst medium becomes $E=\int L_{\rm inj} dt \simeq E_{\rm inj} \left(\frac{t}{t_0}\right)^{1-q}$ with $E_{\rm inj}=L_0t_0/(1-q)$ for $q\neq 1$.

\subsubsection{Thick-shell scenario}

Considering the energy injection scenario, the critical Lorentz factor in a constant-density medium and in stellar wind is given in Eq. \ref{Gc_q}. Hereafter, we adopt the value of energy injection index and timescale $q=0.5$ and $t_0=10\,{\rm s}$, unless otherwise specified.

\paragraph{Constant-density medium.}

 The scaling of the hydrodynamic variables in region 3 advances as $\gamma_3\propto t^{-\frac{5+2q}{16}}$, $n_{3}\propto t^{-\frac{7+6q}{16}}$, $p_{3}\propto t^{-\frac{10+3q}{12}}$ and $N_{\rm e}\propto t^{1-q}$.  For $1<p<2$ ($p > 2$), the evolution of the magnetic field, the minimum and cooling electron Lorentz factors is $B'\propto t^{-\frac{10+3q}{24}}$, $\gamma_{\rm m,r}\propto t^{-\frac{68-3p(5+2q)}{96(p-1)}}\,(t^{-\frac{19-6q}{48}})$, $\gamma_{\rm c,r}\propto t^{\frac{7+18q}{48}}$, respectively.  The synchrotron spectral breaks and the maximum synchrotron flux for $1<p<2$ ($p > 2$) evolve as $\nu^{\rm syn}_{\rm m, r}\propto t^{-\frac{33+20p-6q(2-p)}{48(p-1)}}\,(t^{-\frac{73}{48}})$, $\nu^{\rm syn}_{\rm c, r}\propto t^{\frac{8q-7}{16}}$ and $F^{\rm syn}_{\rm max,r}\propto t^{\frac{13-60q}{48}}$, respectively. 

The characteristic spectral break for $1<p<2$ is

\bary\label{break_thick_aft_k0_q}
h \nu^{\rm ssc}_{\rm m, r} = 
7.2\times 10^{-9}\,{\rm GeV}  \, \left(\frac{1+z}{1.2} \right)^{\frac{137-43p}{48(p-1)}}\,\tilde{g}^{\frac{4}{p-1}}(1.9)\,\chi_{\rm e,-0.3}^{-4} \varepsilon^{\frac{4}{p-1}}_{\rm e_r,-1} \varepsilon^{\frac{3-p}{2(p-1)}}_{\rm B_r,-2}n^{\frac{7-2p}{4(p-1)}}_{-1}\,\Delta_{11.8}^{\frac{5(25+p)}{48(p-1)}}\,\Gamma_{0,2.5}^{\frac{4}{p-1}}\,E^{-\frac{1}{4(p-1)}}_{\rm inj, 53.3}t^{-\frac{101+5p-12q}{48(p-1)}}_{1.5}\,,
\eary

and for $2>p$ is

\bary
h \nu^{\rm ssc}_{\rm m, r} = 8.5\times 10^{-6}\,{\rm GeV}  \, \left(\frac{1+z}{1.2} \right)^{\frac{17}{16}}\,g^4(2.2)\,\chi_{\rm e,-0.3}^{-4}\, \,\varepsilon^4_{\rm e_r,-1} \varepsilon^{\frac12}_{\rm B_r,-2}  n^{\frac{3}{4}}_{-1}\,\Gamma^4_{0,2.5}\Delta_{11.8}^{\frac{45}{16}}\,E^{-\frac{1}{4}}_{\rm inj,53.3}t^{-\frac{37-4q}{16}}_{1.5}\,.
\eary

Taking into account the adiabatic expansion of the flow, the cutoff spectral break with $\nu^{\rm ssc}_{\rm cut, r}=\nu^{\rm ssc}_{\rm c, r}(t_{\rm x})\,\left(\frac{t}{t_{\rm x}} \right)^{-\frac{33}{16}}$ is given by

\bary
h \nu^{\rm ssc}_{\rm cut, r}\simeq 3.0\,{\rm GeV}\, \left(\frac{1+z}{1.2} \right)^{\frac{20q-3}{16}}\,\varepsilon^{-\frac72}_{\rm B_r,-2} n^{-\frac{9}{4}}_{-1}\,   \left(\frac{1+\mathcal{Y_{\rm r}}}{2}\right)^{-4}\,  \Delta^{\frac{9+20q}{16}}_{11.8}\,E^{-\frac{5}{4}}_{\rm inj,53.3}\,t^{-\frac{33}{16}}_{1.5}\,.
\eary

The peak flux of the SSC component becomes 

\bary
F^{\rm ssc}_{\rm max, r}\simeq 1.0\times 10^{-1}\,{\rm mJy}\,\left(\frac{1+z}{1.2} \right)^{\frac{89}{48}}\,G^{-1}(2.2)\,\chi_{e,-0.3}\,   \varepsilon^{\frac12}_{\rm B_r,-2}\,  \Gamma^{-1}_{0,2.5}\,n_{-1}\,  \Delta_{11.8}^{\frac{17}{48}} \,d^{-2}_{\rm z,27.5} \,E^{\frac{3}{2}}_{\rm inj,53.3}\,t^{\frac{31-72q}{48}}_{1.5}. \,\,\,\,\,\,\,\,
\eary

The maximum Lorentz factor and the respective SSC spectral break for $1<p<2$ become

\begin{equation}\nonumber
    \gamma_{\rm max, r}\simeq 2.7\times 10^7\left(\frac{1+z}{1.2} \right)^{-\frac{13}{48}}\,\varepsilon^{-\frac14}_{\rm B_r,-2} n^{-\frac{3}{16}}_{-1}\,\Delta^{-\frac{1}{12}}_{11.8}\,E^{-\frac{1}{16}}_{\rm inj,53.3}\,t^{\frac{10+3q}{48}}_{1.5}\, ,
\end{equation}

and 

\begin{equation}\nonumber
h \nu^{\rm ssc}_{\rm max, r}\simeq3.7\times 10^{15}\,{\rm GeV} \,\left(\frac{1+z}{1.2} \right)^{-\frac{53}{48}}\,\varepsilon^{-\frac12}_{\rm B_r,-2}\,  n_{-1}^{-\frac{1}{2}}\,\Delta_{11.8}^{-\frac{5}{48}}\,t^{\frac{5}{48}}_{1.5}\,, 
\end{equation}
respectively.
In the case $\nu^{\rm ssc}_{\rm c,r} < \nu^{\rm ssc}_{\rm m,r}$, the spectral breaks in the KN regime for $1<p<2$ can be expressed as

\bary
h \nu^{\rm ssc}_{\rm KN, m, r} =
3.3\,{\rm GeV}\,\left(\frac{1+z}{1.2}\right)^{\frac{122-75p}{96(p-1)}} \tilde{g}^{\frac{1}{p-1}}(1.9)\,\chi_{e,-0.3}^{-1}\,\varepsilon^{\frac{1}{p-1}}_{\rm e_r,-1}\,\,\varepsilon^{\frac{2-p}{4(p-1)}}_{\rm B_r,-2}\,n^{\frac{5(2-p)}{16(p-1)}}_{-1}\,E^{\frac{p-2}{16(p-1)}}_{\rm inj,53.3}\Delta^{\frac{62+3p}{96(p-1)}}_{11.8}\,\Gamma_{0,2.5}^{\frac{1}{p-1}}\,t^{-\frac{38+15p-6q(2-p)}{96(p-1)}}_{1.5}
\eary

and for $p > 2$ is

\bary
h \nu^{\rm ssc}_{\rm KN, m, r} =2.0\times 10\,{\rm GeV}\,\left(\frac{1+z}{1.2}\right)^{-\frac{7}{24}} g(2.2)\,\chi_{e,-0.3}^{-1}\,\varepsilon_{\rm e_r,-1}\,\Gamma_{0,2.5}\Delta^{\frac{17}{24}}_{11.8}\,t^{-\frac{17}{24}}_{1.5}\,.
\eary

In the case $\nu^{\rm ssc}_{\rm m, r}<\nu^{\rm ssc}_{\rm c, r}$, the spectral break becomes

\bary
h \nu^{\rm ssc}_{\rm KN, c, r}\simeq 4.7\times 10^2\,{\rm GeV}\,\left(\frac{1+z}{1.2}\right)^{-\frac{13}{12}} \left(\frac{1+\mathcal{Y_{\rm r}}}{2} \right)^{-1}\,\varepsilon^{-1}_{\rm B_r,-2}\,n^{-\frac{3}{4}}_{-1}\,E^{-\frac{1}{4}}_{\rm inj,53.3}\,\Delta^{-\frac{1}{3}}_{11.8}\,t^{\frac{3q-2}{12}}_{1.5}.
\eary

\paragraph{Stellar-wind medium.}

The evolution of the hydrodynamic variables in region 3 is $\gamma_3\propto t^{-\frac{1+2q}{8}}$, $n_{3}\propto t^{-\frac{11-2q}{8}}$, $p_{3}\propto t^{-\frac{4-q}{2}}$ and $N_{\rm e}\propto t^{1-q}$. Therefore, the magnetic field, the breaks of the electron Lorentz factors for $1<p<2$ ($p > 2$) evolve as ${B'}_{\rm r}\propto t^{-\frac{4-q}{4}}$, $\gamma_{\rm m,r}\propto t^{-\frac{24-7p-2q(4-p)}{16(p-1)}}\,(t^{-\frac{5-2q}{8}})$ and $\gamma_{\rm c,r}\propto t^{\frac{9-2q}{8}}$, respectively. The evolution of the synchrotron spectral breaks and the maximum synchrotron flux is $\nu^{\rm syn}_{\rm m, r}\propto t^{-\frac{15+2p-2q(4-p)}{8(p-1)}}\,(t^{-\frac{19-4q}{8}})$, $\nu^{\rm syn}_{\rm c, r}\propto t^{\frac{9-4q}{8}}$ and $F^{\rm syn}_{\rm max,r}\propto t^{-\frac{1+8q}{8}}$, respectively.

The characteristic spectral break for $1<p<2$ is

\bary\label{break_thick_aft_k2_q}
h \nu^{\rm ssc}_{\rm m, r} =
7.0\times 10^{-8}\,{\rm GeV}\,\left(\frac{1+z}{1.2}\right)^{\frac{31-9p}{8(p-1)}} \tilde{g}^{\frac{4}{p-1}}(1.9) \chi_{\rm e,-0.3}^{-4} \varepsilon^{\frac{4}{p-1}}_{\rm e_r,-1} \varepsilon^{\frac{3-p}{2(p-1)}}_{\rm B_r,-2} A^{\frac{7-2p}{2(p-1)}}_{\rm W,-2} \Delta^{\frac{3(5+p)}{8(p-1)}}_{11.8}\,E^{-\frac{4-p}{2(p-1)}}_{\rm inj,53.3} t^{-\frac{39-5p-4q(4-p)}{8(p-1)}}_{1.5}\hspace{1.1cm}{\rm for} \hspace{0.1cm} { 1<p<2 }
\eary

\bary
h \nu^{\rm ssc}_{\rm m, r} =5.7\times 10^{-5}\,{\rm GeV} \left(\frac{1+z}{1.2}\right)^{\frac{13}{8}}\,g^{4}(2.2)\,\chi_{\rm e,-0.3}^{-4}\,\epsilon^4_{\rm e_r,-1} \epsilon^{\frac12}_{\rm B_r,-2} \,A^{\frac{3}{2}}_{\rm W,-2}\,\Delta_{11.8}^{\frac{21}{8}}\, \Gamma^{4}_{0,2.5}\,  E^{-1}_{\rm inj, 53.3}\,t^{-\frac{29-8q}{8}}_{1.5}\hspace{3.1cm}{\rm for} \hspace{0.1cm} { p > 2 }
\eary

For an adiabatically expanding flow, the cutoff spectral breaks with $\nu^{\rm ssc}_{\rm cut, r}=\nu^{\rm ssc}_{\rm c, r}(t_{\rm x})\,\left(\frac{t}{t_{\rm x}} \right)^{-\frac{21}{8}}$ is given by

\bary
h \nu^{\rm ssc}_{\rm cut, r}\simeq 2.5\times 10^{-4}\,{\rm GeV} \left(\frac{1+z}{1.2}\right)^{\frac{21-8q}{8}}\,\left(\frac{1+\mathcal{Y_{\rm r}}}{2} \right)^{-4}  \epsilon^{-\frac72}_{\rm B_r,-2}\,A^{-\frac{9}{2}}_{\rm W,-2}\,\Delta_{11.8}^{\frac{45-8q}{8}}\, E_{\rm inj, 53.3}\, t^{-\frac{21}{8}}\,.
\eary

The maximum flux produced by the SSC mechanism yields

\bary
F^{\rm ssc}_{\rm max,r} \simeq  1.4\,{\rm mJy}\,\left(\frac{1+z}{1.2}\right)^{\frac{19}{8}}\,G^{-1}(2.2)\,\chi_{\rm e,-0.3}\,  \epsilon^{\frac12}_{\rm B_r,-2} \, A^{2}_{W,-2}\, \Gamma^{-1}_{0,2}\,\Delta_{11.8}^{-\frac18}\, d^{-2}_{\rm z,27.5}\,E^{\frac{1}{2}}_{\rm inj, 53.3}\, t^{-\frac{7+4q}{8}}_{1.5}\,,
\eary

The maximum Lorentz factor and the corresponding spectral break for $1<p<2$ are given by

\begin{equation}\nonumber
    \gamma_{\rm max, r}\simeq 1.4\times 10^7\left(\frac{1+z}{1.2} \right)^{-\frac{3}{8}}\,\varepsilon^{-\frac14}_{\rm B_r,-2} A_{W,-2}^{-\frac{3}{8}}\,E^{\frac{1}{8}}_{\rm inj,53.3}\,t^{\frac{4-q}{8}}_{1.5}\, ,
\end{equation}

and

\begin{equation}\nonumber
h \nu^{\rm ssc}_{\rm max, r}\simeq 8.2\times 10^{14}\,{\rm GeV}\,\left(\frac{1+z}{1.2} \right)^{-\frac{11}{8}}\,\varepsilon^{-\frac12}_{\rm B_r,-2}\,  A_{W,-2}^{-1}\,\Delta_{11.8}^{\frac{1}{8}}\,E^{\frac{1}{2}}_{\rm inj, 53.3}\,t^{\frac{7-4q}{8}}_{1.5}\,. 
\end{equation}

respectively. For $\nu^{\rm ssc}{\rm c,r} < \nu^{\rm ssc}{\rm m,r}$, the KN-regime spectral breaks for $1<p<2$ and $p > 2$ are given by

\bary
h \nu^{\rm ssc}_{\rm KN, m, r}\simeq
3.7\,{\rm GeV}\,\left(\frac{1+z}{1.2}\right)^{\frac{26-15p}{16(p-1)}} \tilde{g}^{\frac{1}{p-1}}(1.9) \chi_{\rm e,-0.3}^{-1} \varepsilon^{\frac{1}{p-1}}_{\rm e_r,-1}\,\varepsilon^{\frac{2-p}{4(p-1)}}_{\rm B_r,-2} A^{\frac{5(2-p)}{8(p-1)}}_{\rm W,-2} \Delta^{\frac{3(p+2)}{16(p-1)}}_{11.8}\,\Gamma_{0,2.5}^{\frac{1}{p-1}}\,E^{-\frac{3(2-p)}{8(p-1)}}_{\rm inj,53.3} t^{-\frac{22-5p-6q(2-p)}{16(p-1)}}_{1.5}\,.
\eary

and

\bary
h \nu^{\rm ssc}_{\rm KN, m, r}= 1.9\times 10\,{\rm GeV}\,\left(\frac{1+z}{1.2}\right)^{-\frac14} g(2.2)\,\chi_{e,-0.3}^{-1}\,\varepsilon_{\rm e_r,-1}\,\Gamma_{0,2.5}\,\Delta^{\frac34}_{11.8}\,t^{-\frac34}_{1.5}\,,
\eary

respectively.

For $\nu^{\rm ssc}_{\rm m, r}<\nu^{\rm ssc}_{\rm c, r}$, the spectral break yields

\bary
h \nu^{\rm ssc}_{\rm KN, c, r}&\simeq& 2.8\times 10\,{\rm GeV}\,\left(\frac{1+z}{1.2}\right)^{-\frac32} \left(\frac{1+\mathcal{Y_{\rm r}}}{2} \right)^{-1}\,\varepsilon^{-1}_{\rm B_r,-2}\,A^{-\frac{3}{2}}_{\rm W,-2}\,E^{\frac12}_{\rm inj,53.3}\,t^{\frac{2-q}{2}}_{1.5}\,.
\eary

\subsubsection{Thin-shell scenario}

The shock-crossing time in a constant-density medium and stellar wind is expressed in Eq. \ref{tx_q}. 

\paragraph{Constant-density medium}

The timescales of the hydrodynamic variables in region 3 are $\gamma_3\propto t^{-\frac25}$, $n_{3}\propto t^{-\frac{6}{7}}$, $p_{3}\propto t^{-\frac{8}{7}}$ and $N_{\rm e}\propto t^{1-q}$. For $1<p<2$ ($p > 2$),  the evolution of the post-shock magnetic field and the minimum and cooling electron Lorentz factors becomes ${B'}_{\rm r} \propto t^{-\frac47}$,  $\gamma_{\rm m, r}\propto t^{-\frac{24-7p}{35(p-1)}}\,(t^{-\frac27})$ and $\gamma_{\rm c,r}\propto  t^{\frac{19}{35}}$, respectively. Similarly, for $1<p<2$ ($p > 2$)  the evolution of the synchrotron spectral breaks and the maximum flux is $\nu^{\rm syn}_{\rm m, r}\propto t^{-\frac{2(10p+7)}{35(p-1)}}\,(t^{-\frac{54}{35}})$, $\nu^{\rm syn}_{\rm c, r}\propto t^{\frac{4}{35}}$  and $F^{\rm syn}_{\rm max,r }\propto t^{\frac{1-35q}{35}}$.

The characteristic spectral break for $1<p<2$ is

\bary\label{break_thin_aft_k0_q}
h \nu^{\rm ssc}_{\rm m, r} =
2.6\,\times 10^{-8}\,{\rm GeV} \left(\frac{1+z}{1.2}\right)^{\frac{4(64-13p)+35q(1-p)}{35(p-1)(2+q)}}\,\tilde{g}^{\frac{4}{p-1}}(1.9)\,\chi_{\rm e,-0.3}^{-4}\,\varepsilon^{\frac{4}{p-1}}_{\rm e_r,-1} \varepsilon^{\frac{3-p}{2(p-1)}}_{\rm B_r,-2} \, n^{\frac{2(43-41p)+35q(3-p)}{70(p-1)(q+2)}}_{-1}\Gamma^{-\frac{2(178+24p-35q)}{35(p-1)(2+q)}}_{0,2}\,  E^{\frac{2(3p+31)}{35(p-1)(2+q)}}_{\rm inj,52.3}\cr
\hspace{12cm}t^{-\frac{2(31+3p)}{35(p-1)}}_{2.5}
\eary

\bary\label{break_thin_aft_k0_q}
h \nu^{\rm ssc}_{\rm m, r} =3.3\,\times 10^{-7}\,{\rm GeV} \left(\frac{1+z}{1.2}\right)^{\frac{152-35q}{35(2+q)}}\,g^4(2.2)\,\chi_{\rm e,-0.3}^{-4}\,\varepsilon^4_{\rm e_r,-1} \varepsilon^{\frac12}_{\rm B_r,-2}\,n^{-\frac{78-35q}{70(2+q)}}_{-1}\, \Gamma^{-\frac{2(226-35q)}{35(2+q)}}_{0,2}\,  E^{\frac{74}{35(2+q)}}_{\rm inj,52.3}\,t^{-\frac{74}{35}}_{2.5}
\eary

Under adiabatic expansion of the flow, the cutoff spectral break with $\nu^{\rm ssc}_{\rm cut, r}=\nu^{\rm ssc}_{\rm c, r}(t_{\rm x})\,\left(\frac{t}{t_{\rm x}} \right)^{-\frac{74}{35}}$ can be written as

\bary
h \nu^{\rm ssc}_{\rm cut, r}\simeq 1.4\times 10^{-1}\,{\rm GeV} \left(\frac{1+z}{1.2}\right)^{\frac{3(4+35q)}{35(2+q)}}\,\left(\frac{1+\mathcal{Y_{\rm r}}}{2}\right)^{-4}  \varepsilon^{-\frac72}_{\rm B_r,-2}\,n^{-\frac{358+245q}{70(2+q)}}_{-1} \, \Gamma^{-\frac{2(86+175q)}{35(2+q)}}_{0,2}\, E^{-\frac{66}{35(2+q)}}_{\rm inj,52.3}\, t^{-\frac{74}{35}}_{2.5}\,.
\eary

The peak flux of the SSC component becomes

\bary
F^{\rm ssc}_{\rm max,r} \simeq  2.3\times 10^{-8}\,{\rm mJy}\, \left(\frac{1+z}{1.2}\right)^{\frac{186}{35(2+q)}} \,G^{-1}(2.2)\,\chi_{\rm e,-0.3}\,\varepsilon^{\frac12}_{\rm B_r,-2} \, n^{\frac{86+105q}{70(2+q)}}_{-1} \,\Gamma^{-\frac{286-105q}{35(2+q)}}_{0,2}\,d^{-2}_{\rm z,27.5}E^{\frac{132+35q}{35(2+q)}}_{\rm inj,52.3}\, t^{\frac{8-35q}{35}}_{2.5}\,.
\eary

 For $1<p<2$, the maximum Lorentz factor and the respective spectral break become 

\begin{equation}\nonumber
    \gamma_{\rm max, r}\simeq 3.2\times 10^7 \left(\frac{1+z}{1.2} \right)^{-\frac{6}{7(2+q)}}\,\varepsilon^{-\frac14}_{\rm B_r,-2} n^{-\frac{6+7q}{28(2+q)}}_{-1}\,\Gamma^{\frac{18-7q}{14(2+q)}}_{0,2}\,E_{\rm inj,52.3}^{-\frac{2}{7(2+q)}}\,t^{\frac{2}{7}}_{2.5}\, ,
\end{equation}

and
\begin{equation}\nonumber
h \nu^{\rm ssc}_{\rm max, r}\simeq 2.7\times 10^{15}\,{\rm GeV} \,\left(\frac{1+z}{1.2} \right)^{-\frac{88+35q}{35(2+q)}}\,\varepsilon^{-\frac12}_{\rm B_r,-2}\,  n_{-1}^{-\frac{58+35q}{70(2+q)}}\,\Gamma^{\frac{48}{35(2+q)}}_{0,2}\,E^{-\frac{6}{35(2+q)}}_{\rm inj,52.3}\,t^{\frac{6}{35}}_{2.5}, 
\end{equation}

respectively.

For $\nu^{\rm ssc}_{\rm c, r}<\nu^{\rm ssc}_{\rm m, r}$,  the spectral breaks in the KN regime for $1<p<2$ is

\bary
h\nu^{\rm ssc}_{\rm KN, m, r}\simeq
3.1\,{\rm GeV}  \,\left(\frac{1+z}{1.2}\right)^{\frac{100-49p-35q(p-1)}{35(p-1)(2+q)}}\tilde{g}^{\frac{1}{p-1}}(1.9)\chi_{e,-0.3}^{-1}\,\varepsilon^{\frac{1}{p-1}}_{\rm e_r,-1}\,\epsilon^{\frac{2-p}{4(p-1)}}_{\rm B_r,-2}\,n^{\frac{2(50-49p)+35q(2-p)}{140(p-1)(2+q)}}_{\rm -1} \,\Gamma^{-\frac{160+7p(6-5q)}{70(p-1)(2+q)}}_{0,2}\,E^{\frac{10+7p}{35(p-1)(2+q)}}_{\rm inj,52.3}\cr
\hspace{11.6cm}\,t^{-\frac{10+7p}{35(p-1)}}_{2.5}\hspace{1cm}{\rm for} \hspace{0.1cm} { 1<p<2 }
\eary

and for $p > 2$ is

\bary
h\nu^{\rm ssc}_{\rm KN, m, r}\simeq5.7\,{\rm GeV}  \,\left(\frac{1+z}{1.2}\right)^{\frac{2-35q}{35(2+q)}}g(2.2)\chi_{e,-0.3}^{-1}\,\varepsilon_{\rm e_r,-1}\,n^{-\frac{24}{35(2+q)}}_{\rm -1} \,\Gamma^{-\frac{122-35q}{35(2+q)}}_{0,2}\,E^{\frac{24}{35(2+q)}}_{\rm inj,52.3}\,t^{-\frac{24}{35}}_{2.5},\hspace{3.5cm}{\rm for} \hspace{0.1cm} { p > 2 }
\eary

For $\nu^{\rm ssc}_{\rm m, r}<\nu^{\rm ssc}_{\rm c, r}$, the spectral break in the KN regime becomes

\bary
h\nu^{\rm ssc}_{\rm KN, c, r}\simeq 0.9\times10^2\,{\rm GeV}  \,\left(\frac{1+z}{1.2}\right)^{-\frac{24}{7(2+q)}}\, \left(\frac{1+\mathcal{Y_{\rm r}}}{2} \right)^{-1}\,\epsilon^{-1}_{\rm B_r,-2}\,n^{-\frac{6+7q}{7(2+q)}}_{-1} \,\Gamma^{\frac{2(18-7q)}{7(2+q)}}_{0,2}\,E^{-\frac{8}{7(2+q)}}_{\rm inj, 52.3}\,t^{\frac{1}{7}}_{2.5}.
\eary

%
%
\paragraph{Stellar-wind medium}
The timescales of the hydrodynamic quantities in region 3 are $\gamma_3\propto t^{-\frac13}$, $n_{3}\propto t^{-\frac{2(11-7q)}{7}}$, $p_{3}\propto t^{-\frac{2(37-21q)}{21}}$ and $N_{\rm e}\propto t^{1-q}$.  Therefore,  the magnetic field, the Lorentz factors for the minimum and cooling electrons for $1<p<2$ ($p > 2$) evolve as ${B'}_{\rm r}\propto t^{-\frac{37-21q}{21}}$, $\gamma_{\rm m,r}\propto t^{-\frac{22-7p}{21(p-1)}}\,(t^{-\frac{8}{21}})$ and $\gamma_{\rm c,r}\propto t^\frac{2(10-7q)}{7}$, respectively.  The evolution of synchrotron spectral breaks, and the maximum synchrotron flux for $1<p<2$ ($p > 2$) is $\nu^{\rm syn}_{\rm m, r}\propto t^{-\frac{10p-7q(p-1)}{7(p-1)}}\,(t^{-\frac{20-7q}{7}})$, $\nu^{\rm syn}_{\rm c, r}\propto t^{\frac{76-63q}{21}}$ and $F^{\rm syn}_{\rm max,r}\propto t^{-\frac{23}{21}}$, respectively.

The characteristic spectral break for $1<p<2$ is

\bary\label{break_thin_aft_k2_q}
h \nu^{\rm ssc}_{\rm m, r} = 
2.1\times 10^{-11}\,{\rm GeV}\, \left(\frac{1+z}{1.2}\right)^{-\frac{19-37p-21q(5-3p)}{21q(p-1)}}\,\tilde{g}^{\frac{4}{p-1}}(1.9)\,\chi_{\rm e,-0.3}^{-4}\,\varepsilon^{\frac{4}{p-1}}_{\rm e_r,-1} \varepsilon^{\frac{3-p}{2(p-1)}}_{\rm B_r,-2} \,A^{\frac{2(19-37p)+21q(p+1)}{42q(p-1)}}_{\rm W,-2}\, \Gamma^{\frac{2[2(19-37p)-21q(4-3p)]}{21q(p-1)}}_{0,2}\,  \cr
\hspace{9.1cm} E^{-\frac{19-37p+21q(p-1)}{21q(p-1)}}_{\rm inj,52.3}\, t^{-\frac{4(11+4p)-21q(p-1)}{21(p-1)}}_{2.5}
\eary

and for $2>p$ is

\bary
h \nu^{\rm ssc}_{\rm m, r} =4.5\times 10^{-10}\,{\rm GeV}\, \left(\frac{1+z}{1.2}\right)^{\frac{55-21q}{21q}}\,g^4(2.2)\,\chi_{\rm e,-0.3}^{-4}\,\varepsilon^4_{\rm e_r,-1} \varepsilon^{\frac12}_{\rm B_r,-2}\,A^{-\frac{110-63q}{42q}}_{\rm W,-2}  \, \Gamma^{-\frac{4(55-21q)}{21q}}_{0,2}\,  E^{\frac{55-21q}{21q}}_{\rm inj,52.3}\,t^{-\frac{76-21q}{21}}_{2.5}
\eary

Taking into account the adiabatic expansion of the flow, the cutoff spectral break with $\nu^{\rm ssc}_{\rm cut, r}=\nu^{\rm ssc}_{\rm c, r}(t_{\rm x})\,\left(\frac{t}{t_{\rm x}} \right)^{-\frac{55}{21}}$ are given by

\bary
h \nu^{\rm ssc}_{\rm cut, r}\simeq 3.0\times 10^{-7}\,{\rm GeV} \left(\frac{1+z}{1.2}\right)^{\frac{2(59-42q)}{21q}}\,\left(\frac{1+\mathcal{Y_{\rm r}}}{2}\right)^{-4}  \varepsilon^{-\frac72}_{\rm B_r,-2}\,A^{-\frac{236+147q}{42q}}_{\rm W,-2}  \, \Gamma^{-\frac{4(118-21q)}{21q}}_{0,2}\, E^{\frac{118}{21q}}_{\rm inj,52.3}\, t^{-\frac{55}{21}}_{2.5}\,.
\eary

The peak flux associated with the SSC emission is

\bary
F^{\rm ssc}_{\rm max,r} \simeq  9.2\times 10^{-9}\,{\rm mJy}\, \left(\frac{1+z}{1.2}\right)^{\frac{3+14q}{7q}} \,G^{-1}(2.2)\,\chi_{\rm e,-0.3}\,\varepsilon^{\frac12}_{\rm B_r,-2} \, A^{-\frac{6-35q}{14}}_{\rm W,-2}\, \,\Gamma^{-\frac{12-7q}{7q}}_{0,2}\,d^{-2}_{\rm z,27.5}E^{\frac{3}{7q}}_{\rm inj,52.3}\, t^{-\frac{10}{7}}_{2.5}\,.
\eary

For $1<p<2$, the maximum Lorentz factor, along with the corresponding spectral break, can be expressed as

\begin{equation}\nonumber
    \gamma_{\rm max, r}\simeq 1.3\times 10^7 \left(\frac{1+z}{1.2} \right)^{-\frac{8}{21q}}\,\varepsilon^{-\frac14}_{\rm B_r,-2} A_{W,-2}^{\frac{32-63q}{84q}}\,\Gamma^{\frac{64-63q}{42q}}_{0,2}\,E^{-\frac{16-21q}{42q}}_{\rm inj,52.3}\,t^{\frac{37-21q}{42}}_{2.5}\, ,
\end{equation}

and 

\begin{equation}\nonumber
h \nu^{\rm ssc}_{\rm max, r}\simeq 4.6\times 10^{14}\,{\rm GeV} \,\left(\frac{1+z}{1.2} \right)^{-\frac{3+7q}{7q}}\,\varepsilon^{-\frac12}_{\rm B_r,-2}\,  A_{\rm W,-2}^{\frac{3(2-7q)}{14q}}\,\Gamma^{\frac{2(6-7q)}{7q}}_{0,2}\,E^{\frac{7q-3}{7q}}_{\rm inj,52.3}\,t^{\frac{10-7q}{7}}_{2.5}\,. 
\end{equation}

respectively.

In the case $\nu^{\rm ssc}_{\rm c,r} < \nu^{\rm ssc}_{\rm m,r}$, the spectral breaks in the KN regime for $1<p<2$ can be expressed as

\bary
h\nu^{\rm ssc}_{\rm KN, m, r}\simeq 
0.3\,{\rm GeV}  \,\left(\frac{1+z}{1.2}\right)^{-\frac{4-7p+7q(3p-4)}{14q(p-1)}}\tilde{g}^{\frac{1}{p-1}}(1.9)\chi_{e,-0.3}^{-1}\,\varepsilon^{\frac{1}{p-1}}_{\rm e_r,-1}\,\varepsilon^{\frac{2-p}{4(p-1)}}_{\rm B_r,-2}\,A^{\frac{2(4-7p)+7q(2-p)}{28q(p-1)}}_{\rm W,-2}E^{-\frac{4-7p}{14q(p-1)}}_{\rm inj,52.3}\Gamma^{\frac{4(4-7p)+7q(3p-4)}{14q(p-1)}}_{0,2}\cr
\hspace{11.7cm} t^{-\frac{5}{7(p-1)}}_{2.5}
\eary

and for $p > 2$ is

\bary
h\nu^{\rm ssc}_{\rm KN, m, r} = 0.7\,{\rm GeV}  \,\left(\frac{1+z}{1.2}\right)^{\frac{5-7q}{7q}}g(2.2)\chi_{e,-0.3}^{-1}\,\varepsilon_{\rm e_r,-1}\,A^{-\frac{5}{7q}}_{\rm W,-2} \,\Gamma^{-\frac{20-7q}{7q}}_{0,2}\,E^{\frac{5}{7q}}_{\rm inj,52.3}\,t^{-\frac{5}{7}}_{2.5}
\eary

For $\nu^{\rm ssc}_{\rm m, r}<\nu^{\rm ssc}_{\rm c, r}$, the spectral break in the KN regime becomes
\bary
h\nu^{\rm ssc}_{\rm KN, c, r}\simeq 2.0\,{\rm GeV}  \,\left(\frac{1+z}{1.2}\right)^{-\frac{32}{21q}} \left(\frac{1+\mathcal{Y_{\rm r}}}{2} \right)^{-1}\,\varepsilon^{-1}_{\rm B_r,-2}\,A^{\frac{32-63q}{21q}}_{\rm W,-2}\,\Gamma^{\frac{2(64-63q)}{21q}}_{0,2}\,E^{-\frac{2(16-21q)}{21q}}_{\rm inj,52.3}\,t^{\frac{53-42q}{21}}_{2.5}.
\eary

\subsection{Consistency of cooling regimes in forward and reverse shocks}

The adopted configuration of a fast-cooling forward shock ($\nu_{\rm c, f}^{\rm syn} < \nu_{\rm m, f}^{\rm syn}$) and a slow-cooling reverse shock ($\nu_{\rm m, r}^{\rm syn} < \nu_{\rm c, r}^{\rm syn}$) imposes non-trivial constraints on the microphysical parameters in both regions.\footnote{The subscript ``f" corresponds to quantities derived in the forward-shock region.} In the forward shock, the fast-cooling condition requires efficient energy transfer to relativistic electrons and strong magnetic-field amplification. This is typically achieved for relatively large values of $\varepsilon_{\rm B_f}$ and $\varepsilon_{\rm e_f}$, typically $\varepsilon_{\rm e_f} \sim 0.1$--$0.3$, consistent with expectations from relativistic collisionless shocks. In contrast, maintaining the reverse shock in the slow-cooling regime places constraints on the corresponding parameters $\varepsilon_{\rm e_r}$ and $\varepsilon_{\rm B_r}$. For the reverse shock, the slow-cooling condition can be evaluated using the standard relations between the forward- and reverse-shock quantities \citep{2003ApJ...595..950Z}

\bary
\nu_{\rm m, r}^{\rm syn}&\sim& \hat{\Gamma}^{-2} \mathcal{R}_e^2\,\mathcal{R}_B \nu_{\rm m, f}^{\rm syn}\cr
\nu_{\rm c, r}^{\rm syn}&\sim& \mathcal{R}_B^{-3} \nu_{\rm c, f}^{\rm syn}\,,
\eary

where $\mathcal{R}_e\equiv\frac{\varepsilon_{\rm e_r}}{\varepsilon_{\rm e_f}}$, $\mathcal{R}_B\equiv \left(\frac{\varepsilon_{\rm B_r}}{\varepsilon_{\rm B_f}}\right)^\frac12$ and $\hat{\Gamma}=\frac{\Gamma_{\rm c}^2}{\Gamma_0}$ and $\Gamma_0$ for the thick- and thin-shell regime, respectively.  The constraints on the microphysical parameters in the two regions become

\bary
\frac{\varepsilon_{\rm e_r}}{\varepsilon_{\rm e_f}}&<& 
\frac{\Gamma^2_{\rm c}}{\Gamma_0}\,\frac{\varepsilon_{\rm B_f}}{\varepsilon_{\rm B_r}} \hspace{0.8cm} {\rm for\, thick\, shell} \cr
\frac{\varepsilon_{\rm B_r}}{\varepsilon_{\rm B_f}}&<& 
\Gamma_0\,\frac{\varepsilon_{\rm e_f}}{\varepsilon_{\rm e_r}}\hspace{0.93cm} {\rm for\, thin\, shell}\,.
\eary

Using the standard relations between forward- and reverse-shock quantities, the condition $\nu_{\rm m, r}^{\rm syn} < \nu_{\rm c, r}^{\rm syn}$ implies that the magnetic field strength in the reverse-shock region cannot be too large. In practice, this requires $\varepsilon_{\rm B_r} \lesssim \varepsilon_{\rm B_f}$.   The constraints become more restrictive in the thick-shell regime than in the thin-shell case, as the dynamics are additionally governed by the critical Lorentz factor $\Gamma_{\rm c}$ (Eqs.~\ref{Gc_eps} and \ref{Gc_q}).  The initial bulk Lorentz factor plays a critical role in determining the cooling regime of the reverse shock.  In the thick-shell regime, the reverse shock becomes relativistic, and larger values of $\Gamma_0$ further enhance its strength, increasing $\nu^{\rm syn}_{\rm m,r}$ and driving the system toward the fast-cooling regime.  As a result, maintaining the slow-cooling condition in the reverse shock requires increasingly smaller values of $\varepsilon_{\rm B_r}$ (and, to a lesser extent, $\varepsilon_{e_r}$) for higher $\Gamma_0$.   The electron energy fraction also plays a key role. Larger values of $\varepsilon_{\rm e_r}$ increase the minimum Lorentz factor of electrons and hence $\nu^{\rm syn}_{\rm m,r}$, again favoring fast cooling. Therefore, $\varepsilon_{\rm e_r}$ is not expected to exceed $\varepsilon_{\rm e_f}$ and is likely comparable or smaller.

From a physical perspective, these conditions are consistent with the different environments of the two shocks. The forward shock propagates into the external medium, where plasma instabilities efficiently amplify magnetic fields, whereas the reverse shock propagates into the ejecta, where magnetic field amplification may be less efficient and the microturbulence may decay more rapidly. In general, the combination of a fast-cooling forward shock and a slow-cooling reverse shock is self-consistent within a reasonable region of parameter space characterized by $\varepsilon_{\rm B_r} \lesssim \varepsilon_{\rm B_f}$ and $\varepsilon_{\rm e_r} \lesssim \varepsilon_{\rm e_f}$. This condition is naturally satisfied in the thin-shell regime, while in the thick-shell case it requires more restrictive parameter choices.  This highlights the strong coupling between the dynamical evolution of the ejecta and the microphysical conditions required to maintain distinct cooling regimes in the forward and reverse shocks.

It is worth noting that, although not adopted in this work, a configuration in which both the forward and reverse shocks are in the fast-cooling regime ($\nu_{\rm c,f}^{\rm syn} < \nu_{\rm m,f}^{\rm syn}$ and $\nu_{\rm c,r}^{\rm syn} < \nu_{\rm m,r}^{\rm syn}$) would impose the constraint $1 < \hat{\Gamma}^{-2}\,\mathcal{R}_e^{2}\,\mathcal{R}_B^{4}$ on the microphysical parameters.

\section{SSC Light curves and Closure relations: Discussion and Analysis}\label{sec3}

\subsection{SSC Light curves }
\subsubsection{Radiative scenario}

We derive the temporal evolution of the SSC light curves during the radiative regime, considering SSC spectral breaks and maximum flux (Eqs. \ref{break_thick_aft_k0_e} and \ref{break_thick_aft_k2_e}). This is done for the thick-shell scenario of the reverse shock evolving in the density-constant medium and stellar wind, as detailed in Table \ref{tableReverseThick-radiative}.  The SSC light curves are derived without energy injection for two sets of electron spectral distribution values; $1<p<2$ and $p>2$.  Similarly, we calculate the temporal evolution of the SSC light curves (requiring Eqs. \ref{break_thin_aft_k0_e} and \ref{break_thin_aft_k2_e}) for the thin-shell reverse shock. 
This table shows the temporal index ($F^{\rm ssc}_{\rm \nu, r}\propto t^{-\alpha(\epsilon,p)}\nu^{-\beta}$) as a function of the radiative parameter and the electron spectral index for fast- and slow-cooling regime.   The SSC flux in the adiabatic regime is restored as $\epsilon\to 0$. Deviations from $\epsilon=0$ influence the evolution of SSC spectral breaks, leading to noticeable variations in SSC flux. It is important to note that in the thin-shell case, the SSC flux does not depend on this parameter; i.e. $\epsilon=0$.  This could be explained by taking into account that the shock-crossing time is larger than the fast-to-slow transition time. It is important to note that once the shock-accelerated electrons enter the slow-cooling regime, the adiabatic approximation $\epsilon\to 0$ can be applied to describe the hydrodynamics \citep{2000ApJ...529..151M}.  When the reverse shock lies in the thick-shell case and the electron distribution has a spectral index of $p=1.9$,  the SSC flux under the cooling condition $\nu^{\rm ssc}_{\rm m,r} < \nu < \nu^{\rm ssc}_{\rm cut,r}$ for $\epsilon=0\,(\epsilon=1)$ evolves as $\alpha=2.69\,(2.68)$ for stellar wind and $\alpha=1.88\,(2.47)$ for ISM, and when the spectral index is $p=2.2$,  the SSC flux varies as $\alpha=2.95\,(2.92)$ for wind and $\alpha=2.09\,(2.67)$ for ISM. Similarly, for the thin-shell case and an electron spectral index of $p=1.9$, the SSC flux evolves as $\alpha=1.82$ for ISM and  $\alpha=2.75$ for stellar wind, and with $p=2.2$, the SSC flux is characterized by $\alpha=2.04$ for ISM and  $\alpha=3.0$ for stellar wind. The GRB tail emission serves as a crucial marker, signaling the end of the prompt phase and the transition into the afterglow \citep{2006ApJ...642..354Z}.  This emission, which occurs from a dozen to a hundred seconds after the trigger time with typical decay ($\sim 3$), indicates whether the prompt emission and afterglow originate from distinct components or emitting regions. This steep decay at the end of the prompt episode could be modeled with the reverse-shock emission evolving in the thick-shell domain ($t_x\approx T_{90}$), which has a slope ($\sim 3$), as shown. 

On the other hand, the SSC flux in the cooling condition ($F^{\rm ssc}_{\rm \nu, r}\propto t^{-\alpha(\epsilon,p)}\nu^{\frac13}$) with $\epsilon=0\,(\epsilon=1)$ and $p=1.9$ evolves as $\alpha=0.09\,(0.78)$ for k=0 and  $\alpha=0.40\,(0.69)$ for k=2, and with $p=2.2$ it evolves as  $\alpha=0.17\,(0.85)$ for k=0) and  $\alpha=0.50\,(0.78)$ for k=2 when the reverse shock lies in the thick-shell regime. Taking into account the similar cooling condition, for $p=1.9$, the SSC flux evolves as $\alpha=0.01$ for k=0 and  $\alpha=0.45$ for k=2, and with $p=2.2$,  $\alpha=0.06$ for k=0 and  $\alpha=0.56$ for k=2 when the reverse shock lies in the thick-shell regime.  As discussed, this scenario in different cooling conditions illustrates the plateau phase and flares in both circunstellar medium; a constant-density medium and in a stellar wind.

\paragraph{Radiative parameter ($\epsilon$)}

The ratio between the radiated and dissipated energy that defines the radiative parameter indicates which fraction of the internal energy is released as radiation. For example, most of this energy is emitted when both the radiated and dissipated energies are comparable. This process leads to the afterglow phase within the radiative regime \citep{2000ApJ...529..151M, 2014MNRAS.445.1625N, 2024MNRAS.534.3783F}. The radiative parameter can be written as $\epsilon \equiv \varepsilon_{\rm e,r} \zeta$
%
%
where $\zeta$ represents the amount of energy emitted. This parameter is defined throught the breaks of electron Lorentz factors $\zeta=
\frac{\gamma_{\rm m,r}}{\gamma_{\rm c,r}}\frac{p-2}{3-p}\left[\frac{1}{p-2} \left(\frac{\gamma_{\rm c,r}}{\gamma_{\rm m,r}} \right)^{3-p} -1 \right]$ for $\gamma_{\rm m,r}\leq \gamma_{\rm c,r}$ and $1$ for $\gamma_{\rm c,r} < \gamma_{\rm m,r}$.  In other words, the radiative parameter is constant $\epsilon=\varepsilon_e$ during the fast cooling regime, and evolves during the slow-cooling regime.
%


It is worth noting that some works \citep[e.g., see][]{2000ApJ...532..281B, 2005ApJ...619..968W,2010MNRAS.403..926G} define the equivalent kinetic energy as a function of ${\rm s}$-parameter $E_0\left(\frac{t}{t_{\rm dec}}\right)^{-s}$ instead of the $\epsilon$-parameter $ E_0 \left( \frac{\Gamma}{\Gamma_0}\right)^\epsilon$, with $t_{\rm dec}$ the deceleration time. The relation between both parameters become 
\begin{eqnarray}
(3+s)&=&\frac{24}{(8-\epsilon)},\,\,\,\,\,\,\,\,\,\,\,{\rm for\,\,\,\, k=0}\cr
(1+s)&=&\frac{4}{(4-\epsilon)},\,\,\,\,\,\,\,\,\,\,\,{\rm for\,\,\,\, k=2}\,.
\end{eqnarray}
The ${\rm s}$-parameter  takes values in the range $0\leq~s \leq 3/7$ for ${\rm k=0}$ and $0\leq~s \leq 1/3$ for ${\rm k=2}$.

\subsubsection{Energy injection scenario}

We calculate the temporal evolution of the SSC light curves, considering energy injection during the adiabatic regime. We consider the SSC spectral breaks and maximum flux (Eqs. \ref{break_thick_aft_k0_q} and \ref{break_thick_aft_k2_q}) for the thick-shell scenario of the reverse shock, which is evolving in the interstellar medium (ISM) and stellar wind, as detailed in Table \ref{tableReverseThick-injection}. In this case, we derive the SSC light curves during the adiabatic regime for two sets of electron spectral distribution values; $1<p<2$ and $p > 2$.  Similarly, we calculate the temporal evolution of the SSC light curves (Eqs. \ref{break_thin_aft_k0_q} and \ref{break_thin_aft_k2_q}) for the thin-shell reverse shock.   This table details the temporal index ($F^{\rm ssc}_{\rm \nu, r}\propto t^{-\alpha(q,p)}\nu^{-\beta}$) as a function of the energy injection index and the electron spectral index for fast- and slow-cooling regime.   The most significant result is the explicit dependence of the temporal decay index $\alpha$ on the energy injection parameter $q$. For instance, in the thick-shell ISM case for $2 < p$ and $\nu_{\rm m,r}^{\rm ssc} < \nu < \nu_{\rm cut,r}^{\rm ssc}$, the flux decays as $\alpha = -\frac{173 - 111p + 12q(p-13)}{96}$; for a typical spectral value of $p=2.2$, this becomes $\alpha\approx0.74+1.35q$. In the standard scenario with no injection ($q=1$) we obtain $\alpha\approx0.74$, while for an injection scenario ($q=0.5$) we have $\alpha\approx1.42$  This demonstrates that an injection index $q < 1$ produces a steeper decay than the standard case. We also note that this formalism smoothly connects to the standard adiabatic case when $q=1$. The expressions in Table~\ref{tableReverseThick-injection} for this index are consistent with well-established adiabatic reverse shock models.  On the other hand, the addition of energy injection preserves the differences between the thick and thin-shell cases, which allows the shell type to be discriminated, independently of the value of $q$. In particular, the regime where $\nu < \nu_{\rm m,r}^{\rm ssc}$ is highly sensitive to injection: in the thick-shell wind case for $2 < p$, $\alpha = -\frac{2 - 5q}{6}$, meaning that an injection index of $q \approx 0.4$ can transition the light curve from a slow decay to a slow rise. In fact, for such an energy injection parameter, we may obtain a constant flux density, which directly links the observed plateau phases in afterglows to the energy injection.

Although the energy-injection scenario is developed in the adiabatic limit ($\epsilon \simeq 0$) for analytical simplicity, this assumption is not fundamental to the model. The formalism can be extended to arbitrary radiative efficiencies, allowing the effects of continuous energy injection and radiative losses to be treated simultaneously. Considering the injection scenario assuming radiative evolution for a constant-density medium, the characteristic spectral break for $1<p<2$ and $p>2$ scales as $h \nu^{\rm ssc}_{\rm m, r} \propto t^{-\frac{808-96q+5p(8-\epsilon)-125\epsilon}{48(p-1)(8-\epsilon)}}$ and $h \nu^{\rm ssc}_{\rm m, r} \propto t^{-\frac{296-32q-45\epsilon}{16(8-\epsilon)}}$.  The cutoff spectral break and The peak flux associated with the SSC emission evolve as $h \nu^{\rm ssc}_{\rm cut, r} \propto t^{-\frac{33}{16}}$ and $F^{\rm ssc}_{\rm max, r} \propto t^{\frac{248-576q-175\epsilon}{48(8-\epsilon)}}$, respectively. In the case of stellar-wind environment, the characteristic spectral break for $1<p<2$ and $p>2$ scales as  $h \nu^{\rm ssc}_{\rm m, r} \propto t^{\frac{(39-5p)(\epsilon-4)+16q(4-p)}{8(p-1)(4-\epsilon)}}$ and $h \nu^{\rm ssc}_{\rm m, r} \propto t^{-\frac{116-32q-29\epsilon}{8(4-\epsilon)}}$, respectively. The cutoff spectral break and The peak flux associated with the SSC emission evolve as $h \nu^{\rm ssc}_{\rm cut, r} \propto t^{-\frac{21}{8}}$ and $F^{\rm ssc}_{\rm max,r} \propto t^{-\frac{4(7+4q)-7\epsilon}{8(4-\epsilon)}}$.

\paragraph{Injection parameter ($q$)}

The injection parameter characterizes the temporal evolution of the energy injected into the blast wave following the prompt episode. Smaller values of this parameter indicate that the engine continues to inject energy for a longer duration, whereas larger values of ${\rm q}$ suggest a rapid decrease in injection.  A notable re-energization capable of flattening the afterglow light curve has been investigated concerning the continued activity of the central engine, specifically a magnetar with ${\rm q}=0$ or an accreting black hole with ${\rm q}=0.5$ \citep{2001ApJ...552L..35Z, 2006ApJ...642..354Z, 2005astro.ph.11699D, 2006Sci...311.1127D}. In the ${\rm q}\approx 1$ regime, the engine continues to inject energy, albeit at a gradually decreasing rate.  The energy provided by the progenitor compensates for the energy lost through radiation and the adiabatic expansion of the blast wave, as suggested in the shell-interaction scenario. Slower shells, characterized by a power-law distribution, catch up with external shocks \citep{1998ApJ...496L...1R}. It is worth mentioning that the stardard SSC snecario is recovered for ${\rm q=1}$ \citep{2024MNRAS.534.3783F}.

\subsection{SSC Closure Relations and Second Fermi-LAT Catalog}

The analysis sample consists of 86 GRBs with measured redshifts from 2FLGC, observed from 2008 until August 4, 2018. 
These selected bursts with high-energy emissions above $\geq 100\,{\rm MeV}$ exhibited long-lasting emissions ranging from 31 to 34,366 s, modeled with a PL and a broken power law (BPL).  In our analysis, we use these temporal indices, either a PL ($\alpha_{\rm L}$) or a BPL ($\alpha_{\rm L_i}$ with i=1 or 2), and the spectral index $\beta_{\rm L}$ obtained from the photon index $\Gamma_L=\beta_{\rm L} + 1$, all reported in this catalog.    The analysis of CRs, including the uncertainties in each result, is carried out based on the strategy outlined in \cite{2021PASJ...73..970D, 2021ApJS..255...13D, 2020ApJ...903...18S, 2022ApJ...934..188F, 2023MNRAS.525.1630F}. In the following, we apply this methodology to the CRs obtained in the radiative and energy injection scenarios.

\subsubsection{Radiative scenario}

 Table \ref{Table-radiative-CRs} shows the CRs of SSC reverse-shock scenario ($F^{\rm ssc}_{\rm \nu,r} \propto t^{-\alpha(\beta, \epsilon, p)} \nu^{-\beta}$) in the thick- and thin-shell case, as well as for ISM and stellar wind under each cooling condition. This table displays the SSC CRs during the radiative regime and without energy injection for two sets of electron spectral distribution values; $1<p<2$ and $p>2$.\\

In Figures~\ref{fig1:CRs_eps_k0_PL}--\ref{fig4:CRs_eps_k2_BPL}, we compare the observed spectral and temporal indices from the 2FLGC sample (green points) with the SSC CRs predicted for the reverse shock evolving in the radiative regime ($\epsilon=1$) without energy injection (purple curves). Figures~\ref{fig1:CRs_eps_k0_PL} and~\ref{fig2:CRs_eps_k_0_BPL} show results for a homogeneous ISM, with the first using PL fits and the second BPL fits to the LAT spectra. Figures~\ref{fig3:CRs_eps_k2_PL} 
and~\ref{fig4:CRs_eps_k2_BPL} present the corresponding comparisons for a stellar-wind environment ($k=2$). 
In all cases, the top and bottom rows distinguish between thick- and thin-shell dynamics, and 
columns correspond to different SSC cooling regimes. In all Figures, we highlight the coincidences with purple points.  The comparison shown in Figures~\ref{fig1:CRs_eps_k0_PL} to \ref{fig4:CRs_eps_k2_BPL} indicates that the LAT indices derived from both the PL and BPL models are only partially consistent with the radiative SSC reverse-shock closure relations. Notably, the highest number of matches—two coincidences—occurs in the thin-shell scenario under the cooling condition where $\nu^{\rm ssc}_{\rm m,r} < \nu < \nu^{\rm ssc}_{\rm cut,r}$, as illustrated in Figures~\ref{fig1:CRs_eps_k0_PL} and~\ref{fig2:CRs_eps_k_0_BPL}. This corresponds to a constant interstellar medium scenario.  Furthermore, in the case of $\nu<\nu^{\rm ssc}_{\rm m,r}$, we find no matches in any of the Figures and for $\nu^{\rm ssc}_{\rm cut,r}<\nu$ we only find one match in any Figure, which does not allow to discriminate between scenarios. Thus, the reverse shock evolving in the radiative regime cannot be considered a viable explanation for the LAT afterglow observations. While there appears to be a slight preference for the ISM thin-shell case, the evidence is too limited to support this conclusion. Overall, the fact that there are not many matches across the cooling regimes highlights that the radiative reverse shock scenario is not viable to account for the observed distributions.

Table ~\ref{tab:CR-Results_epsilon} provides a statistical analysis that compares the theoretical predictions of the SSC reverse shock model, within the radiative regime, against the observed data from the 2FLGC. This table illustrates the results for the purely radiative case ($\epsilon = 1.0$) and the purely adiabatic case ($\epsilon=0$). The analysis in this table is separated into thick-shell or thin-shell emission, and into constant ISM or stellar-wind environments.  Table ~\ref{tab:CR-Results_epsilon} lists the number and percentage of GRBs in the catalog whose measured spectral and temporal indices match the model's predictions for various cooling conditions. We consider a sample of bursts, some fitted with a simple PL and others requiring a BPL. For PL, we consider a sample of 65 bursts, while for BPL, we have 21. 

First, we observe that the total number of coincidences between the model and the data is relatively low. This indicates that the SSC emission from the reverse shock is not the primary mechanism for most of the LAT-detected GRBs. However, there is a subset of GRBs that do align with the predictions, highlighting the significance of this process in specific cases. In general, the derived CRs show much greater consistency with bursts that are fitted using a BPL compared to those fitted with a simple PL. The most successful scenario overall is found for a thin shell evolving in a constant-density medium during the slow-cooling condition $\mathbf{\nu_{\rm m,r}^{\rm ssc} < \nu < \nu_{\rm cut,r}^{\rm ssc}}$, which in the radiative scenario, the PL coincidences corresponds to 3.08\% (2 GRBs) and the BPL to 9.52\% (2 GRBs). A constant-density medium is preferred over a stellar-wind environment.  In most cases, the constant-density medium yields at least one more coincidence than the wind scenario.  The condition where the observed frequency lies above the cutoff break ($\mathbf{\nu_{\rm cut,r}^{\rm ssc} < \nu}$) shows a minimal but consistent agreement, explaining 1.54\% (1 GRB) of PL fits and 4.76\% (1 GRB) of BPL fits across various regimes. This suggests that the highest-energy LAT photons for a small subset of bursts may originate from this cooling regime. Finally, we note that no coincidences were found for the spectral regime below the minimum frequency ($\mathbf{\nu < \nu_{\rm m,r}^{\rm ssc}}$) in any of the scenarios, media, or shell types considered. 

The change from an adiabatic ($\epsilon=0$) to a radiative ($\epsilon=1$) reverse-shock evolution has only a minor effect on the total number of single PL coincidences but produces a dramatic reduction in the number of BPL coincidences. Numerically, PL matches drop only from 10 to 9 total coincidences (a 10\% decrease), while BPL matches fall from 29 to 6 - a nearly 80\% reduction. This contrasting behavior indicates that the radiative model primarily affects cases where a spectral break is required to match the observed indices.

For the PL model, the only difference occurs in the thin shell with constant density regime, where the radiative case is favored. For the more complex BPL model, the adiabatic case is consistently preferred across multiple regimes, with the most significant reductions occurring in wind medium scenarios (both thick and thin shells) and the thin shell ISM case. This pattern suggests that the spectral breaks observed in BPL cases are much better explained by adiabatic evolution, possibly because radiative losses alter the electron energy distribution in ways inconsistent with the observed spectral shapes.

\subsubsection{Energy injection scenario}

Table \ref{Table-radiative-CRs_q} shows the CRs of SSC reverse-shock scenario ($F^{\rm ssc}_{\rm \nu,r} \propto t^{-\alpha(\beta, q, p)} \nu^{-\beta}$) in the thick- and thin-shell case and ISM and stellar wind for each cooling condition. This table displays the SSC CRs during the adiabatic regime with energy injection for two sets of electron spectral distribution values; $1<p<2$ and $p > 2$.\\

Figures~\ref{fig5:CRs_q_k0_PL}--\ref{fig8:CRs_q_k2_BPL} present the case where the reverse shock evolves in the adiabatic regime with continuous energy injection ($q=0.5$). Figures \ref{fig5:CRs_q_k0_PL} and~\ref{fig6:CRs_q_k0_BPL}~illustrate the comparisons for a homogeneous medium using PL and BPL spectral indices, respectively, while Figures~\ref{fig7:CRs_q_k2_PL} and~\ref{fig8:CRs_q_k2_BPL} show the corresponding wind-environment ($k=2$) results. As in the case for the figures related to the radiative case, the rows separate thick- and thin-shell evolution, and the columns indicate different SSC cooling conditions. 

In general, the comparison in Figures~\ref{fig5:CRs_q_k0_PL}--\ref{fig8:CRs_q_k2_BPL} shows that the LAT indices derived from PL and BPL exhibit improved consistency with the SSC reverse shock CRs when energy injection is included. In Figure~\ref{fig5:CRs_q_k0_PL}, we find three coincidences for both the thick- and thin-shell cases, while in Figure~\ref{fig6:CRs_q_k0_BPL} there are two coincidences. In contrast, Figure~\ref{fig7:CRs_q_k2_PL} displays only one coincidence in the thick-shell case, and Figure~\ref{fig8:CRs_q_k2_BPL} shows a single match, restricted to the $\nu^{\rm ssc}_{\rm cut,r} < \nu$ regime for both shell types. As in the radiative scenario, we again find no coincidences for $\nu < \nu^{\rm ssc}_{\rm m,r}$ in any figure.  Taken together, these results indicate that the adiabatic reverse shock with moderate energy injection provides a more favorable match to the observations than the purely radiative case, particularly in the ISM environment, where multiple coincidences are found in both shell regimes. Nevertheless, the wind environment offers little overlap, and the limited number of matches overall prevents us from making stronger conclusions. The preference thus lies with the ISM adiabatic thin- and thick-shell cases.

Table~\ref{tab:CR-Results_q} presents a statistical analysis that compares the theoretical predictions of the SSC reverse shock model with the energy injection, and the observed data from the 2FLGC. This table focuses on the scenario with nontrivial energy injection ($q = 0.5$), while also providing the case without energy injection ($q=0$) for comparison. The data presented in this table are categorized based on thick- or thin-shell cases and further divided into constant-density medium and stellar-wind environments. It presents the quantity and percentage of GRBs in the catalog that exhibit spectral and temporal indices that align with the predictions of the model under different cooling conditions. A sample of bursts is analyzed, some fitting using a single PL and others requiring a BPL. As for the radiative case, in the case of PL, a sample size of 65 bursts is analyzed, whereas for BPL, the sample consists of 21 bursts. 

It should be noted that the total number of coincidences between the radiative scenario and the data is relatively low, indicating that this mechanism  with energy injection for $q=0.5$ does not satisfy the evolution of spectral and temporal indices for most LAT-detected bursts. Nonetheless, a subset of bursts demonstrates alignment, indicating the significance of this process in particular instances. This scenario exhibits a notably greater alignment with bursts fitted with a BPL in comparison to those fitted with a simple PL. The most successful scenario overall is found for a thin shell evolving in a constant-density ISM during the slow-cooling condition $\mathbf{\nu_{\rm m,r}^{\rm ssc} < \nu < \nu_{\rm cut,r}^{\rm ssc}}$, which in the energy injection case accounts for 6.15\% (4 GRBs) of PL fits and 14.29\% (3 GRBs) of BPL fits.  A constant-density medium is favored in comparison to a stellar-wind environment. In the majority of events, the scenario involving a constant density medium results in a minimum of one additional coincidence compared to the wind scenario.  The condition where the observed frequency lies above the cutoff break ($\mathbf{\nu_{\rm cut,r}^{\rm ssc} < \nu}$) shows a minimal but consistent agreement, explaining 1.54\% (1 GRB) of PL fits and 4.76\% (1 GRB) of BPL fits across various regimes.  This indicates that the highest-energy LAT photons for a limited number of bursts may derive from this cooling regime. Subsequently, we see that no coincidences were identified for the spectral regime below the minimum frequency ($\mathbf{\nu < \nu_{\rm m,r}^{\rm ssc}}$) in any of the situations, media, or shell types examined. This suggests that the detected emission beyond 100 MeV from Fermi-LAT should derive from spectral segments.

The introduction of energy injection ($q=0.5$) produces contrasting effects on PL and BPL coincidences that are opposite to what was observed for the radiative parameter. For the PL model, energy injection substantially increases the total number of coincidences from 10 to 13, representing a 30\% enhancement. On the other hand, for the BPL model, energy injection dramatically reduces coincidences from 29 to 10, which amounts to a 65\% reduction. This pattern indicates that energy injection better models cases with simple PL spectra while it is strongly disfavored for BPL cases.

For the PL model, energy injection shows particularly strong enhancement in two key regimes. In both the thick and thin shell with ISM medium within the $\nu^{\rm ssc}_{\rm m, r} < \nu < \nu^{\rm ssc}_{\rm cut, r}$ spectral regime, coincidences increase substantially from 1 to 4. This suggests that the energy injection model is most compatible with constant-density environments for power-law spectra in the critical frequency range between the characteristic and cooling frequencies.

For the BPL model, the reduction in coincidences happens across multiple scenarios. The thick shell with ISM medium shows reductions in both the $\nu^{\rm ssc}_{\rm m, r} < \nu < \nu^{\rm ssc}_{\rm cut, r}$ regime (from 4 to 3) and above the cooling frequency (from 4 to 1). Wind medium scenarios are also affected, with both the thick and thin shell wind regimes dropping from 2 to 0 coincidences. The thin shell with ISM medium in the $\nu^{\rm ssc}_{\rm m, r} < \nu < \nu^{\rm ssc}_{\rm cut, r}$ regime also decreases from 5 to 3.


Figure \ref{fig9:Counts_PL} illustrates the number of bursts as a function of the parameter ($q$) in compliance with the constraints of the SSC reverse-shock model during the adiabatic scenario ($\epsilon=0$). This analysis is presented for both thick- and thin-shell cases, where the spectral and temporal indices of the 2FLGC are represented using a PL function.  The upper panels correspond to the deceleration jet in a stellar wind, and the lower panels correspond to a density-constant medium.   When the jet is decelerated in the stellar wind, during the thick-shell regime, the number of bursts is uniformly distributed, whereas in the thick-shell regime, the number of bursts that satisfy the SSC Crs is $q\geq 0.8$.  When the jet is decelerated in a constant-density medium, the number of counts is larger in the thick shell than in the thin shell.  In both cases, the number of bursts that satisfy the SSC CRs corresponds to $q\geq 0$.

Figure \ref{fig10:Counts_BPL} is the same as Figure \ref{fig9:Counts_PL},  but considering the spectral and temporal indices, they are described with a 2FLGC BPL function. We note that no bursts satisfy the CRs in stellar wind. In the homogeneous medium, the larger number of bursts that satisfies the SSC CRs corresponds to $q\geq 0$.

\subsection{Circumburst environment}

The circumburst medium is a crucial factor in studying the properties of GRBs and their associated afterglows. This medium refers to the environment surrounding the progenitor of the burst, into which the relativistic outflow extends after the initial burst, leading to the production of the afterglow. The surrounding medium serves as the target material for the external shock that governs the afterglow, while also influencing its absorption, cooling, and emission characteristics.   GRBs with long duration ($> 2\,{\rm s}$) occur after the collapse of a huge, fast spinning star into a black hole (BH) or magnetar
\citep{1993ApJ...405..273W,1998ApJ...494L..45P, 2011MNRAS.413.2031M}. Before death takes place, the star produces powerful stellar winds that are expelled modifying the circumburst medium.   Consequently, a transition or discontinuity between the stellar wind and the ISM has been predicted \citep{1975ApJ...200L.107C, 1977ApJ...218..377W, 2006ApJ...643.1036P} and observed in some afterglows \citep[e.g., GRB 030226, 050319, 081109A, 140423A, 160626B, and 190114C;][]{2003ApJ...591L..21D, 2007ApJ...664L...5K, 2009MNRAS.400.1829J, 10.1111/j.1365-2966.2009.15886.x, 2020ApJ...900..176L, 2017ApJ...848...15F, 2019ApJ...879L..26F}.  GRBs with short durations arise from the merging of compact objects; BH and a neutron star (NS) or two NSs \citep{1992ApJ...392L...9D, 1992Natur.357..472U, 1994MNRAS.270..480T, 2011MNRAS.413.2031M}.  Mergers occur at considerable distances from star-forming places, usually after prolonged periods of inspiral. (${\rm \sim Gyr}$) \citep{1992ApJ...392L...9D, 1992Natur.357..472U, 1994MNRAS.270..480T}. The merger emerges in a low-density interstellar medium, generally located in the galactic halo or outer boundaries \citep[e.g., GRB 170817A][]{PhysRevLett.119.161101,2041-8205-848-2-L12}.

\subsubsection{Radiative scenario}\label{subsec:radiative_environment}
Table~\ref{tab:CR-Results_epsilon} provides the number and fraction of bursts consistent with the SSC reverse-shock CRs when modeled with a single PL and with a BPL. We consider both the thick- (top) and thin-shell (bottom) regimes for ISM and stellar-wind environments, assuming a fully radiative scenario ($\epsilon = 1.0$). For the thick-shell regime, most of the coincidences occur in the intermediate ($\nu^{\rm ssc}_{\rm m, r} < \nu < \nu^{\rm ssc}_{\rm cut, r}$) and high-frequency ($\nu^{\rm ssc}_{\rm cut, r}<\nu$) segments, with the exception of the wind case for BPL, where there are no matches. The ISM and wind configurations have the same fractions ($\sim 1.5$) for the PL fits, while only the ISM scenario has a coincidence for BPL. In contrast, the thin-shell regime shows slightly higher percentages, with the ISM scenario reaching up to $\sim 3\%$ (PL) and $\sim 9\%$ (BPL) in the intermediate frequency range, indicating a slight preference for this configuration. 

The wind environment in the thin-shell case shows either the same number of matches or fewer compared to the ISM. Overall, most coincidences align with the intermediate-frequency light curve in the thin-shell ISM model. In contrast, the lowest-frequency light curve (for frequencies $\nu < \nu^{\rm ssc}_{\rm m, r}$) is disfavored in all cases. These findings suggest a preference for the SSC reverse-shock  contribution at intermediate energies within a thin-shell structure.

\subsubsection{Energy injection scenario}
Table~\ref{tab:CR-Results_q} is analogous to Table~\ref{tab:CR-Results_epsilon}, but here we assume an energy injection index of $q = 0.5$ (and adiabatic evolution $\epsilon=0$). For the thick-shell regime, most of the coincidences occur in the intermediate ($\nu^{\rm ssc}_{\rm m, r} < \nu < \nu^{\rm ssc}_{\rm cut, r}$) frequency segment, with a smaller number at high frequencies ($\nu^{\rm ssc}_{\rm cut, r}<\nu$). The ISM scenario shows the highest fraction of matches, reaching $\sim 6\%$ for PL and $\sim 14\%$ for BPL in the intermediate segment. The wind environment is not favored in comparison, as it only has $\sim 1.5\%$ coincidences for PL at intermediate and high frequencies, and no BPL matches at intermediate frequencies. The thin-shell regime has similar trends, with the ISM configuration having exactly the same number of matches as the thick-shell case across all cooling conditions and PL fits. The wind configuration is once again subdominant, contributing only at high frequencies with $\sim 1.5\%$ for both PL and BPL fits, and no coincidences at lower or intermediate frequencies. In summary, the majority of coincidences align with the intermediate-frequency segment in the ISM scenario for both thick- and thin-shell regimes. The low frequency has no matches in all cases. These results indicate that in the presence of energy injection, the SSC reverse-shock emission in an ISM environment is the most compatible with the observations (even compared to the radiative scenario from section~\ref{subsec:radiative_environment}). Nevertheless, the total fraction of matches remains small, pointing towards the general inability of SSC reverse-shock emission to explain the data.

\subsection{The synchrotron reverse-shock limit}

Synchrotron emission is generated by the movement of relativistic electrons as they move across magnetic fields in a helical path.   Nonetheless, a limitation exists for the energy of the released photons.  Electrons accelerated in a shock acquire energy; however, they simultaneously dissipate energy by synchrotron radiation inside the local magnetic field.  The synchrotron limit from the reverse-shock region is estimated by comparing the cooling timescale with the acceleration timescale. The synchrotron limit from the reverse-shock zone can be evaluated by juxtaposing the cooling and acceleration timescales. The largest achievable Lorentz factor for electrons is $\gamma_{\rm max, r}=\left(3q_e/\xi\sigma_T B'_{\rm r}\right)^{\frac12}$, with $q_e$ the elementary electron charge, $\sigma_T$ the Thomson cross section, and $\xi\sim 1$ the parameter in the Bohm limit. 

\subsubsection{Radiative scenario}


The synchrotron limit during the radiative regime in the reverse-shock region is written as 

{\small
\begin{eqnarray}\label{break_thin_aft_k2}
h \nu^{\rm syn}_{\rm max, r} &=& \begin{cases}
3.3\times 10^2\,{\rm MeV}\, \left(\frac{1+z}{1.2}\right)^{-\frac{3(24-5\epsilon)}{16(8-\epsilon)}}\,n^{-\frac{1}{8-\epsilon}}\,\Delta_{11.8}^{\frac{1}{16}}\,\Gamma_{\rm 0,2.5}^{-\frac{\epsilon}{8-\epsilon}}\,E^{\frac{1}{8-\epsilon}}_{0,53.3}t_{1.5}^{-\frac{56-\epsilon}{16(8-\epsilon)}}\hspace{3cm}{\rm for} \hspace{0.1cm} {\rm  k=0 }\cr
9.7\times 10^2\,{\rm MeV}\, \left(\frac{1+z}{1.2}\right)^{-\frac{20-7\epsilon}{8(4-\epsilon)}}\,A^{-\frac{1}{4-\epsilon}}_{W,-2}\,\Delta_{11.8}^{\frac{1}{8}}\,\Gamma_{\rm 0,2.5}^{-\frac{\epsilon}{4-\epsilon}}\,E^{\frac{1}{4-\epsilon}}_{0,53.3}t_{1.5}^{-\frac{12-\epsilon}{8(4-\epsilon)}}\hspace{3.15cm}{\rm for} \hspace{0.1cm} {\rm  k=2 }\cr
\end{cases},
\eary
}
and
{\small
\begin{eqnarray}\label{break_thin_aft_k2}
h \nu^{\rm syn}_{\rm max, r} &=& \begin{cases}
1.6\times10^3\,{\rm MeV}\, \left(\frac{1+z}{1.2}\right)^{-\frac{3}{5}}\,n^{-\frac{2}{15}}\,\Gamma_{0,2}^{-\frac{1}{15}}\,E^{\frac{2}{15}}_{0,52.3}t_{2.5}^{-\frac{2}{5}}\hspace{3.6cm}{\rm for} \hspace{0.1cm} {\rm  k=0 }\cr
1.7\times10^3\,{\rm MeV}\, \left(\frac{1+z}{1.2}\right)^{-\frac{2}{3}}\,A^{-\frac{1}{3}}_{W,-2}\,\Gamma_{0,2}^{-\frac{1}{3}}\,E^{\frac{1}{3}}_{0,52.3}t_{2.5}^{-\frac{1}{3}}\hspace{3.55cm}{\rm for} \hspace{0.1cm} {\rm  k=2 }\cr
\end{cases}
\end{eqnarray}
}
for the thick and thin regime, respectively.

Figure \ref{fig:max_eps_thick}
shows all the high-energy photons $>100\,{\rm MeV}$ and  probabilities of $>90$\% of being associated to each burst together with the maximal photon energies released by the synchrotron reverse-shock scenario during the radiative regime for $\epsilon=0.7$ and $\epsilon=0.3$.\footnote{We require the \texttt{gtsrcprob} tool to retrieve photons with a probability greater than 90$\%$ and correlate them to each burst} This figure considers the reverse shock in the thick-shell case evolving in the stellar wind and density-constant medium. We require the values of $E_0= 2\times 10^{53}\,{\rm erg}$, $\Gamma_0=10^{2.5}$, $\Delta=10^{11.8}\,{\rm cm}$ and $z=0.2$.

In the thick-shell regime, the maximum synchrotron energy \( h \nu^{\text{syn}}_{\text{max}, r} \) demonstrates a clear dependence on the index \( \epsilon \). For both constant density media (with \( k=0 \)) and wind-like media (with \( k=2 \)), an increase in \( \epsilon \) generally leads to a more negative PL index for the density \( n \) in the \( k=0 \) case, as well as for the wind parameter \( A_W \) in the \( k=2 \) case, and the initial Lorentz factor \( \Gamma_0 \). This implies an overall decrease in synchrotron energy. However, the energy \( E_0 \) becomes more positive as \( \epsilon \) increases, which suggests that for energetic bursts, we would expect an increase in the synchrotron limit.

In the thick-shell regime, the maximum synchrotron energy $h \nu^{\rm syn}_{\rm max, r}$ exhibits a clear dependence on the index $\epsilon$. For both constant density $k=0$ and wind-like $k=2$ media, increasing $\epsilon$ generally causes the PL index of the density $n$ for $k=0$, wind parameter $A_W$ for $k=2$, and the initial Lorentz factor $\Gamma_0$ to become more negative, implying a smaller overall synchrotron energy. However, the energy $E_0$ becomes more positive as $\epsilon$ increases, so for energetic bursts, we would expect an increase in the synchrotron limit.  In terms of temporal evolution, the energy decreases more rapidly with larger $\epsilon$. For $k=0$, the temporal index goes from $-0.44$ for $\epsilon=0$ to $-0.49$ for $\epsilon=1$, while for $k=2$ it goes from $-0.375$ to $-0.458$.  This indicates that the effects of the radiative regime are more pronounced in wind-like media. In contrast, the thin-shell regime shows no dependence on $\epsilon$ for either $k=0$ or $k=2$ regarding time dependence.

\subsubsection{Energy injection scenario}

In this scenario, the synchrotron limit due to energy injection in the reverse-shock region becomes

{\small
\begin{eqnarray}\label{break_thin_aft_k2}
h \nu^{\rm syn}_{\rm max, r} &=& \begin{cases}
5.2\times10^3\,{\rm MeV}\,  \left(\frac{1+z}{1.2}\right)^{-\frac{9}{16}}\,n^{-\frac{1}{8}}_{-1}\,\Delta_{11.8}^{\frac{1}{16}}\,E^{\frac{1}{8}}_{\rm inj,53.3}t_{1.5}^{-\frac{5+2q}{16}}\hspace{3.4cm}{\rm for} \hspace{0.1cm} {\rm  k=0 }\cr
4.2\times10^3\,{\rm MeV}\,  \left(\frac{1+z}{1.2}\right)^{-\frac{5}{8}}\,A^{-\frac{1}{4}}_{W,-2}\,\Delta_{11.8}^{\frac{1}{8}}\,E^{\frac{1}{4}}_{\rm inj,53.3}t_{1.5}^{-\frac{1+2q}{8}}\hspace{3.15cm}{\rm for} \hspace{0.1cm} {\rm  k=2 }\cr
\end{cases},
\eary
}
and
{\small
\begin{eqnarray}\label{break_thin_aft_k2}
h \nu^{\rm syn}_{\rm max, r} &=& \begin{cases}
2.7\times10^3\,{\rm MeV}\,  \left(\frac{1+z}{1.2}\right)^{-\frac{4+5q}{5(2+q)}}\,n^{-\frac{2}{5(2+q)}}_{-1}\,\Gamma_{0,2}^{-\frac{6-5q}{5(2+q)}}\,E^{\frac{2}{5(2+q)}}_{\rm inj,52.3}t_{2.5}^{-\frac{2}{5}}\hspace{2.6cm}{\rm for} \hspace{0.1cm} {\rm  k=0 }\cr
2.8\times \,10^3\,{\rm MeV}\,  \left(\frac{1+z}{1.2}\right)^{\frac{1-3q}{3q}}\,A^{-\frac{1}{3q}}_{W,-2}\,\Gamma_{0,2}^{-\frac{4-3q}{3q}}\,E^{\frac{1}{3q}}_{\rm inj,52.3}t_{2.5}^{-\frac{1}{3}}\hspace{3.4cm}{\rm for} \hspace{0.1cm} {\rm  k=2 }\cr
\end{cases}
\end{eqnarray}
}
for the thick and thin regime, respectively.  Figure \ref{fig:max_q_thick} exhibits the maximal photon energies produced by the synchrotron reverse-shock scenario with energy injection for $q=0.5$ and $q=0$ and all the LAT-detected photons $>100\,{\rm MeV}$ associated to each burst. This figure considers the reverse shock in the thick-shell case evolving in the stellar wind and density-constant medium. We require the values of $E_{\rm inj}= 2\times 10^{53}\,{\rm erg}$, $\Gamma_0=10^{2.5}$, $\Delta=10^{11.8}\,{\rm cm}$ and $z=0.2$.  For this scenario, we use the density of constant medium ${\rm n=10^{-1}\,cm^{-3}}$ and the density of stellar-wind parameter $A_{\rm W}=10^{-2}$.  The synchrotron model adequately explains the high-energy photons in GRB 091127, 110731A, and 131108A under the thick-shell condition. In all other bursts, it is necessary to use an alternative method to clarify the high-energy photons.

In comparison to the radiative scenario, the modification of the synchrotron limit due to energy injection is only apparent in the time dependence for the thick-shell regime. For $q=1$, we recover the standard expectation where the temporal index is $-0.44$ for $k=0$ and $-0.38$ for $k=2$. For decreasing $q$, the drop becomes flatter, as we obtain $-0.38$ and $-0.25$, respectively, for $q=0.5$. On the other hand, the thin-shell regime shows no dependence on $q$ on time dependence. The dependence is now in the other parameters and is very similar to the radiative thick-shell case. For both constant density $k=0$ and wind-like $k=2$ media, decreasing $q$ causes the PL index of the density and the initial Lorentz factor to become more negative, while the energy index becomes more positive. So we have an interplay that either reduces or increases the synchrotron limit depending on whether the burst is very energetic (increase) or the medium is very dense (decrease).

 In Figures 11 and 12, we plot the maximum photon energy expected from the reverse-shock synchrotron emission, corresponding to $h\nu_{\rm max}^{\rm syn}$ as discussed in this section. We note that the maximum photon energy associated with the SSC component is not shown. This is because, for the parameter space explored in this work, the SSC maximum energy is significantly higher than the synchrotron limit, typically extending well beyond the energy range covered by the LAT observations.

As a result, including $h\nu_{\rm max,r}^{\rm ssc}$ would not provide additional constraints for the comparison presented in these figures. Therefore, we focus on the synchrotron maximum energy, which represents the relevant limiting factor for assessing whether the observed high-energy photons can be explained within the synchrotron framework.

\subsection{GRBs with atypical temporal and spectral indexes ($\beta_{\rm L} \lesssim 0.5$, $\alpha_{\rm L}>2$ and $1< p < 2$)}
%

\subsubsection{Temporal index for gamma-rays ($\alpha_{\rm L}>2$)}

The afterglow emission with a decay index greater than two presents challenges for descriptions using synchrotron forward-shock models. However, we can offer a suitable interpretation for this scenario. 
Once the reverse shock crosses the shell ($t>t_{\rm x}$), the observed flux begins to decrease following the profile of the SSC light curves.  At the moment that the SSC spectral break ($\nu^{\rm ssc}_{\rm cut,r}$) crosses the Fermi-LAT band ($ \nu_{\rm LAT}$), the SSC emission from the reverse shock ceases, and the photons emitted at large angles relative to the jet axis start to arrive with $F_\nu\propto t^{-(\beta+2)}\nu^{-\beta}$
\citep[i.e., the curvature effect;][]{2000ApJ...543...66P, 2003ApJ...597..455K} hiding this cooling condition. \\

The sample of GRBs from 2FLGC used in this analysis includes 14 GRBs (090510, 090902B, 090926A, 100116A, 100414A, 150627A, 160625B, 170115B, 170214A, 171010A and 180720B), which were modeled with a PL and BPL with one of the temporal index higher than two. In the case of a BPL, two temporal indices ({\small$\alpha_{\rm L_1}$} and {\small$\alpha_{\rm L_2}$}) were required. For instance,  GRB 180720B  was modeled with temporal indices {\small$\alpha_{\rm L_1}=1.5\pm0.2$} and {\small$\alpha_{\rm L_2}=3.2\pm 0.6$}, and  GRB 150627A	with temporal indices {\small$\alpha_{\rm L_1}=3.0\pm0.2$} and {\small$\alpha_{\rm L_2}=0.41\pm0.1$}.  The variation of the temporal index from soft to hard can be described as the result of one component that crosses a spectral break or two components from different processes.  In one-component scenario, we consider the magnetar as the central engine ($q=0$) and the thin-shell reverse shock evolving in the stellar wind. Therefore,  the spectral break evolves as $\nu^{\rm ssc}_{\rm m,r}\propto t^{-\frac{76}{21}}$, and the SSC flux as $F^{\rm ssc}_{\rm \nu, r}\propto t^{-\frac29}$ for $ \nu < \nu^{\rm ssc}_{\rm m,r}$, and  $F^{\rm ssc}_{\rm \nu, r}\propto t^{\frac{2(4-19p)}{21}}$ for $ \nu^{\rm ssc}_{\rm m,r} < \nu < \nu^{\rm ssc}_{\rm cut,r}$. When the spectral break ($\nu^{\rm ssc}_{\rm m,r}$) crosses the LAT band ($\nu_{\rm LAT}$), the SSC flux varies from $F^{\rm ssc}_{\rm \nu, r}\propto  t^{-3.2}$ to $\propto  t^{-0.2}$ for $p\sim 2$. In the two-component scenario, the initial LAT component with a temporal index exceeding $\alpha_{\rm L}> 2$ would be attributed to the SSC process from the reverse-shock region, whereas the subsequent one would be associated with the synchrotron or SSC mechanism from the forward-shock region. Initially, the emission from the reverse shock predominates, followed by the dominance of the emission from the forward shock.

\subsubsection{Spectral index for gamma-rays ($\beta_{\rm L} \lesssim 0.5$)}

 Table \ref{table:beta_1}  displays a sample of GRBs reported in the 2FLGC with atypical spectral indices, which can hardly be described with the standard synchrotron afterglow model. Compared to the CRs presented in Tables \ref{Table-radiative-CRs} and \ref{Table-radiative-CRs_q}, it is evident that regardless of what regime the shell evolves in and the characteristics of the circumburst medium, these bursts can be characterized by the SSC flux under cooling conditions $F^{\rm ssc}_{\rm \nu, r}\propto t^{-\alpha}\nu^{-\beta}$ with $\beta=\frac{p-1}{2}$ for $1<p<2$ and $p > 2$. For instance, as $\beta < 0.5$, $p$ would be in the range $1<p<2$.

\subsubsection{A hard spectral index for electrons ($1 < p < 2$)}

The results derived from the multiwavelength GRB observations indicate that the electron spectral index in relativistic shocks is not constant, instead of varying from each other.  Relativistic shocks depend on diverse physical conditions such as magnetization, shock obliquity, upstream/downstream plasma conditions, etc. For instance, a hard value of $p$ is expected in magnetic reconnections.  Several authors have shown that it follows a normal distribution with minimum and maximum values between 1.5 and 3.5, respectively \citep{2002ApJ...581.1248P, 2008MNRAS.388..144R, 2015ApJ...811...83Z, 2006MNRAS.371.1441S, 2008ApJ...672..433S, 2009MNRAS.395..580C, 2019ApJ...883..134T}. \cite{2002ApJ...581.1248P} examined time-resolved BATSE\footnote{The Burst and Transient Source Experiment} prompt spectra and inquired whether standard optically thin synchrotron can replicate the observed photon indexes of the Band function. They considered 5021 spectra from the first BATSE GRB catalog.  Through exploring the use of many power-law configurations of the electron spectral index, they illustrated via histograms that, regardless of the power-law configuration, a significant proportion of spectra exhibited a spectral index within the range of $1<p<2$. \cite{2006MNRAS.371.1441S} showed that, assuming the synchrotron scenario and its respective cooling conditions, there is no uniform value for the electron spectral index in a substantial sample of GRB afterglows. The authors analyzed 13, 14, and 28 afterglow observations obtained from the HETE-2, BeppoSAX, and Swift satellites, respectively, revealing that an important part of the spectral data was characterized by values $1<p<2$. With hard values of spectral indexes; $p=1.47^{+0.004}_{-0.003}$, $1.40^{+0.007}_{-0.004}$ and $1.29 - 1.32$, \cite{2008MNRAS.388..144R} modeled GRB 010222, GRB 020813 and GRB 041006, respectively, exhibiting consistency with the multiwavelength observations.  \cite{2015ApJ...811...83Z} modeled from NIR/optical to X-ray afterglow observations in GRB 091127.   They found evidence of a hard value of the electron spectral index ($p=1.5\pm 0.01$) and fast spectral evolution.   Through closure relations,  \cite{2019ApJ...883..134T} research methodically examines whether Fermi-LAT ($\geq$ 100 MeV) temporally extended emission in GRBs is compatible with the synchrotron forward shock scenario. They found that irrespective of the presence of energy injection, the most likely scenario for synchrotron emission under $1<p<2$ conditions was a stellar wind environment.  \cite{2022ApJ...934..188F} outlined the closure relations of the SSC forward-shock model within both the adiabatic and radiative regime, and the scenario in which the central engine continuously injects energy into the blast wave to analyze the evolution of the spectral and temporal indexes of the bursts reported in 2FLGC. The authors found that a small fraction of the bursts satisfied the closure relations with $1<p<2$.

\paragraph{Radiative regime}\label{subsec:radiative_regime}
In the context of the SSC reverse shock model with a hard electron spectral index $1 < p < 2$, the temporal decay index $\alpha$ of the light curve depends on the nature of the evolution between radiative and adiabatic by the parameter $\epsilon$. Tables~\ref{tableReverseThick-radiative} and~\ref{Table-radiative-CRs} provide the CRs for SSC emission in both thick- and thin-shell regimes in homogeneous and stellar-wind media. When $\epsilon=0$, the evolution is completely adiabatic. In this case, for the thick-shell regime in a homogeneous medium (Table~\ref{tableReverseThick-radiative}), the temporal index $\alpha$ in the slow-cooling regime ($\nu_{\rm m,r}^{\mathrm{ssc}} < \nu < \nu_{\rm cut,r}^{\mathrm{ssc}}$) is given by $\alpha = \frac{5p + 171}{96}$. For $p = 1.95$, this gives $\alpha \approx 1.88$, indicating a rather steep decay. In the stellar-wind case, the expression becomes $\alpha = \frac{45 - p}{16}$, which for $p = 1.95$ gives $\alpha \approx 2.691$, a much steeper decay. This benchmark scenario highlights the dependence of the light curve slope on the density profile of the circumburst medium when the afterglow evolution is adiabatic.

To note the effect of the transition to the radiative scenario, we consider the maximal case when $\epsilon=1$, that is, the purely radiative case. In the same thick-shell ISM case, the temporal index becomes $\alpha = \frac{1593+35p}{672}$. Substituting $p = 1.95$ yields $\alpha \approx 2.47$, which is steeper than the constant-parameter case. This steepening occurs because radiative losses become increasingly dominant, reducing the available kinetic energy and accelerating the decline of the flux. A rather interesting trend appears in the wind case, where the expression is $\alpha = \frac{127+p}{48}$, which gives $\alpha \approx 2.686$ for the same parameters, which is almost the same as in the adiabatic case. In fact, it is slightly smaller. This behavior suggests that in a stellar-wind environment, the transition from adiabatic to radiative evolution does not significantly modify the temporal slope when the electron spectrum is hard. The slower variation of $\alpha$ in this case arises from competing effects between density stratification ($n \propto r^{-2}$) and the enhanced cooling efficiency in the radiative regime, which partially offset each other.

\paragraph{Energy injection}
In this case, the temporal decay index $\alpha$ of the light curve is also modified by the presence of continuous energy injection from the central engine, parameterized by the injection index $q$. Tables~\ref{tableReverseThick-injection} and~\ref{Table-radiative-CRs_q} provide the CRs for SSC emission in both thick- and thin-shell regimes in homogeneous and stellar-wind media with energy injection. For the benchmark case with no energy injection ($q=1$), the evolution reverts to the standard adiabatic scenario. In the thick-shell regime for a homogeneous medium (Table~\ref{tableReverseThick-injection}), the temporal index $\alpha$ in the slow-cooling regime ($\nu_{\rm m,r}^{\rm{ssc}}<\nu<\nu_{\rm cut,r}^{\rm{ssc}}$) is given by $\alpha = \frac{39 + 5p + 132q}{96}$. For $p=1.95$ and the standard noninjection case ($q=1$), this yields $\alpha \approx 1.88$, identical to the adiabatic value in Section~\ref{subsec:radiative_regime}. However, for a moderate injection scenario with $q=0.5$, the index becomes $\alpha \approx 1.19$, while for $q=0$, it is $\alpha\approx0.51$, indicating that the decay becomes shallower as the injection of energy increases. This flattening of the light curve occurs because the continuous energy input counteracts the deceleration and radiative losses.

A similar trend as in the radiative case is observed in the stellar wind case, where the expression is $\alpha = \frac{53 - 5p + 4q(p-2)}{16}$. For $p=1.95$ and $q=1$, this gives $\alpha \approx 2.69$, again matching the adiabatic case. Introducing energy injection with $q=0.5$ results in $\alpha \approx 2.697$, while for $q=0$ we obtain $\alpha=2.7$, both corresponding to a very small increase compared to the standard scenario.

\subsubsection{Simultaneous Fermi-LAT and early optical observations}

In addition to satisfying the CRs, the SSC interpretation requires that the synchrotron seed photon flux is consistent with observations. The SSC flux is directly related to the synchrotron emission through the maximum synchrotron flux. In the SSC reverse-shock scenario, the seed synchrotron photons are expected to peak at optical energies. Among the bursts showing temporal agreement with the SSC reverse-shock CRs, GRBs 130427A, 140102A, 160625B, and 180720B also exhibited bright optical counterparts commonly interpreted as synchrotron emission from the reverse shock, whereas their LAT emission was attributed to the SSC component. The coexistence of these components provides observational support for the synchrotron seed-photon population required in the SSC scenario. In this framework, the optical synchrotron photons produced in the reverse-shock region are naturally up-scattered to GeV energies, yielding LAT emission without requiring extreme energetics or unrealistically large seed-photon fluxes. 

\paragraph{GRB 140102A}

The Burst Alert Telescope (BAT) aboard the \textit{Neil Gehrels Swift Observatory} detected GRB 140102A on 2014 January 2 at 21:17:37 UT. \citep{2014GCN.15653....1H} The \textit{Fermi}-GBM light curve exhibits two bright overlapping peaks with a duration of $3.6\pm0.1\,{\rm s}$ in the 50--300 keV energy range. During the interval from 0.4 to 4 s after the trigger, \textit{Fermi}-GBM measured an energy fluence of $(1.78\pm 0.03)\times 10^{-5}\,{\rm erg cm^{-2}}$ in the 0.01--10 MeV energy range. The \textit{Fermi}-LAT detected GRB 140102A at 21:17:37.64 UT \citep{2014GCN.15659....1S}, reporting an energy flux of $(5.96\pm 2.37)\times 10^{-10}\,{\rm erg\, cm^{-2}\,s^{-1}}$ in the 0.1--10 GeV energy range \citep{2021MNRAS.505.4086G}. Follow-up observations were carried out by the BOOTES-4/MET robotic telescope in Lijiang, China, beginning 28 s after the trigger time. Additional detections were also reported by \textit{Swift}-UVOT and the 1.3 m ARIES telescope using multiple filters \citep{2014GCN.15690....1H, 2021MNRAS.505.4086G}.  The photometric redshift was found to be $2.02^{+0.05}_{-0.05}$ \citep{2021MNRAS.505.4086G}.

\paragraph{GRB 130427A}

GRB 130427A was detected by the {\itshape Fermi}/GBM instrument at 07:47:06.42 UTC on 2013 April 27 \citep{2013GCN.14473....1V}.  The exceptional brightness of this event motivated an extensive multiwavelength follow-up campaign involving both space- and ground-based observatories, including ultraviolet, optical, and X-ray telescopes \citep{2014Sci...343...48M}, SPI-ACS/\textit{INTEGRAL} \citep{2013GCN.14484....1P}, \textit{AGILE} \citep{2013GCN.14515....1V}, \textit{Konus-Wind} \citep{2013GCN.14487....1G}, \textit{NuSTAR} \citep{2013ApJ...779L...1K}, \textit{RHESSI} \citep{2013GCN.14590....1S}, MAXI/GSC \citep{2013GCN.14462....1K}, and VLT/X-shooter \citep{2013GCN.14491....1F}. The prompt emission measured by {\itshape Fermi}/GBM had a duration of $T_{90}=138.24\,{\rm s}$ \citep{Ajello_2019}, with an isotropic-equivalent gamma-ray energy of $(1.7\pm 0.2)\times 10^{52}\, {\rm erg}$ \citep{Ajello_2019}. In the 10 keV--100 GeV energy range, the total fluence and isotropic-equivalent energy were $4.9\times 10^{-3}\,{\rm erg\, cm^{-2}}$ and $1.4\times 10^{54}\,{\rm erg}$, respectively \citep{2014Sci...343...42A}.
The Rapid Telescopes for Optical Response (RAPTOR) observed a prominent early optical flash from GRB 130427A in the R-band, reaching a peak magnitude of ${\rm  7.03\pm0.03\, mag}$ between 9.31 and 19.31 s after the GBM trigger \citep{2014Sci...343...38V}.  Optical spectroscopy performed with Gemini-North by \citet{2013GCN.14455....1L} measured a redshift of z=0.34.

\paragraph{GRB 160625B}

GRB 160625B was detected and localized by the {\itshape Fermi}/GBM instrument on 2016 June 25 at 22:40:16.28 UT \citep{2016GCN..19581...1B}. Shortly afterward, the burst was also detected by {\itshape Fermi}/LAT at 22:43:24.82 UT \citep{2016GCN..19586...1D}. The prompt emission measured by GBM had a duration of $T_{90}=453.38\,{\rm s}$, with a fluence of  $(2.5\pm 0.3)\times 10^{-5}\,{\rm erg\,cm^{-2}}$ and $E_{\rm \gamma, iso}=(1.5\pm 0.1)\times 10^{53}\, {\rm erg}$ \citep{Ajello_2019}. The analysis of optical spectroscopy led to a redshift of z=1.406 \citep{2016GCN..19600...1X}.  The prompt optical flash was recorded by the MASTER-IAC robotic telescope in Tenerife, Spain, providing high-time-resolution data starting only 48 seconds after the Fermi GBM trigger \citep{2018NatAs...2...69Z}.
Extensive optical follow-up observations were also carried out by several ground-based instruments/telescopes \citep{2018NatAs...2...69Z,2017Natur.547..425T}.

\paragraph{GRB 180720B}

GRB 180720B was detected by the GBM and LAT instruments aboard the {\itshape Fermi} satellite \citep{2018GCN.22981....1R,2018GCN.22980....1B}. The {\itshape Swift}/BAT instrument triggered on this burst on 2018 July 20 at 14:21:44 UT. The prompt emission measured by {\itshape Fermi}/GBM had a duration of $T_{90}=48.90\,{\rm s}$ , corresponding to a fluence of $(0.19\pm 0.05)\times 10^{-5}\,{\rm erg\,cm^{-2}}$ and an isotropic-equivalent gamma-ray energy of $(0.39\pm 0.09)\times 10^{52}\, {\rm erg}$ \citep{Ajello_2019}.  Follow-up observations in the optical and near-infrared (NIR) bands began on 2018 July 20 at 14:22:57 UT, approximately 73 s after the trigger time \citep{2018GCN.22977....1S}. The optical counterpart was first detected and monitored by the Kanata 1.5 m telescope \citep{2024NatAs...8..134A}. Subsequent observations with the VLT/X-shooter spectrograph detected the optical afterglow and measured a redshift of z=0.654 \citep{2018GCN.22996....1V}.

Figure \ref{fig:othergrbs} shows a sample of simultaneous Fermi-LAT (purple) and early optical (red) data with the best fit curves of GRB 130427A, GRB 140102A, GRB 160625B and GRB 180720B. The best-fitting curves of Fermi-LAT and optical data are obtained with SSC and synchrotron reverse-shock model. The early optical and LAT data were modeled using a $\chi^2$ minimization method implemented with the \texttt{LMFIT} Python package for non-linear least-squares fitting.
\citep{newville_2025_15014437}.

All panels in this figure show that the optical synchrotron flux required to account for the observed \textit{Fermi}-LAT emission via the SSC process is consistent with observations. In addition, the temporal correlation between the optical and LAT fluxes suggests that both components originate from the same emission region and electron population. 

Using our estimates of the SSC flux, we infer the corresponding synchrotron flux needed to reproduce the LAT emission and find that it lies within the typical range observed during the early optical afterglow phase. For representative parameters, the inferred synchrotron flux is $F^{\rm syn}_{\rm \nu, r} \sim 10$--$10^3$ mJy, consistent with optical observations. Therefore, the synchrotron emission required by the SSC interpretation does not violate observational constraints and remains physically plausible.

\section{Summary}\label{sec4}

To analyze the evolution of the spectral and temporal indices of bursts documented in the 2FLGC, we focused on a subset of bursts with photon energies exceeding $>$ 100 MeV. We derived the light curves and the CRs of the SSC reverse-shock model in both adiabatic and radiative scenarios, while taking into account continuous energy injection from the central engine into the blast wave. In both scenarios, we considered two spectral indices for the electron population: one in the range of $1<p<2$ and $p > 2$. We also examined the evolution of the reverse shock in the thick- and thin-shell cases, where the outflow is decelerated within a stratified environment created by the progenitor's stellar wind and in a homogeneous medium.\\


We have illustrated the temporal evolution of the SSC light curves during the radiative regime, where the flux $F^{\rm ssc}_{\rm \nu, r}\propto t^{-\alpha(\epsilon,p)}$ depends on the radiative parameter and the electron spectral index. Significant fluctuations in the spectrum and temporal characteristics of afterglow emission occur due to radiative losses only when the parameter $\epsilon$ is substantial and approaches unity. In contrast, when the radiative parameter is close to zero, the differences in temporal characteristics are minimal.

By illustrating the evolution of the SSC light curves, we determined the expected flux given by $F^{\rm ssc}_{\rm \nu, r}\propto t^{-\alpha(q,p)}$, which depends on the energy injection index and the electron spectral index. We found that deviations from the SSC reverse shock occur for $q=1$. The temporal analysis of the Swift-XRT data yielded a value of $q=0.5$, which is consistent with the standard synchrotron forward-shock scenario.   Based on these findings, the flux of the SSC process evolving in a medium of constant density, under the cooling condition $ \nu^{\rm ssc}_{\rm m,r} < \nu < \nu^{\rm ssc}_{\rm cut,r}$, decreases as $\propto t^{-1.42}$ for the thick shell case and $\propto t^{-1.54}$ for the thin shell case, assuming $p=2.2$. Under the same conditions, but considering a magnetar as the central engine, the expected SSC flux decreases as $\propto t^{-0.74}$ for the thick shell case and $\propto t^{-1.05}$ for the thin shell case.


We conducted an analysis comparing the CRs of the SSC reverse-shock model across cooling conditions with the descriptions of the spectral and temporal characteristics of bursts documented in 2FLGC. In the radiative scenario, we consider the cases of fully radiative ($\epsilon=1$) or partially radiative ($0 <\epsilon < 1$) regimes, and
in the energy injection scenario, we consider the cases $q=0.5$, which are consistent with Swift-XRT data, and $q=0$ with a magnetar or pulsar spin down. Initially, we observe that the overall number of coincidences between the theoretical model and Fermi-LAT observations is limited, suggesting that SSC emission from the reverse shock is not the dominant process at high energies. We note that the closure relations exhibit much greater concordance with bursts fitted with a BPL than with those fitted using a simple PL. A constant-density medium is favored over a stellar-wind environment. This choice remains consistent in both radiative and injection circumstances. The analysis revealed that regardless of the scenarios (radiative/adiabatic or energy injection), the cooling condition $\mathbf{ \nu_{\rm m,r}^{\rm ssc} < \nu <  \nu_{\rm cut,r}^{\rm ssc} }$ is favored. In the radiative scenario, the thin-shell case is favored over the thick one, and in the energy injection scenario, the thick and thin-shell cases have the same coincidences.


We have derived all the high-energy photons $>100\,{\rm MeV}$ and probabilities of $>90$\% of being associated with each burst together with the maximal photon energies released by the synchrotron reverse-shock scenario during the radiative regime for $\epsilon=0.7$ and $\epsilon=0.3$ and with energy injection $q=0.5$ and $q=0$. We have found that the synchrotron scenario can only explain the high-energy photons in GRB 091127, 110731A, and 131108A for the thick-shell case.   In all other bursts, another mechanism, such as SSC emission, must be invoked to account for the high-energy photons. \cite{1999ApJ...520..641S} analyzed the forward and reverse shocks to investigate the early afterglow over several multiwavelengths. The reverse shock, at its peak flux, contains energy comparable to that of the GRB; however, it has a far lower temperature than the forward shock, leading to emission at much lower frequencies. The reverse shock dominates the first optical emission, and an optical flash ($>15^{\rm th}\,{\rm mag}$) is expected to be simultaneous with the peak of the forward shock in X-rays or gamma-rays. Consequently, the predicted synchrotron radiation strength at the reverse shock region is minimal at MeV-GeV levels, making it unlikely that synchrotron emission contributes to the emissions measured by Fermi-LAT.\\

Several (14) bursts in 2FLGC were described with a BPL with one of the temporal indices larger than two, as is the case of  GRB 150627A ({\small$\alpha_{\rm L_1}=3.0\pm0.2$} and {\small$\alpha_{\rm L_2}=0.41\pm0.1$}).   The variation of the temporal index from soft to hard is typically attributed to two components. The initial LAT component with a temporal index $\alpha_{\rm L} > 2$ could be explained by the SSC process from the reverse-shock region. In contrast, the subsequent one would be associated with the synchrotron or SSC mechanism in the forward-shock region. Initially, the emission from the reverse shock prevails, followed by the forward shock.   Observable fluxes at the end of the prompt phase and at the onset or during the afterglow with a decay index exceeding two (2) pose difficulties for interpretation using the synchrotron forward-shock model. Nevertheless, an appropriate interpretation can be provided in the SSC reverse-shock scenario in the high-latitude emission.  On the other hand, bursts exhibiting an unusual spectral index ($\beta_{\rm L}\lesssim 0.5$) are challenging to characterize using the synchrotron afterglow approach, as the Fermi-LAT band must be situated below both spectral breaks ($\nu_{\rm LAT} < \nu^{\rm syn}_{\rm m,f}< \nu^{\rm syn}_{\rm cut,f}$ or $\nu_{\rm LAT} < \nu^{\rm syn}_{\rm cut,f}< \nu^{\rm syn}_{\rm m,f}$).  It should be noted that, regardless of what regime the shell evolves and the characteristics of the circumburst medium, these bursts may be characterized by the SSC flux under cooling condition $F^{\rm ssc}_{\rm \nu, r}\propto t^{-\alpha}\nu^{-\beta}$ with $\beta=\frac{p-1}{2}$ for $1<p<2$ and $p > 2$. For instance, as $\beta < 0.5$, $p$ would be in the range $1<p<2$.

 It is possible to identify emission processes, such as synchrotron or SSC, by analyzing the evolution of the spectral index and the temporal index in the CRs. Although synchrotron photons in the reverse-shock region are limited to keV energies, the SSC process can extend these photons to GeV energies \citep{2001ApJ...559..110Z, 2009ApJ...706L.138A, 2011MNRAS.412..522B, 2019ApJ...883..162F}. The characteristics of CRs depend on various factors, including the circumstellar environment, energy injection, and adiabatic or radiative regimes, among others.

\begin{acknowledgments}
We thank the anonymous referee for insightful and constructive comments that have significantly improved the clarity, consistency, and overall quality of this manuscript. NF is grateful to UNAM-DGAPA-PAPIIT for the funding provided by grant IN112525. BBK is supported by IBS under the project code IBS-R018-D3. AG is grateful to UNAM-DGAPA-PAPIIT for the funding provided by grant CJIC/CTIC/1121/2024. M.G.D. acknowledges funding from the AAS Chretienne Fellowship and the MINIATURA2 grant. 
\end{acknowledgments}

\facilities{Fermi(LAT)}

\software{Python \citep{python}}





\newpage
\begin{table}
\centering \renewcommand{\arraystretch}{1.6}\addtolength{\tabcolsep}{2pt}
\caption{Temporal index as a function of the radiative parameter ($\epsilon$) and the electron spectral index ($1<p<2$ and $p > 2$) of SSC reverse-shock light curves ($F^{\rm ssc}_{\rm \nu, r}\propto t^{-\alpha(\epsilon,p)}\nu^{-\beta}$) for the slow-cooling regime. The evolution of the reverse shock in ISM and stellar wind for the thick- and thin-shell case is considered for $1<p<2$ and $p > 2$.}
\label{tableReverseThick-radiative}
\begin{tabular}{cccccc}
\hline
\hline
     &               & ISM     & ISM      & wind    & wind      \\

     &               & $1<p<2$     & $p > 2$       & $1<p<2$     & $p > 2$       \\
\hline

 & $\beta$    & $\alpha$    & $\alpha$    & $\alpha$    & $\alpha$    \\
\hline
Thick shell &    &    &    &    &    \\
\hline
$ \nu < \nu^{\rm ssc}_{\rm m,r}$    &    $-\frac13$         &     $-\frac{8(106+25\epsilon)-p(472+265\epsilon)}{72(p-1)(8-\epsilon)}$        &      $\frac{16+55\epsilon}{12(8-\epsilon)}$       &       $-\frac{112-30\epsilon-p(68-13\epsilon)}{12(p-1)(4-\epsilon)}$      &      $\frac{6+\epsilon}{3(4-\epsilon)}$       \\
$ \nu^{\rm ssc}_{\rm m,r} < \nu < \nu^{\rm ssc}_{\rm cut,r}$    &    $\frac{p-1}{2}$           &       $\frac{1368+225\epsilon+5p(8-\epsilon)}{96(8-\epsilon)}$      &       $-\frac{136-485\epsilon-9p(88-15\epsilon)}{96(8-\epsilon)}$      &      $\frac{180-53\epsilon-p(4-5\epsilon)}{16(4-\epsilon)}$       &      $\frac{4+15\epsilon+p(84-29\epsilon)}{16(4-\epsilon)}$       \\

\hline
Thin shell &  &     &     &    &    \\
\hline
$ \nu < \nu^{\rm ssc}_{\rm m,r}$    &    $-\frac13$         &     $-\frac{143-75p}{105(p-1)}$        &      $\frac{1}{15}$       &       $-\frac{5(31-19p)}{63(p-1)}$      &      $\frac{5}{9}$       \\
$ \nu^{\rm ssc}_{\rm m,r} < \nu < \nu^{\rm ssc}_{\rm cut,r}$    &    $\frac{p-1}{2}$          &       $\frac{58+3p}{35}$      &       $-\frac{10-37p}{35}$      &      $\frac{5(25-p)}{42}$       &      $\frac{5(1+11p)}{42}$       \\

\hline
\end{tabular}
\end{table}

\begin{table}
\centering \renewcommand{\arraystretch}{1.6}\addtolength{\tabcolsep}{2pt}
\caption{The same as Table \ref{tableReverseThick-radiative}, but the temporal power index of the SSC reverse-shock scenario ($F^{\rm ssc}_{\rm \nu, r}\propto t^{-\alpha(q,p)}\nu^{-\beta}$) depends on the energy injection index ($q$) instead of radiative parameter.}
\label{tableReverseThick-injection}
\begin{tabular}{cccccc}
\hline
\hline
     &               & ISM     & ISM      & wind    & wind      \\

     &               & $1<p<2$     & $p > 2$       & $1<p<2$     & $p > 2$       \\
\hline

 & $\beta$    & $\alpha$    & $\alpha$    & $\alpha$    & $\alpha$    \\
\hline
Thick shell &    &    &    &    &    \\
\hline
$ \nu < \nu^{\rm ssc}_{\rm m,r}$    &    $-\frac13$         &     $-\frac{4+49p+6q(17-18p)}{72(p-1)}$        &      $-\frac{17-19q}{12}$       &       $-\frac{30-13p-2q(1+2p)}{12(p-1)}$      &      $-\frac{2-5q}{6}$       \\
$ \nu^{\rm ssc}_{\rm m,r} < \nu < \nu^{\rm ssc}_{\rm cut,r}$    &    $\frac{p-1}{2}$           &       $\frac{39+5p+132q}{96}$      &       $-\frac{173-111p+12q(p-13)}{96}$      &      $\frac{53-5p+4q(p-2)}{16}$       &      $-\frac{15-29p+8q(p-2)}{16}$       \\

\hline
Thin shell &  &     &     &    &    \\
\hline
$ \nu < \nu^{\rm ssc}_{\rm m,r}$    &    $-\frac13$         &     $-\frac{2(19+15p)-105q(p-1)}{105(p-1)}$        &      $-\frac{14-15q}{15}$       &       $-\frac{134-74p-21q(p-1)}{63(p-1)}$      &      $\frac{2+3q}{9}$       \\
$ \nu^{\rm ssc}_{\rm m,r} < \nu < \nu^{\rm ssc}_{\rm cut,r}$    &    $\frac{p-1}{2}$          &       $\frac{23+3p+35q}{35}$      &       $-\frac{45-37p-35q}{35}$      &      $\frac{8(13+2p)-21q(p-1)}{42}$       &      $-\frac{4(4-19p)+21q(p-1)}{42}$       \\

\hline
\end{tabular}
\end{table}

\begin{table}
\centering \renewcommand{\arraystretch}{1.85}\addtolength{\tabcolsep}{1.5pt}
\caption{CRs of SSC reverse-shock scenario ($F^{\rm ssc}_{\rm \nu,r}\propto t^{-\alpha(\beta, \epsilon,p)}\nu^{-\beta}$) in the radiative regime for each cooling condition.  The evolution of the reverse shock in ISM and stellar wind for the thick- and thin-shell case is considered for $1<p<2$ and $p > 2$.}
\label{Table-radiative-CRs}
\begin{tabular}{c c c  c c c}
 \hline \hline
&\hspace{0.5cm}     &\hspace{0.5cm}   ISM &\hspace{0.2cm}   ISM  & \hspace{0.5cm}   wind   & \hspace{0.5cm}  wind \\ 
&\hspace{0.5cm}     &\hspace{0.5cm} $1 < p <2$   &\hspace{0.5cm}   $p > 2$  & \hspace{0.2cm}    $1 < p <2$   & \hspace{0.5cm}  $p > 2$ \\
\hline
 &\hspace{0.5cm} $\beta$    &\hspace{0.5cm} $\alpha(\beta)$   &\hspace{0.5cm}   $\alpha(\beta)$  & \hspace{0.2cm}    $\alpha(\beta)$   & \hspace{0.5cm}  $\alpha(\beta)$    \\ \hline

Thick shell &\hspace{0.5cm}     &\hspace{0.2cm}    &\hspace{0.5cm}     & \hspace{0.5cm}       & \hspace{0.5cm}  \\ \hline 	
 $ \nu < \nu^{\rm ssc}_{\rm m, r}$   	        & \hspace{0.5cm}  $-\frac{1}{3}$           &\hspace{0.5cm}  $\frac{[8(106+25)\epsilon-p(472+265\epsilon)]\beta}{24(p-1)(8-\epsilon)}$        &      $-\frac{[16+55\epsilon]\beta}{4(8-\epsilon)}$       &       $\frac{[112-30\epsilon-p(68-13\epsilon)]\beta}{4(p-1)(4-\epsilon)}$      &      $-\frac{(6+\epsilon)\beta}{4-\epsilon}$	\\	
 $ \nu^{\rm ssc}_{\rm m, r} < \nu < \nu^{\rm ssc}_{\rm cut, r}$   	                & \hspace{0.5cm}  $\frac{p-1}{2}$  &\hspace{0.5cm}  $\frac{2(352+55\epsilon)+5\beta(8-\epsilon)}{48(8-\epsilon)}$      &       $\frac{328+175\epsilon+9\beta(88-15\epsilon)}{48(8-\epsilon)}$      &      $\frac{8(11-3\epsilon)-\beta(4-5\epsilon)}{8(4-\epsilon)}$       &      $\frac{44-7\epsilon+\beta(84-29\epsilon)}{8(4-\epsilon)}$ \\ 	
 $\nu^{\rm ssc}_{\rm cut,r} < \nu$   	                                 & \hspace{0.5cm}  $\frac{p}{2}$     &\hspace{0.5cm}  $\beta+2$	                    &\hspace{0.2cm}  $\beta+2$ &\hspace{0.5cm}  $\beta+2$ &\hspace{0.5cm}  $\beta+2$\\ 
\hline	
Thin shell &\hspace{0.5cm}     &\hspace{0.5cm}    &\hspace{0.5cm}     & \hspace{0.5cm}       & \hspace{0.5cm}  \\ \hline 	
 $ \nu < \nu^{\rm ssc}_{\rm m, r}$   	        & \hspace{0.5cm}  $-\frac{1}{3}$           &\hspace{0.5cm}  $\frac{(143-75p)\beta}{35(p-1)}$        &      $-\frac{\beta}{5}$       &       $\frac{5(31-19p)\beta}{21(p-1)}$      &      $-\frac{5\beta}{3}$	\\	
 $ \nu^{\rm ssc}_{\rm m, r} < \nu < \nu^{\rm ssc}_{\rm cut, r}$   	                & \hspace{0.5cm}  $\frac{p-1}{2}$  &\hspace{0.5cm}  $\frac{61+6\beta}{35}$      &       $\frac{27+74\beta}{35}$      &      $\frac{5(12-\beta)}{21}$       &      $\frac{5(6+11\beta)}{21}$ \\ 	
 $\nu^{\rm ssc}_{\rm cut,r} < \nu$   	                                 & \hspace{0.5cm}  $\frac{p}{2}$     &\hspace{0.5cm}  $\beta+2$	                    &\hspace{0.5cm}  $\beta+2$ &\hspace{0.5cm}  $\beta+2$ &\hspace{0.5cm}  $\beta+2$\\

%
%
\hline
\end{tabular}
\end{table}

\begin{table}
\centering
\caption{The quantity and percentage of GRBs that adhere to each of the CRs of the SSC scenario from the reverse-shock region developing in a thick- and thin-shell regime. We consider the values of spectral and temporal indices of a PL and a BPL reported in 2FLGC for the radiative parameter $\epsilon=1$ ($\epsilon=0$).}
\label{tab:CR-Results_epsilon}
\resizebox{\columnwidth}{!}{%
\begin{tabular}{cccccccccc}\cline{1-10}
\multicolumn{2}{c}{} &
  PL &
  $\beta$ &
  \begin{tabular}[c]{@{}c@{}}$\alpha(\beta)$\\ $1 < p < 2$\end{tabular} &
  \begin{tabular}[c]{@{}c@{}}$\alpha(\beta)$\\  $2 < p$\end{tabular} &
  \begin{tabular}[c]{@{}c@{}}Coincidences \\  PL \end{tabular} &
  \begin{tabular}[c]{@{}c@{}}Percent (\%) \\  PL \end{tabular} &
  \begin{tabular}[c]{@{}c@{}}Coincidences \\  BPL \end{tabular} &
  \begin{tabular}[c]{@{}c@{}}Percent (\%) \\  BPL \end{tabular} \\ \cline{1-10} 
\multicolumn{1}{c|}{\multirow{6}{*}{Thick shell}} & \multirow{3}{*}{ISM}  & $ \nu < \nu^{\rm ssc}_{\rm m, r}$ & $-\frac{1}{3}$ & $\frac{[8(106+25)\epsilon-p(472+265\epsilon)]\beta}{24(p-1)(8-\epsilon)}$ &  $-\frac{[16+55\epsilon]\beta}{4(8-\epsilon)}$                                   & 0\,(0) & 0.00\,(0.00) & 0\,(0) & 0.00\,(0.00) \\
\multicolumn{1}{c|}{}                       &                             & $ \nu^{\rm ssc}_{\rm m, r} < \nu < \nu^{\rm ssc}_{\rm cut, r}$ & $\frac{p-1}{2}$ & $\frac{2(352+55\epsilon)+5\beta(8-\epsilon)}{48(8-\epsilon)}$ & $\frac{328+175\epsilon+9\beta(88-15\epsilon)}{48(8-\epsilon)}$ & 1\,(1) & 1.16\,(1.16) & 1\,(4) & 4.76\,(4.65) \\
\multicolumn{1}{c|}{}                       &                             & $\nu^{\rm ssc}_{\rm cut,r} < \nu$ & $\frac{p}{2}$ & $\beta+2$ & $\beta+2$                                                                                                                                         & 1\,(1) & 1.16\,(1.16) & 1\,(4) & 4.76\,(4.65) \\ \cline{2-10} 
\multicolumn{1}{c|}{}                       & \multirow{3}{*}{Wind}       & $ \nu < \nu^{\rm ssc}_{\rm m, r}$ & $-\frac{1}{3}$ & $\frac{[112-30\epsilon-p(68-13\epsilon)]\beta}{4(p-1)(4-\epsilon)}$ & $-\frac{(6+\epsilon)\beta}{4-\epsilon}$                                                & 0\,(0) & 0.00\,(0.00) & 0\,(0) & 0.00\,(0.00) \\
\multicolumn{1}{c|}{}                       &                             & $ \nu^{\rm ssc}_{\rm m, r} < \nu < \nu^{\rm ssc}_{\rm cut, r}$ & $\frac{p-1}{2}$ & $\frac{8(11-3\epsilon)-\beta(4-5\epsilon)}{8(4-\epsilon)}$ & $\frac{44-7\epsilon+\beta(84-29\epsilon)}{8(4-\epsilon)}$         & 1\,(1) & 1.16\,(1.16) & 0\,(2) & 0.00\,(2.33) \\
\multicolumn{1}{c|}{}                       &                             & $\nu^{\rm ssc}_{\rm cut,r} < \nu$ & $\frac{p}{2}$ & $\beta+2$ & $\beta+2$                                                                                                                                         & 1\,(1) & 1.16\,(1.16) & 0\,(4) & 0.00\,(4.65)\\ \hline
\multicolumn{1}{c|}{\multirow{6}{*}{Thin shell}}  & \multirow{3}{*}{ISM}  & $ \nu < \nu^{\rm ssc}_{\rm m, r}$ & $-\frac{1}{3}$ & $\frac{(143-75p)\beta}{35(p-1)}$ & $-\frac{\beta}{5}$                                                                                                        & 0\,(2) & 0.00\,(2.33) & 0\,(0) & 0.00\,(0.00) \\
\multicolumn{1}{c|}{}                       &                             & $ \nu^{\rm ssc}_{\rm m, r} < \nu < \nu^{\rm ssc}_{\rm cut, r}$ & $\frac{p-1}{2}$ & $\frac{61+6\beta}{35}$ & $\frac{74\beta+27}{35}$                                                                               & 2\,(1) & 2.33\,(1.16) & 2\,(5) & 9.52\,(5.81)\\
\multicolumn{1}{c|}{}                       &                             & $\nu^{\rm ssc}_{\rm cut,r} < \nu$ & $\frac{p}{2}$ & $\beta+2$ & $\beta+2$                                                                                                                                         & 1\,(1) & 1.16\,(1.16) & 1\,(4) & 4.76\,(4.65) \\ \cline{2-10} 
\multicolumn{1}{c|}{}                       & \multirow{3}{*}{Wind}       & $ \nu < \nu^{\rm ssc}_{\rm m, r}$ & $-\frac{1}{3}$ & $\frac{5\beta(31-19p)}{21(p-1)}$ & $-\frac{5\beta}{3}$                                                                                                       & 0\,(0) & 0.00\,(0.00) & 0\,(0) & 0.00\,(0.00) \\
\multicolumn{1}{c|}{}                       &                             & $ \nu^{\rm ssc}_{\rm m, r} < \nu < \nu^{\rm ssc}_{\rm cut, r}$ & $\frac{p-1}{2}$ & $\frac{5(12-\beta)}{21}$ & $\frac{5(11\beta+6)}{21}$                                                                           & 1\,(1) & 1.16\,(1.16) & 0\,(2) & 0.00\,(2.33) \\
\multicolumn{1}{c|}{}                       &                             & $\nu^{\rm ssc}_{\rm cut,r} < \nu$ & $\frac{p}{2}$ & $\beta+2$ & $\beta+2$                                                                                                                                         & 1\,(1) & 1.16\,(1.16) & 1\,(4) & 4.76\,(4.65) \\ \hline
\end{tabular}%
}
\end{table}

\begin{table}
\centering \renewcommand{\arraystretch}{1.85}\addtolength{\tabcolsep}{1.5pt}
\caption{The same as Table \ref{Table-radiative-CRs}, but CRs of the SSC reverse-shock scenario ($F^{\rm ssc}_{\rm \nu,r}\propto t^{-\alpha(\beta,q,p)}\nu^{-\beta}$) depends on the energy injection index ($q$) instead of radiative parameter.}
\label{Table-radiative-CRs_q}
\begin{tabular}{c c c  c c c}
 \hline \hline
&\hspace{0.5cm}     &\hspace{0.5cm}   ISM &\hspace{0.2cm}   ISM  & \hspace{0.5cm}   wind   & \hspace{0.5cm}  wind \\ 
&\hspace{0.5cm}     &\hspace{0.5cm} $1 < p <2$   &\hspace{0.5cm}   $p > 2$  & \hspace{0.2cm}    $1 < p <2$   & \hspace{0.5cm}  $p > 2$ \\
\hline
 &\hspace{0.5cm} $\beta$    &\hspace{0.5cm} $\alpha(\beta)$   &\hspace{0.5cm}   $\alpha(\beta)$  & \hspace{0.2cm}    $\alpha(\beta)$   & \hspace{0.5cm}  $\alpha(\beta)$    \\ \hline

Thick shell &\hspace{0.5cm}     &\hspace{0.2cm}    &\hspace{0.5cm}     & \hspace{0.5cm}       & \hspace{0.5cm}  \\ \hline 	
 $ \nu < \nu^{\rm ssc}_{\rm m, r}$   	        & \hspace{0.5cm}  $-\frac{1}{3}$           &\hspace{0.5cm}  $\frac{[4+49p+6q(17-18p)]\beta}{24(p-1)}$        &      $\frac{(17-19q)\beta}{4}$       &       $\frac{[30-13p-2q(1+2p)]\beta}{4(p-1)}$      &      $\frac{(2-5q)\beta}{2}$	\\	
 $ \nu^{\rm ssc}_{\rm m, r} < \nu < \nu^{\rm ssc}_{\rm cut, r}$   	                & \hspace{0.5cm}  $\frac{p-1}{2}$  &\hspace{0.5cm}  $\frac{5\beta+22(3q+1)}{48}$	                    &\hspace{0.2cm}  $\frac{111\beta-31+12q(6-\beta)}{48}$ &\hspace{0.5cm}  $\frac{24-5\beta+2q(2\beta-1)}{8}$ &\hspace{0.5cm}  $\frac{29\beta+7+4q(1-2\beta)}{8}$ \\ 	
 $\nu^{\rm ssc}_{\rm cut,r} < \nu$   	                                 & \hspace{0.5cm}  $\frac{p}{2}$     &\hspace{0.5cm}  $\beta+2$	                    &\hspace{0.2cm}  $\beta+2$ &\hspace{0.5cm}  $\beta+2$ &\hspace{0.5cm}  $\beta+2$\\ 
\hline	
Thin shell &\hspace{0.5cm}     &\hspace{0.5cm}    &\hspace{0.5cm}     & \hspace{0.5cm}       & \hspace{0.5cm}  \\ \hline 	
 $ \nu < \nu^{\rm ssc}_{\rm m, r}$   	        & \hspace{0.5cm}  $-\frac{1}{3}$           &\hspace{0.5cm}  $\frac{[2(19+15p)-105q(p-1)]\beta}{35(p-1)}$        &      $\frac{(14-15q)\beta}{5}$       &       $\frac{[134-74p-21q(p-1)]\beta}{21(p-1)}$      &      $-\frac{(2+3q)\beta}{3}$	\\	
 $ \nu^{\rm ssc}_{\rm m, r} < \nu < \nu^{\rm ssc}_{\rm cut, r}$   	                & \hspace{0.5cm}  $\frac{p-1}{2}$  &\hspace{0.5cm}  $\frac{2(3\beta + 13) + 35q}{35}$	                    &\hspace{0.5cm}  $\frac{2(37\beta-4) + 35q}{35}$ &\hspace{0.5cm}  $\frac{4(4\beta + 15) - 21\beta q}{21}$ &\hspace{0.5cm}  $\frac{2(38\beta+15)- 21\beta q}{21}$ \\ 	
 $\nu^{\rm ssc}_{\rm cut,r} < \nu$   	                                 & \hspace{0.5cm}  $\frac{p}{2}$     &\hspace{0.5cm}  $\beta+2$	                    &\hspace{0.5cm}  $\beta+2$ &\hspace{0.5cm}  $\beta+2$ &\hspace{0.5cm}  $\beta+2$\\

%
%
\hline
\end{tabular}
\end{table}


\begin{table}
\centering
\caption{The quantity and percentage of GRBs that adhere to each of the CRs of the SSC scenario from the reverse-shock region developing in a thick- and thin-shell regime. We consider the values of spectral and temporal indices of a PL and a BPL reported in 2FLGC for the the energy injection index $q=0.5$ ($q=1.0$).}
\label{tab:CR-Results_q}
\resizebox{\columnwidth}{!}{%
\begin{tabular}{cccccccccc}\cline{1-10}
\multicolumn{2}{c}{} &
  PL &
  $\beta$ &
  \begin{tabular}[c]{@{}c@{}}$\alpha(\beta)$\\ $1 < p < 2$\end{tabular} &
  \begin{tabular}[c]{@{}c@{}}$\alpha(\beta)$\\  $p > 2$\end{tabular} &
  \begin{tabular}[c]{@{}c@{}}Coincidences \\  PL \end{tabular} &
  \begin{tabular}[c]{@{}c@{}}Percent (\%) \\  PL \end{tabular} &
  \begin{tabular}[c]{@{}c@{}}Coincidences \\  BPL \end{tabular} &
  \begin{tabular}[c]{@{}c@{}}Percent (\%) \\  BPL \end{tabular} \\ \cline{1-10} 
\multicolumn{1}{c|}{\multirow{6}{*}{Thick shell}} & \multirow{3}{*}{ISM}  & $ \nu < \nu^{\rm ssc}_{\rm m, r}$ & $-\frac{1}{3}$ & $\frac{[4+49p+6q(17-18p)]\beta}{24(p-1)}$ &  $\frac{(17-19q)\beta}{4}$                                        & 0\,(0) & 0.00\,(0.00) & 0\,(0) & 0.00\,(0.00) \\
\multicolumn{1}{c|}{}                       &                             & $ \nu^{\rm ssc}_{\rm m, r} < \nu < \nu^{\rm ssc}_{\rm cut, r}$ & $\frac{p-1}{2}$ & $\frac{5\beta+22(3q+1)}{48}$ & $\frac{111\beta-31+12q(6-\beta)}{48}$            & 4\,(1) & 4.65\,(1.16) & 3\,(4) & 14.29\,(4.65) \\
\multicolumn{1}{c|}{}                       &                             & $\nu^{\rm ssc}_{\rm cut,r} < \nu$ & $\frac{p}{2}$ & $\beta+2$ & $\beta+2$                                                                                          & 1\,(1) & 1.16\,(1.16) & 1\,(4) & 4.76\,(4.65) \\ \cline{2-10} 
\multicolumn{1}{c|}{}                       & \multirow{3}{*}{Wind}       & $ \nu < \nu^{\rm ssc}_{\rm m, r}$ & $-\frac{1}{3}$ & $\frac{[30-13p-2q(1+2p)]\beta}{4(p-1)}$ & $\frac{(2-5q)\beta}{2}$                                             & 0\,(0) & 0.00\,(0.00) & 0\,(0) & 0.00\,(0.00) \\
\multicolumn{1}{c|}{}                       &                             & $ \nu^{\rm ssc}_{\rm m, r} < \nu < \nu^{\rm ssc}_{\rm cut, r}$ & $\frac{p-1}{2}$ & $\frac{24-5\beta+2q(2\beta-1)}{8}$ & $\frac{29\beta+7+4q(1-2\beta)}{8}$         & 1\,(2) & 1.16\,(2.33) & 0\,(2) & 0.00\,(2.33) \\
\multicolumn{1}{c|}{}                       &                             & $\nu^{\rm ssc}_{\rm cut,r} < \nu$ & $\frac{p}{2}$ & $\beta+2$ & $\beta+2$                                                                                          & 1\,(1) & 1.16\,(1.16) & 1\,(4) & 4.76\,(4.65)\\ \hline
\multicolumn{1}{c|}{\multirow{6}{*}{Thin shell}}  & \multirow{3}{*}{ISM}  & $ \nu < \nu^{\rm ssc}_{\rm m, r}$ & $-\frac{1}{3}$ &  $\frac{[2(19+15p)-105q(p-1)]\beta}{35(p-1)}$ & $\frac{(14-15q)\beta}{5}$                                     & 0\,(2) & 0.00\,(2.33) & 0\,(0) & 0.00\,(0.00) \\
\multicolumn{1}{c|}{}                       &                             & $ \nu^{\rm ssc}_{\rm m, r} < \nu < \nu^{\rm ssc}_{\rm cut, r}$ & $\frac{p-1}{2}$ & $\frac{2(3\beta + 13) + 35q}{35}$ & $\frac{2(37\beta-4) + 35q}{35}$             & 4\,(1) & 4.65\,(1.16) & 3\,(5) & 14.29\,(5.81)\\
\multicolumn{1}{c|}{}                       &                             & $\nu^{\rm ssc}_{\rm cut,r} < \nu$ & $\frac{p}{2}$ & $\beta+2$ & $\beta+2$                                                                                          & 1\,(1) & 1.16\,(1.16) & 1\,(4) & 4.76\,(4.65) \\ \cline{2-10} 
\multicolumn{1}{c|}{}                       & \multirow{3}{*}{Wind}       & $ \nu < \nu^{\rm ssc}_{\rm m, r}$ & $-\frac{1}{3}$ & $\frac{[134-74p-21q(p-1)]\beta}{21(p-1)}$ & $-\frac{(2+3q)\beta}{3}$	                                       & 0\,(0) & 0.00\,(0.00) & 0\,(0) & 0.00\,(0.00) \\
\multicolumn{1}{c|}{}                       &                             & $ \nu^{\rm ssc}_{\rm m, r} < \nu < \nu^{\rm ssc}_{\rm cut, r}$ & $\frac{p-1}{2}$ & $\frac{4(4\beta + 15) - 21\beta q}{21}$ & $\frac{2(38\beta+15)- 21\beta q}{21}$ & 0\,(1) & 0.00\,(1.16) & 0\,(2) & 0.00\,(2.33) \\
\multicolumn{1}{c|}{}                       &                             & $\nu^{\rm ssc}_{\rm cut,r} < \nu$ & $\frac{p}{2}$ & $\beta+2$ & $\beta+2$                                                                                          & 1\,(1) & 1.16\,(1.16) & 1\,(4) & 4.76\,(4.65) \\ \hline
\end{tabular}%
}
\end{table}


\begin{table*}
\centering \renewcommand{\arraystretch}{1.5}\addtolength{\tabcolsep}{2pt}

\caption{Events from the 2FLGC with spectral indices $\beta_{\rm L}\lesssim  0.5$.}
\label{table:beta_1}
\begin{tabular}{lcc}
\hline
Events (GRB) & $\beta_{\rm L} \pm \delta_{\beta_{\rm L}}$ & $\alpha_{\rm L} \pm \delta_{\alpha_{\rm L}}$  \\
\hline
170522A	&	$0.50	\pm	0.30$ &	$-$	\\
160829A	&	$0.30	\pm	0.30$ &	$-$	\\
160702A	&	$0.40	\pm	0.40$ &	$-$	\\
160521B	&	$0.40	\pm	0.30$ &	$1.3\pm 0.2 $	\\

160422A	&	$0.20	\pm	0.30$ &	$-$	\\

150514A	&	$0.10	\pm	0.40$ &	$-$	\\
141221B	&	$0.50	\pm	0.40$ &	$- $	\\


140810A	&	$0.50	\pm	0.20$ &	$0.8\pm 0.2 $	\\


120729A	&	$0.50	\pm	0.40$ &	$-$	\\

110213A	&	$0.50	\pm	0.40$ &	$ -$	\\

101227B	&	$0.50	\pm	0.50$ &	$-$	\\
101014A	&	$0.30	\pm	0.30$ &	$-0.2\pm0.3$	\\
091127	&	$0.20	\pm	0.40$ &	$-$	\\
090427A	&	$-0.20	\pm	0.70$ &	$-$	\\


\hline
\end{tabular}
\end{table*}


\begin{figure}
{ \centering
\resizebox*{\textwidth}{0.6\textheight}
{\includegraphics{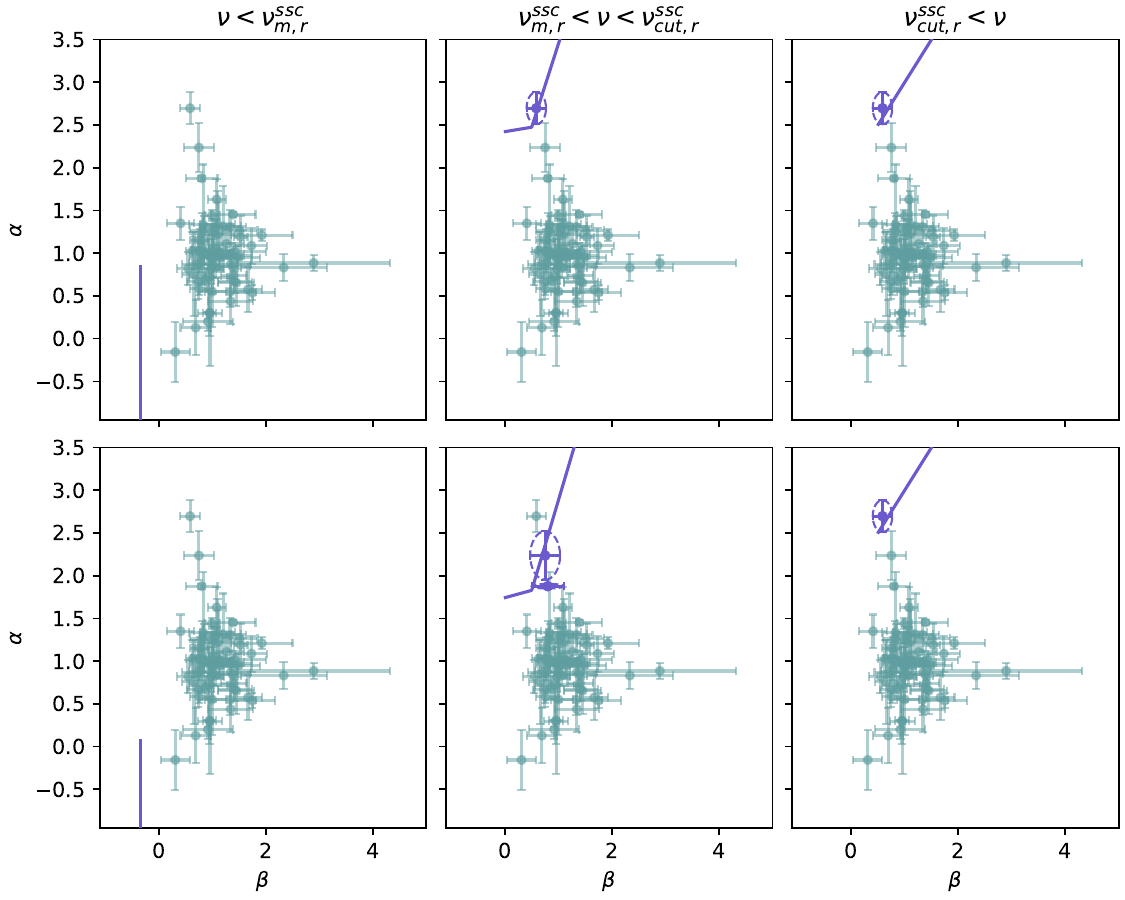}
}}
\caption{The panels correspond to spectral and temporal indexes modeled with a PL function from 2FLGC (green) and the SSC CRs (blue) from the reverse shock in the radiative scenario ($\epsilon=1$) and no energy injection. The deceleration of the outflow in a constant density medium for each cooling condition as indicated in superior part of each column is considered. The first row corresponds to the evolution of the reverse shock in the thick-shell case and the second row to the evolution of the reverse shock in the thin-shell case. The error bars shown in the panels correspond to the $1\sigma$ uncertainties. Because the temporal ($\alpha$) and spectral ($\beta$) indices are correlated, the dashed ellipses represent their joint confidence regions. } 
 \label{fig1:CRs_eps_k0_PL}
\end{figure}

\begin{figure}
{ \centering
\resizebox*{\textwidth}{0.6\textheight}
{\includegraphics{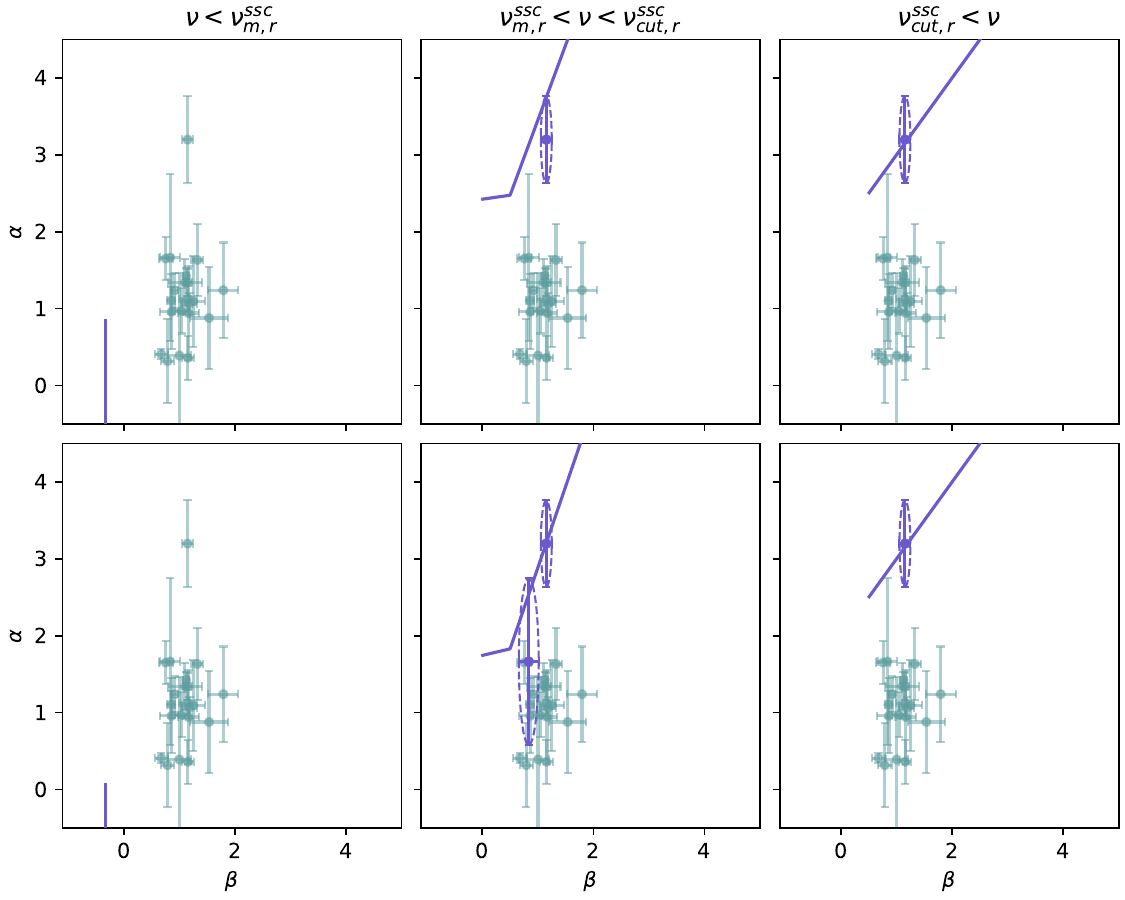}
}}
\caption{The same as Figure \ref{fig1:CRs_eps_k0_PL}, but considering the spectral and temporal indexes modeled with a BPL function from 2FLGC.}
 \label{fig2:CRs_eps_k_0_BPL}
\end{figure}

\begin{figure}
{ \centering
\resizebox*{\textwidth}{0.6\textheight}
{\includegraphics{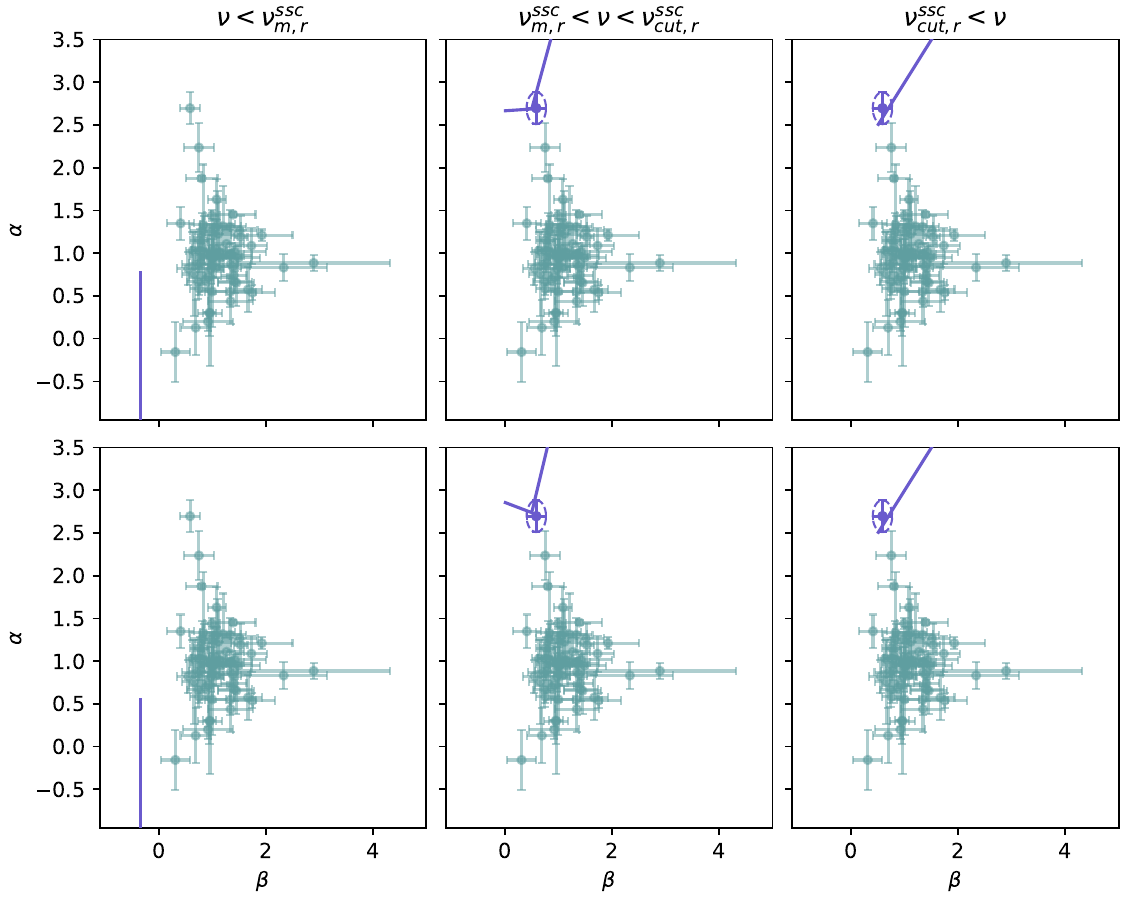}
}}
\caption{The same as Figure \ref{fig1:CRs_eps_k0_PL}, but considering a stellar-wind environment ($k=2$).}
 \label{fig3:CRs_eps_k2_PL}
\end{figure}

\begin{figure}
{ \centering
\resizebox*{\textwidth}{0.6\textheight}
{\includegraphics{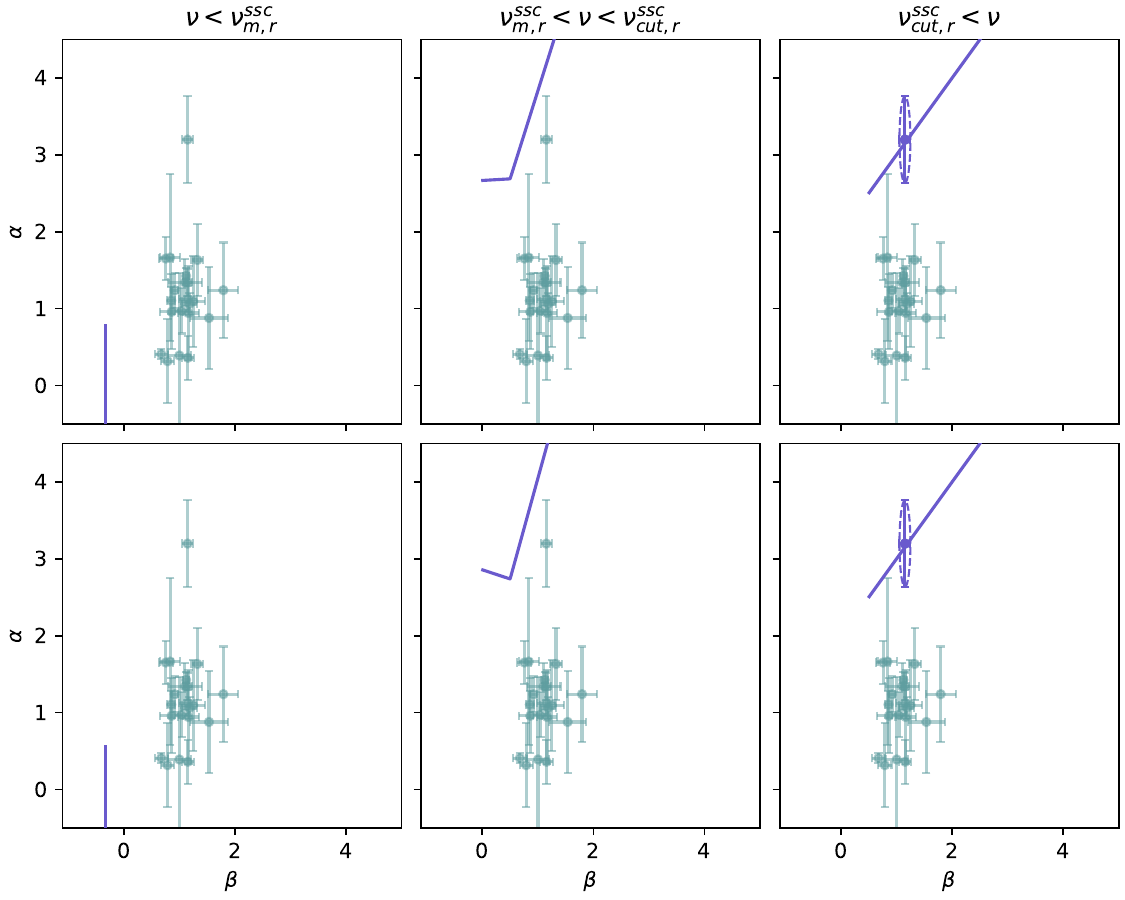}
}}
\caption{The same as Figure \ref{fig2:CRs_eps_k_0_BPL}, but considering a stellar-wind environment ($k=2$).}
 \label{fig4:CRs_eps_k2_BPL}
\end{figure}

\begin{figure}
{ \centering
\resizebox*{\textwidth}{0.6\textheight}
{\includegraphics{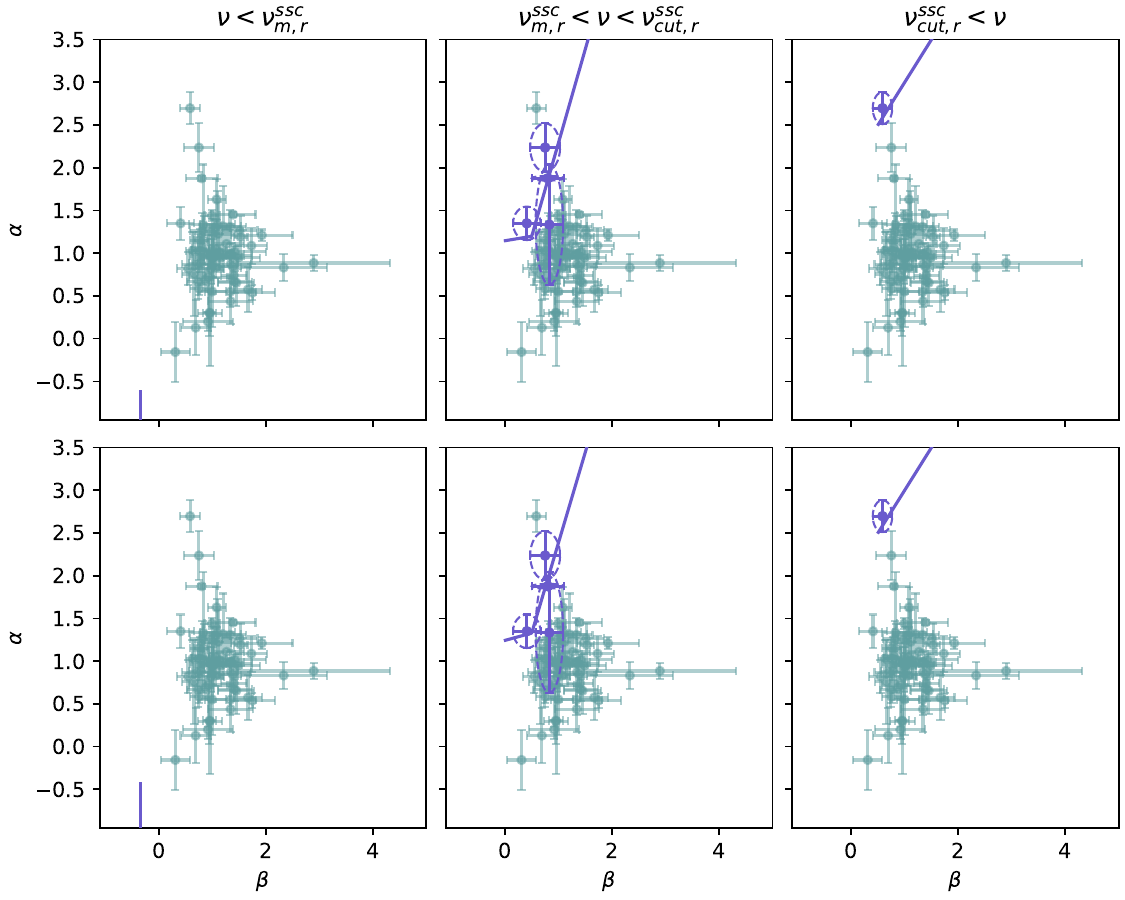}
}}
\caption{The same as Figure \ref{fig1:CRs_eps_k0_PL}, but the reverse shock lies in the adiabatic regime ($\epsilon=0$) with energy injection ($q=0.5$).}
 \label{fig5:CRs_q_k0_PL}
\end{figure}

\begin{figure}
{ \centering
\resizebox*{\textwidth}{0.6\textheight}
{\includegraphics{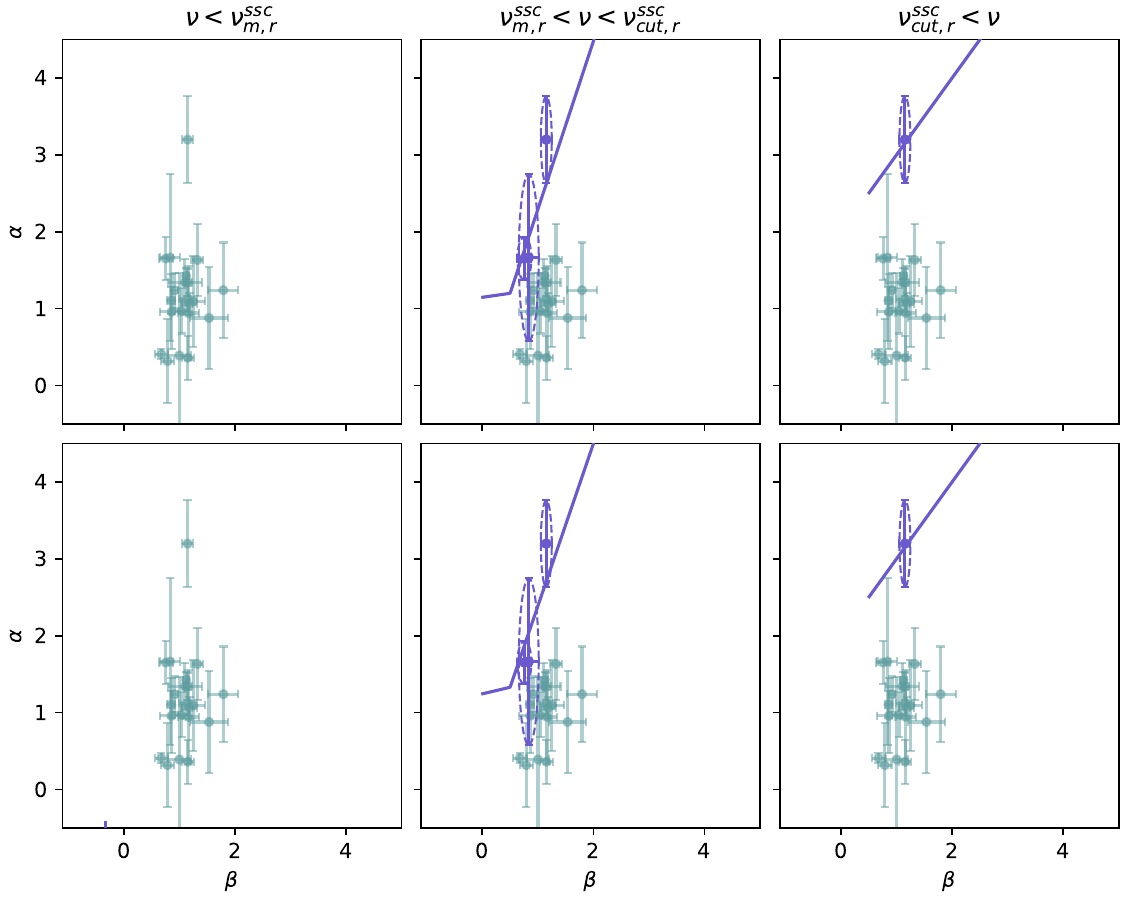}
}}
\caption{The same as Figure \ref{fig2:CRs_eps_k_0_BPL}, but the reverse shock lies in the adiabatic regime ($\epsilon=0$) with energy injection ($q=0.5$).}
 \label{fig6:CRs_q_k0_BPL}
\end{figure}

\begin{figure}
{ \centering
\resizebox*{\textwidth}{0.6\textheight}
{\includegraphics{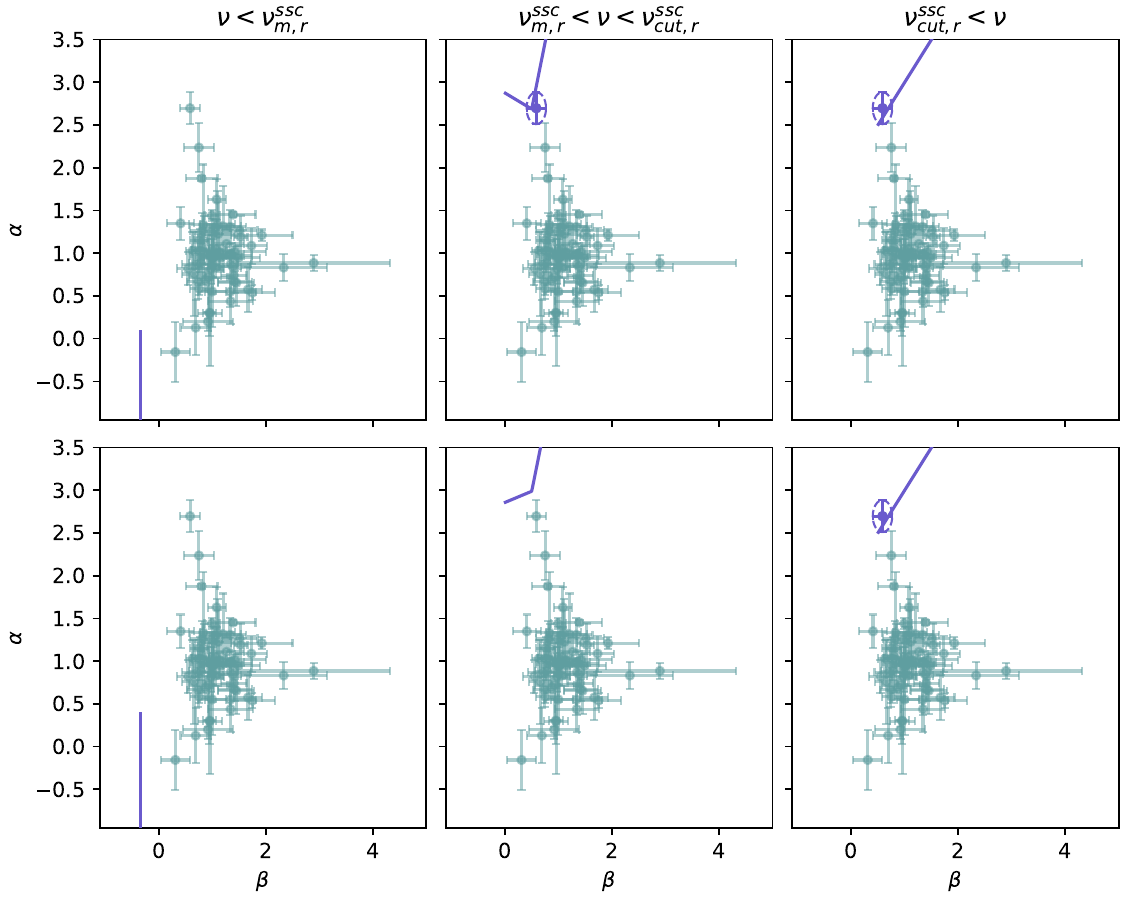}
}}
\caption{The same as Figure \ref{fig3:CRs_eps_k2_PL}, but the reverse shock lies in the adiabatic regime ($\epsilon=0$) with energy injection ($q=0.5$).}
 \label{fig7:CRs_q_k2_PL}
\end{figure}

\begin{figure}
{ \centering
\resizebox*{\textwidth}{0.6\textheight}
{\includegraphics{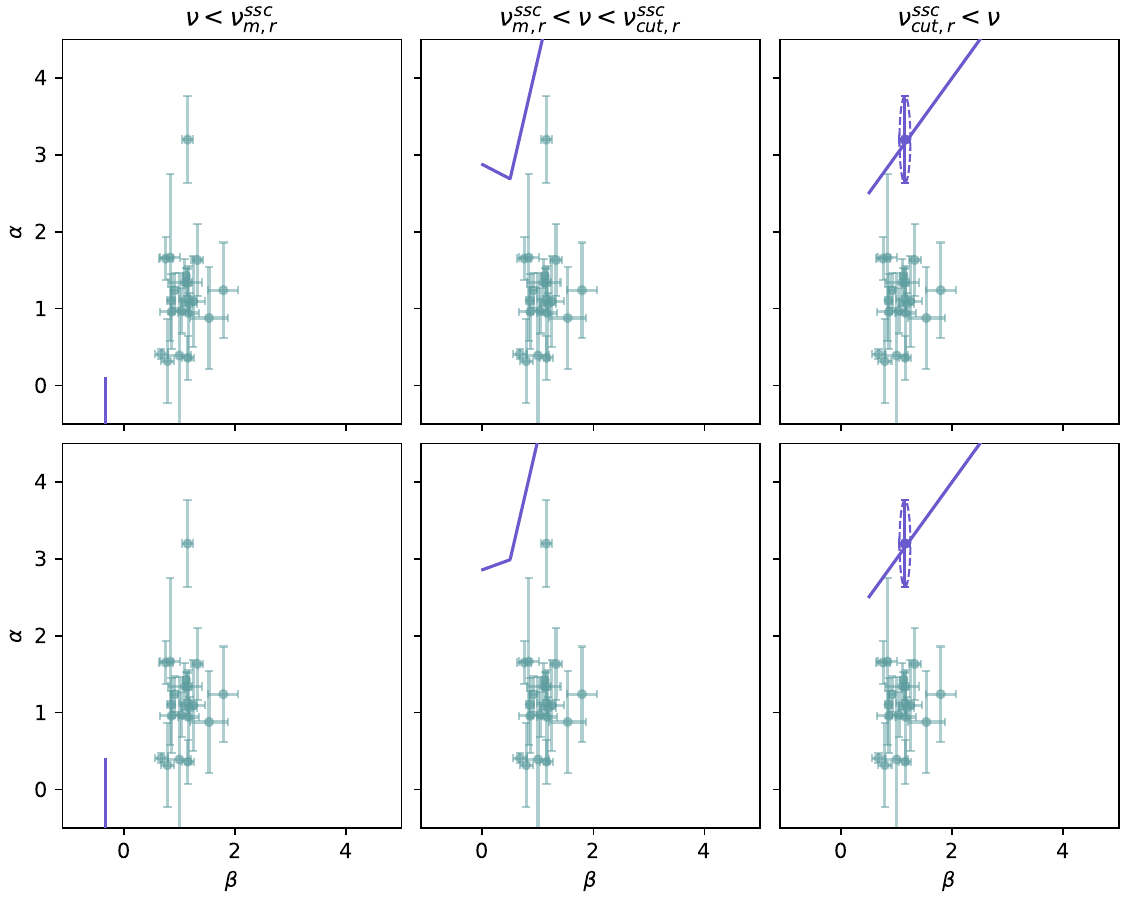}
}}
\caption{The same as Figure \ref{fig4:CRs_eps_k2_BPL}, but the reverse shock lies in the adiabatic regime ($\epsilon=0$) with energy injection ($q=0.5$).}
 \label{fig8:CRs_q_k2_BPL}
\end{figure}

\begin{figure}
{ \centering
\resizebox*{\textwidth}{0.6\textheight}
{\includegraphics{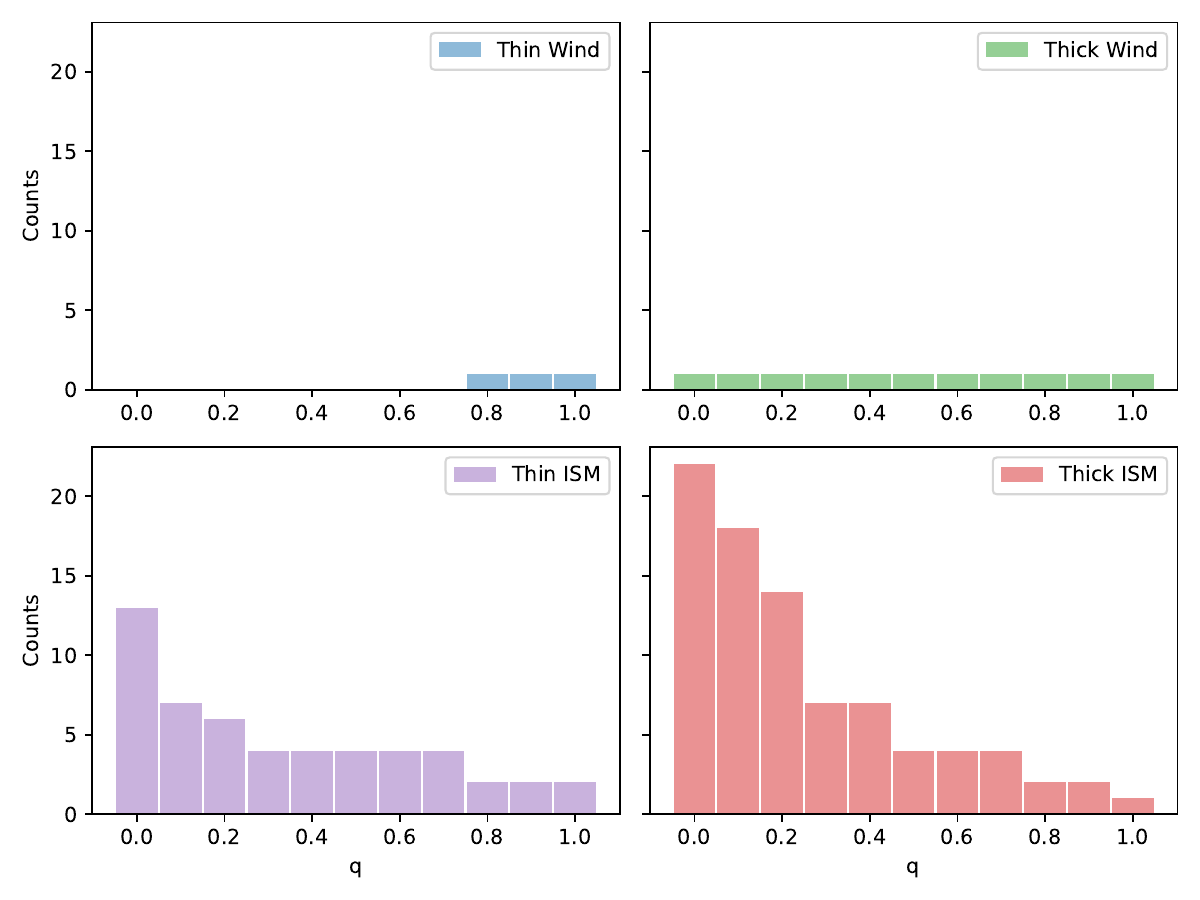}}
}
\caption{Histogram of events as a function of energy injection parameter ($q$).  We consider the spectral and temporal indexes modeled with a PL function from 2FLGC, the SSC CRs from the reverse shock evolving in the thick- and thin-shell case during the adiabatic scenario ($\epsilon=0$) and the deceleration of the outflow in a stellar wind and constant density medium.}
 \label{fig9:Counts_PL}
\end{figure}

\begin{figure}
{ \centering
\resizebox*{\textwidth}{0.6\textheight}
{\includegraphics{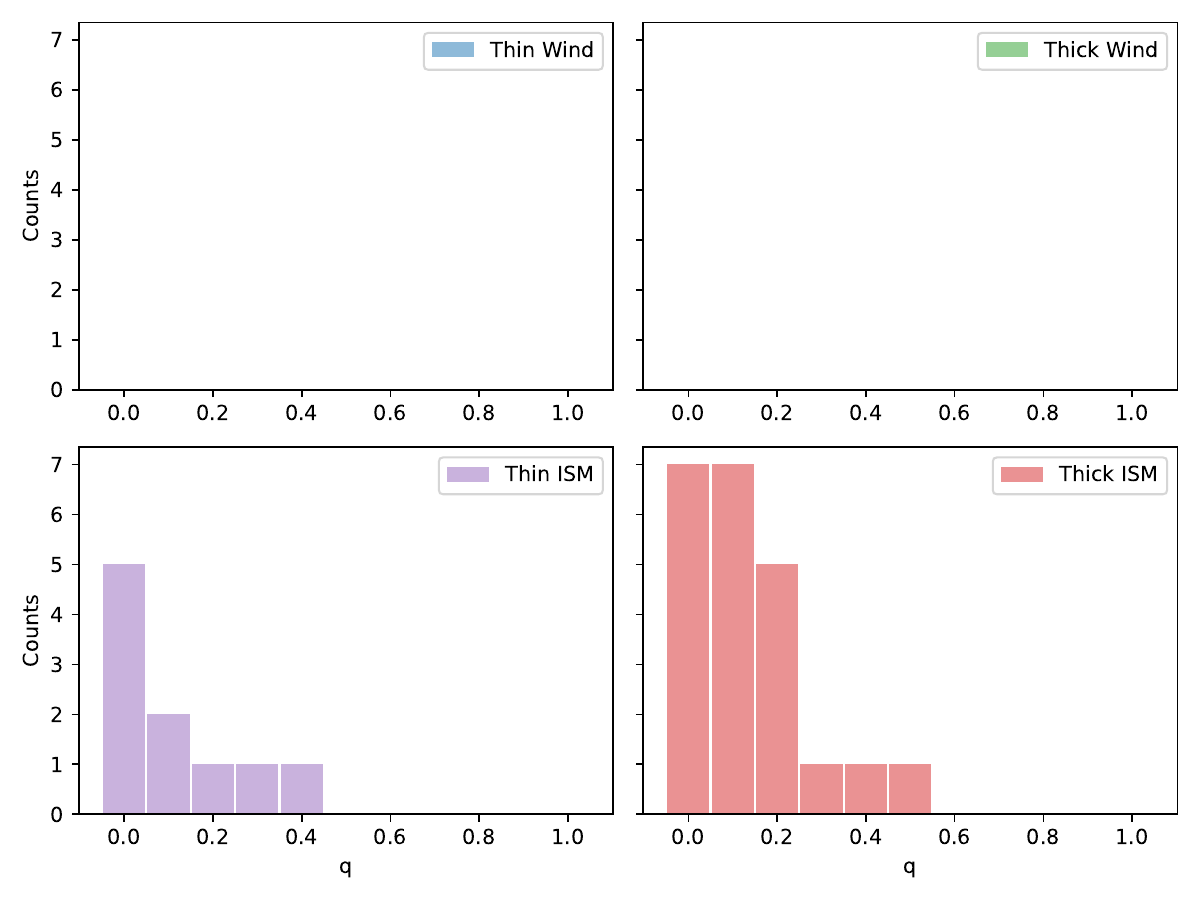}}
}
\caption{The same as Figure \ref{fig9:Counts_PL},  but considering the spectral and temporal indexes modeled with a BPL function from 2FLGC.}
 \label{fig10:Counts_BPL}
\end{figure}


\begin{figure}
{ \centering
\resizebox*{\textwidth}{0.935\textheight}
{\includegraphics{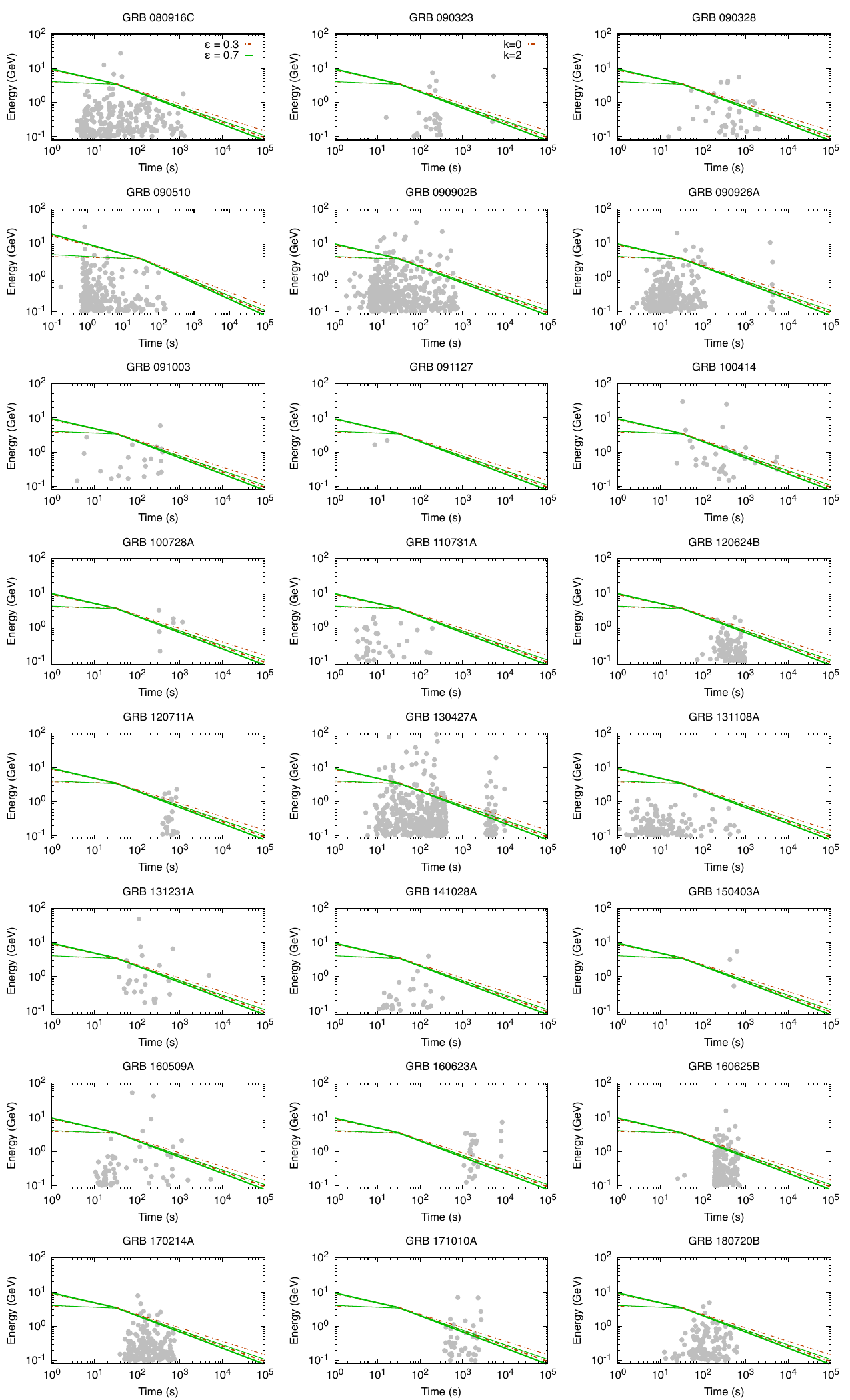}}}
\caption{Each burst in our sample is associated with high-energy photons above 100 MeV and a probability of $>90$\% to be associated. The maximal photon energies released by the synchrotron reverse-shock scenario during the radiative regime with $\epsilon=0.7$ and $\epsilon=0.3$. The reverse shock  that evolves in the thick-shell regime is considered for stellar ($A_{\rm W}=10^{-2}$) wind and ISM (${\rm n=1\,cm^{-3}}$). The values used are $E_0= 2\times 10^{53}\,{\rm erg}$, $\Gamma_0=10^{2.5}$, $\Delta=10^{11.8}\,{\rm cm}$ and $z=0.2$. Data sets from the Fermi-LAT instrument were obtained from the publicly accessible database at the Fermi website.} 
\label{fig:max_eps_thick}
\end{figure}


\begin{figure}
{ \centering
\resizebox*{\textwidth}{0.965\textheight}
{\includegraphics{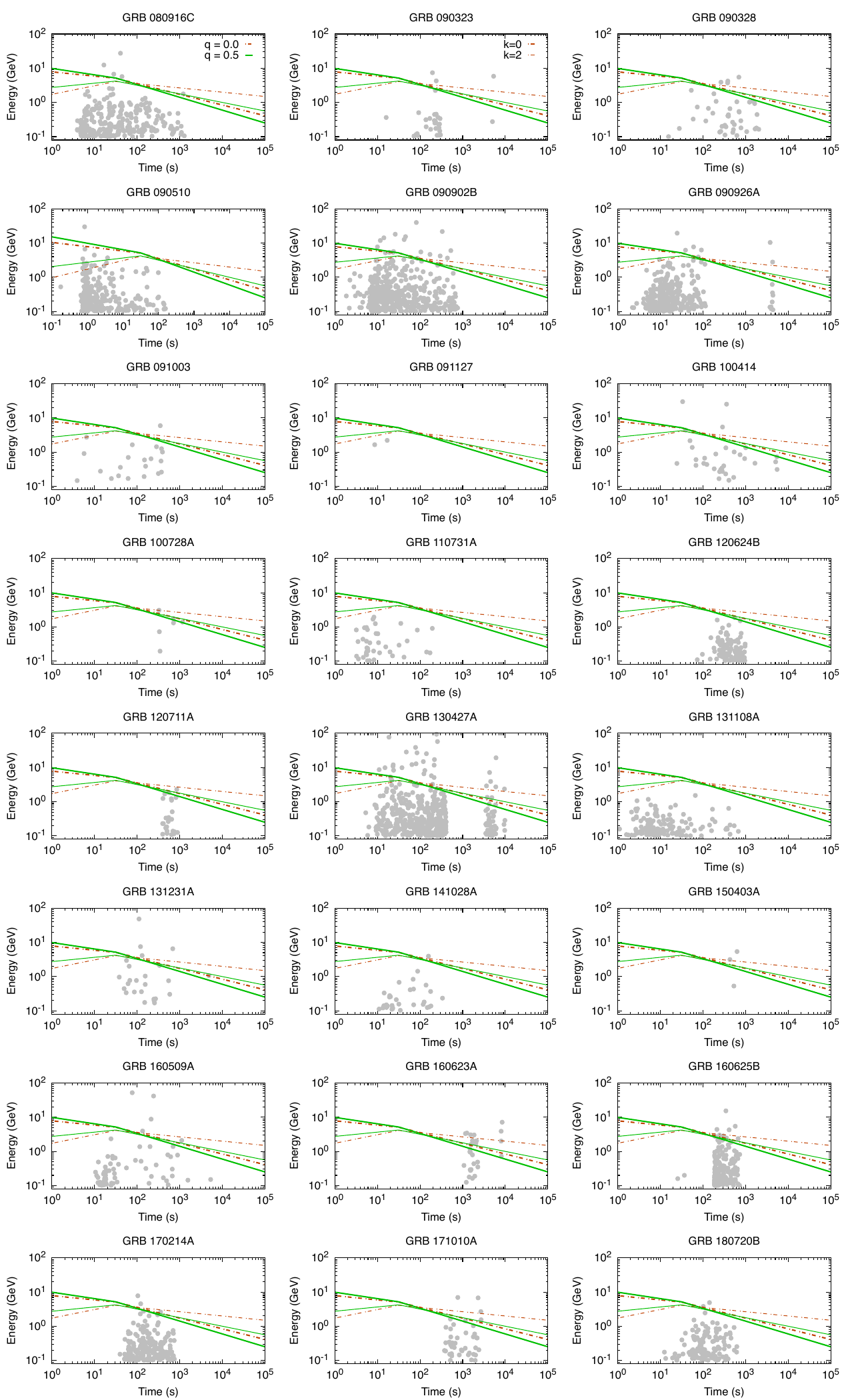}}
}
\caption{The same as Figure \ref{fig:max_eps_thick} but with energy injection ($q=0.5$ and $q=0$) during the adiabatic regime ($\epsilon=0$). The reverse shock  that evolves in the thick-shell regime is considered for stellar ($A_{\rm W}=10^{-2}$) wind and ISM (${\rm n=10^{-1}\,cm^{-3}}$). The values used are $E_0= 2\times 10^{53}\,{\rm erg}$, $\Gamma_0=10^{2.5}$, $\Delta=10^{11.8}\,{\rm cm}$ and $z=0.2$.}
 \label{fig:max_q_thick}
\end{figure}

\begin{figure}
{ \centering
\resizebox*{\textwidth}{0.75\textheight}
{\includegraphics{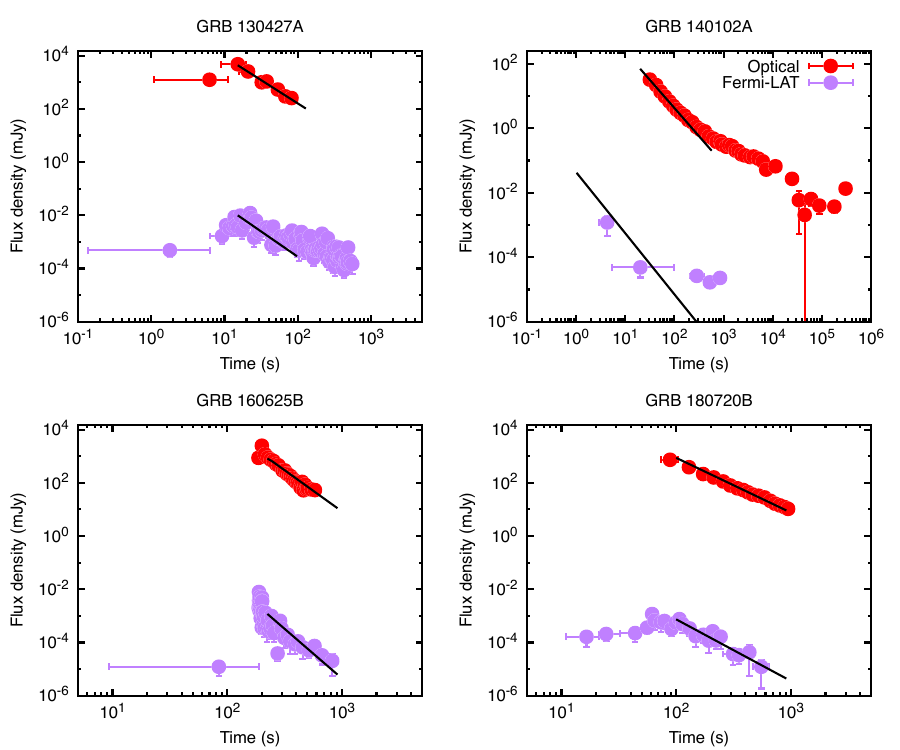}}
}
\caption{Sample of simultaneous \textit{Fermi}-LAT (purple) and early-time optical (red) observations for GRB 130427A, GRB 140102A, GRB 160625B, and GRB 180720B. The solid lines show the best-fit models, where the LAT emission is described by the SSC reverse-shock component and the optical emission by the synchrotron reverse-shock component.}
 \label{fig:othergrbs}
\end{figure}



\clearpage
\newpage

\bibliography{sample701}{}
\bibliographystyle{aasjournalv7}



\end{document}